\documentclass{article}
\usepackage{arxiv}

\usepackage[utf8]{inputenc} 
\usepackage[T1]{fontenc}    
\usepackage[hidelinks]{hyperref}     
\usepackage{url}            
\usepackage{booktabs}       
\usepackage{amsfonts}       
\usepackage{nicefrac}       
\usepackage{microtype}      
\usepackage{lipsum}
\usepackage{bm}
\usepackage{graphicx}
\usepackage{subcaption}
\usepackage{amsmath}
\usepackage{placeins}
\graphicspath{ {./images/} }

\title{Singularities at the vertex of connected angular inhomogeneities under thermal and elastic loading}

\author{
 Yuanpeng Yang \\
  Shanghai Institute of Applied Mathematics \\and Mechanics
  ,Shanghai University\\
  Shanghai, 200044 \\
  \texttt{3142270895@shu.edu.cn} \\
  \And
  Huiming Yin \\
  Department of Civil Engineering and \\Engineering Mechanics,
  Columbia University\\
  New York, NY, 10027 \\
  \texttt{yin@civil.columbia.edu} \\
    \And
  Chunlin Wu \thanks{Corresponding Author} \\
  Shanghai Institute of Applied Mathematics and Mechanics\\
  Shanghai University\\
  Shanghai, 200044 \\
  \texttt{chunlinwu@shu.edu.cn} \\
}

\begin{document}
\maketitle
\begin{abstract}
    This paper investigates the singularities at the vertex of multiply connected angular inhomogeneities for heat conduction and elastic deformation. With the aid of Eshelby's equivalent inclusion method (EIM), each inhomogeneity is simulated as an equivalent inclusion, exhibiting the same material properties as the matrix but containing a continuously distributed eigen-field with potential singularities at the vertices and edge lines. Specifically, the eigen-temperature-gradient (ETG) and eigenstrain are utilized to simulate material mismatch of thermal conductivity and stiffness, respectively. Using the separation of variables, the eigen-fields can be formulated in terms of distance to vertices and opening angles, and disturbed thermal/elastic fields are evaluated by domain integrals of Green's function multiplied by eigen-fields, which form Fredholm's integral equation of the second kind. The boundary value problem is reduced to solve for eigenvalues, which are used to determine the order  of singularity. The present solution is versatile - by placing two identical inhomogeneities together, it recovers the classic solutions for a single wedge in a bimaterial media or infinite domain.  The general and analytical formulae take full consideration of interactions of multiple inhomogeneities and reveal the effects of opening angles and material properties on the thermal and elastic singularities. 
\end{abstract}


\section{Introduction}
Singularities may form in the neighborhood of the vertex of a wedge for heat transfer or elastic problems. When a concentrated load is applied in the infinite medium, the stress/flux exhibit $r^{-2}$ and $r^{-1}$ singular distributions for three-dimensional and two-dimensional problems \cite{Barber1992,Evans2010}, respectively. Extensive efforts have been devoted to derive fundamental solutions in half-spaces \cite{Mindlin1936, Hwu2023}, bimaterial \cite{Walpole1996, Zhou2020}, and multi-layered media \cite{Yue1995}, which revealed that the material interfaces strongly affect the distribution of fields and singularity order. In contrast, when the field point coincides with an interface, the order of singularity changes, as analyzed by Bogy for edge-bonded bimaterial wedges \cite{Bogy1968}. In general, the singularity order depends on  material mismatch (Dundurs parameters) \cite{dundurs1969discussion,Wu2024}, geometric shapes \cite{Dempsey1979}, and loading conditions \cite{Dundurs1989}. The analytical solutions not only provide a systematic understanding of singular behaviors, but also serve as the reference for the development of numerical methods with enriched interpolations \cite{Sukumar2004,Cheng2014,Huang2020}.

In the literature, stress and flux singularities have been extensively studied. Williams \cite{Williams1952} first combined Airy's stress function and separation of variables, where the elastic fields are expressed in terms of distance (to the tip) and angles. Williams successfully identified the celebrated $1 / \sqrt{r}$ singularity of the crack tip and extended the solution for a wedge with various geometric settings. Subsequently, Williams \cite{Williams1956} reached the same conclusion through the complex variable approach. The two pioneering works paved the way for subsequent research on complicated geometries. For instance, Dempsey and Sinclair \cite{Dempsey1979} further extended Williams' approach on a composite wedge. Specifically, the authors proposed a general scheme to consider four types of continuity equations among wedges, including the slipping and bonded cases. Moreover, Dempsey and Sinclair \cite{dempsey1981singular} applied their general scheme in a bimaterial wedge problem, and the authors concluded the sufficient and necessary conditions for logarithmic singularities. Bogy \cite{Bogy1968} employed the Mellin transform to investigate the classic bimaterial wedge, and Bogy summarized that the singularity level will be dependent on the extent of material mismatches, i.e., the ratio of shear moduli. Following Bogy's work, Dundurs \cite{dundurs1969discussion} pointed out that the singularity level of the bimaterial wedge is governed by two dimensionless parameters, namely the Dundurs parameters. 

Regarding heat flux singularities, Sih \cite{Sih1962} first analyzed heat flux singularities of an opening crack using the complex variable approach, and the author concluded that the heat flow does not introduce additional singularities. Using the separation of variables, Chen and Huang \cite{Chen1992} investigated the heat flux singularity of the crack tip in a bimaterial wedge, where the crack is subjected to prescribed temperature or insulated boundary conditions. They established two key results: (i) unlike the bimaterial wedge problem in elasticity, the heat flux singularity parameter is real, which does not exhibit oscillatory features; (ii) the singularity order of the crack (parallel to the bimaterial interface) remains constant, and it is entirely independent on the ratio of thermal conductivity between two wedge components, which is confirmed and extended in the current paper for a single void as the inhomogeneity in a bimaterial matrix. Mantic et al. \cite{Mantic2003} followed \cite{Clements1981} and analyzed heat flux singularities of anisotropic multi-connected corners, in which the authors employed an integral transform to address the orthotropic thermal conductivity. In addition to analytical methods, integral methods have been applied to quantify singularities, such as the celebrated J-integral \cite{Rice1968}, and the heat flux integration method \cite{Atluri1986}. 

While the above papers studied the singularities of wedges, the singularities of angular inhomogeneities are of practical significance and can be solved by Eshelby's equivalent inclusion method (EIM) \cite{Eshelby1957},  which replaces inhomogeneities by the matrix containing continuously distributed eigen-fields to simulate the material mismatch. Treating cracks as special inhomogeneities with zero stiffness, Mura \cite{Mura1987}  simulated slit-like and penny-shaped cracks by adjusting the aspect ratio of the three semi-axes of the ellipsoid. Although Mura successfully recovered the $r^{-1/2}$ singularity of the slit-like crack, varying the aspect ratios does not change the properties of classic Eshelby's tensor. Specifically, Eshelby's tensor is constant over the ellipsoidal inclusion; therefore, even for the slit-like crack, the eigenstrain exhibits uniform distribution. This strategy, adjusting aspect ratio of axes, is of limited application, which can seldom be extended to angular inhomogeneities because: (i) true corners cannot be simulated by quadratic/smooth geometry of an ellipse/ellipsoid; (ii) Eshelby's tensors of polygonal/polyhedral inclusions are no longer constant as ellipsoidal inclusions \cite{Rodin1996,Wu2021a,Wu2021b}. Consequently, accurate analysis of angular inhomogeneities requires more effort in addition to the classic Eshelby's framework for ellipsoidal inhomogeneities. 

The subsequent extension of Eshelby's framework to various shapes requires two aspects: (i) the proper assumption of eigen-fields; and (ii) domain integrals of Green's function multiplied by eigen-fields over the equivalent inclusion. Many studies assume a uniform eigen-field, and extensive efforts have been devoted to addressing domain integrals of Green's function over inclusions. Some works focus on special geometries, such as the cuboid \cite{Chiu1980}, the Star of David \cite{Mura1997}, or the weakly non-circular case \cite{Huang2009}. Rodin \cite{Rodin1996} constructed transformed coordinates based on the observing point and its projected plane, and the author derived Eshelby's tensors for polygonal/polyhedral inclusions. To address the weakly singular integral, Nozaki and Taya \cite{Nozaki1997, Nozaki2000} proposed to create a unit circular circumference around the field point, where the integral limits can be well defined by connecting the field point and the vertices of polygons. Considering the lengthy expressions in Rodin's work, Rosati's group \cite{Trotta2017, Trotta2018} derived Eshelby's tensors only based on the vertices of polygonal/polyhedral inclusions and the field point. However, the above works are based on the uniform distribution of the eigen-field. Unfortunately, for actual angular inhomogeneities, the eigen-field is not uniform. 

To approximate actual distributions of eigen-fields, Liu and Gao \cite{Liu2013} derived strain-gradient Eshelby's tensors for polygonal inclusions under transformed coordinate system. Subsequently, Wu et al. \cite{Wu2021a, Wu2021b} derived Eshelby's tensors for polygonal and polyhedral inclusions containing polynomial-form eigenstrain. Despite that higher-order polynomials of eigenstrain can improve the accuracy of predictions, because polynomials are smooth function while the actual eigenstrain is singular, obvious discrepancies near vertices were observed, see triangular inhomogeneity \cite{Wu2021a} and tetrahedral inhomogeneity \cite{Wu2021b}. Recently, Wu and Yin \cite{Wu2024} employed Airy's stress function with separation of variables, expressing eigenstrain in separable functions of radius and angle. By enforcing the equivalent stress conditions, the eigenstrain is governed by Fredholm's integral equation of the second kind, and the singularity parameter is determined by solving the linear eigenvalue problem. The analysis demonstrates that: (i) the eigenstrain exhibits $r^{-\lambda}$ (with possible $\ln r$ factors) singularity; and (ii) the stress fields, including exterior and interior parts, inherit the same singularity level as the eigenstrain. Consequently, this work reveals that using polynomial-form eigenstrain alone can seldom capture the drastic variations in the vicinity of vertices. 

This paper extends Eshelby's EIM framework to more general configurations, and achieved three contributions on singular behaviors of inhomogeneity problems in plane-strain elasticity and steady-state heat conduction: (i) it derives the formulation for interactions among multiple connected triangular inhomogeneities, which does not require setting up auxiliary continuity conditions and solve the corresponding boundary value problem; and (ii) it provides a straightforward and simple framework to investigate inhomogeneities embedded in the half-space and bimaterial media through adjusting of opening angles and material properties, which retains the advantage of infinite Green's function and avoids switching to half-space/bimaterial Green's functions; and (iii) it rigorously proves that the both interior/exterior stress/strain fields must share the same singularity parameter as the eigenstrain, and analogous conclusion holds for temperature gradient/flux and the eigen-temperature-gradient (ETG). To the best of the authors' knowledge, it is the first work in the literature to systematically investigate the singular behaviors of inhomogeneities embedded in the half-space and bimaterial media within the EIM framework. 

In the following, Section 2 formulates the boundary value problem (BVP) and states interface continuity equations along the interfaces. Section 3 provides a brief review of Eshelby's equivalent inclusion method in steady-state heat conduction, and uses the separation of variables to obtain thermal fields and ETG. Domain integrals of Green's function and ETG, interactions between inhomogeneities are evaluated with explicit formulae. In parallel to the thermal analysis, Section 4 investigates the elastic problem and derives the explicit formulae as well. Section 5 verifies our formulation against classic solutions of single-material/bimaterial wedges in heat conduction and elasticity. Section 6 applies the formulae to analyze singularities for inhomogeneities embedded in a bimaterial medium, including inhomogeneities oriented parallel and perpendicular to the bimaterial interface. Finally, some conclusive remarks are provided in Section 7.

\section{Problem statement}
\begin{figure}
    \centering
    \includegraphics[width=0.6\linewidth,keepaspectratio]{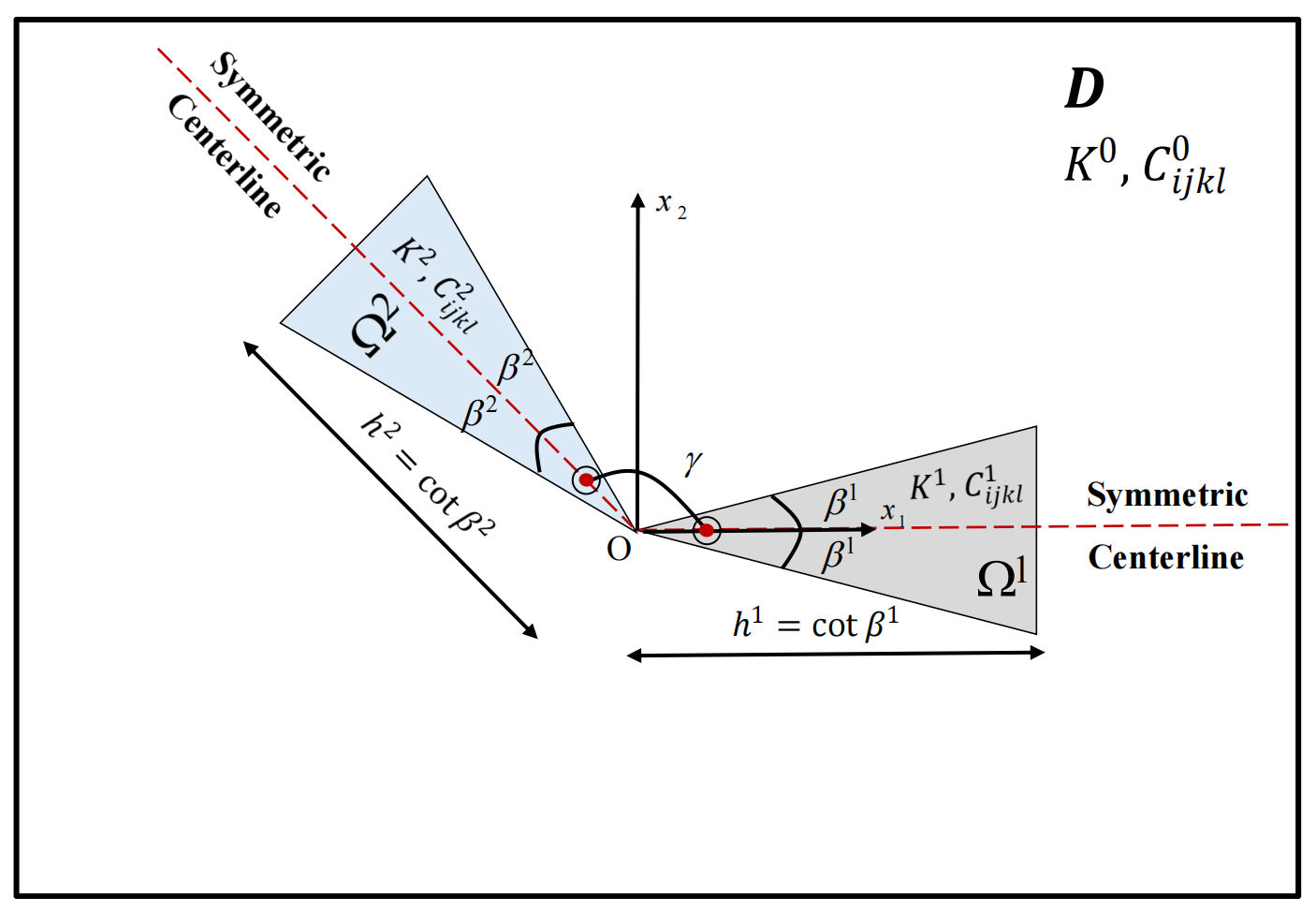}
    \caption{Schematic illustration of an infinite domain $\mathcal{D}$ embedded with two isosceles triangular inhomogeneities ($\Omega^1, \Omega^2)$ with opening angles $2 \beta^1, 2\beta^2$ and height $h^1, h^2$, respectively. The angle between symmetric lines of two subdomains is $\gamma$. The matrix exhibits thermal conductivity $K^0$ and $C_{ijkl}^0$, while subdomains exhibits dissimilar properties $K^I, C^I_{ijkl}$. }
    \label{fig:problem}
\end{figure}

Consider an infinite, homogeneous, isotropic domain $\mathcal{D}$ embedded with multiple triangular cylindrical inhomogeneities, whose cross-sections lie in the $ x_1-x_2$ plane. Fig. \ref{fig:problem} plots two isosceles triangular subdomains ($\Omega^1$ and $\Omega^2$), which are characterized by their width $b^I$, opening angles $2 \beta^I$, heights $h^I$, and their symmetric lines. Each triangular subdomain shares a common vertex at the origin of the Cartesian coordinate system. The symmetric line of the $\Omega^1$ coincides with the $x_1$ axis, and the angle between the symmetric lines of the first and second subdomain is denoted as $\gamma$. 

In general, the matrix ($\mathcal{D} - \sum_{I=1}^{N_I} \Omega^I$) and each triangular subdomain $\Omega^I$ exhibit different thermal conductivity and stiffness. Specifically, the $I^{th}$ subdomain exhibits thermal conductivity $K^I$ and stiffness $C_{ijkl}^I$, while the matrix exhibits thermal conductivity $K^0$ and stiffness $C_{ijkl}^0$. For isotropic stiffness, $C^I_{ijkl} = \lambda^I \delta_{ij} \delta_{kl} + \mu^I (\delta_{il} \delta_{jk} + \delta_{ik} \delta_{jl})$, where $\lambda^I$ and $\mu^I$ are two lam\`e constants of the $I^{th}$ subdomain, while $\nu^I$ represents Poisson's ratio of the $I^{th}$ subdomain. 

This paper assumes perfect-bonding interfaces among the matrix and each subdomain. therefore, interface continuity conditions on temperature/flux and displacement/stress satisfy the following:

\begin{equation}
\begin{aligned}
    &T(\textbf{x}^+) = T(\textbf{x}^-) \quad \text{and} \quad q_i(\textbf{x}^+) n_i^+ = q(\textbf{x}^-) n_i^-, \quad \text{On $\partial \Omega^I$}, \quad \text{Thermal} \\ 
    &u_i(\textbf{x}^+) = u_i(\textbf{x}^-) \quad \text{and} \quad \sigma_{ij}(\textbf{x}^+) n_j^+ = \sigma_{ij}(\textbf{x}^-) n_j^-, \quad \text{On $\partial \Omega^I$}, \quad \text{Elastic}
\end{aligned}
    \label{eq:continuity}
\end{equation}
where $\partial \Omega^I$ represents the interface between the $I^{th}$ inhomogeneity and the matrix; the ``$\pm$'' refers to the outer and inner surface of the interface  $\partial \Omega^I$, respectively; $\textbf{n}$ indicates the interfacial unit normal vector of the surface; $T$ and $\textbf{q}$ refer to temperature and heat flux, respectively; $\textbf{u}$ and $\bm{\sigma}$ stand for displacement and stress, respectively. Based on the constitutive relations, $q_i = -K T_{,i}$ and $\sigma_{ij} = C_{ijkl} \varepsilon_{kl}$, where $T_{,i}$ and $\varepsilon_{ij}$ are temperature gradient and strain, respectively. Note that Eq. (\ref{eq:continuity}) formulates the boundary value problem (BVP), which provides auxiliary equations to determine unknown coefficients. 
Alternatively, this paper employs Eshelby's equivalent inclusion method to simulate inhomogeneities, in which eigen-fields are source terms in the governing equation. Hence, the above interface continuity conditions are satisfied without introducing additional interface constraints.

Since this paper aims to convey Eshelby's equivalent inclusion method for the determination of singularities, only cylindrical inhomogeneities (plane strain) are considered, which can be straightforwardly extended for plane stress problems through modification of material constants. Although Fig. \ref{fig:problem} and subsequent derivations only cover two inhomogeneities, the formulation has been generalized to multiple inhomogeneities. Section 3 and Section 4 discuss the extensions in detail, in which the interactions among inhomogeneities can be acquired by switching material properties, opening angles, and angles between symmetric lines. 

Based on the dual equivalent inclusion method (DEIM) \cite{Wu2023_IJSS}, the thermoelastic influence induced by the mismatch of thermal conductivity can be handled by domain integrals of the first-order partial derivatives of the thermoelastic Green's function multiplied by ETG. Because the thermoelastic Green's function is biharmonic and the singularity level of ETG is less than $r^{-1}$, their domain integrals cannot introduce any new singularity into the composite system. Therefore, the heat flux singularity and stress singularity can be analyzed separately in Sections 3 and 4, respectively.

\section{Formulation of heat flux singularity}
\subsection{Governing equation and fundamental solution}
The heat equation governs steady-state heat conduction in the domain $\mathcal{D}$ that (in the absence of heat source), 

\begin{equation}
    -\frac{\partial}{\partial x_i} \left[ K(\textbf{x}) \thinspace  T_{,i}(\textbf{x}) \right] = 0
    \label{eq:heat_gov_ori}
\end{equation}
where $K(\textbf{x})$ refers to position-dependent thermal conductivity. Because this paper utilizes eigen-temperature-gradient (ETG) to simulate mismatch of thermal conductivity, Eq. (\ref{eq:heat_gov_ori}) can be rewritten using the matrix conductivity together with ETG, 

\begin{equation}
    -K^0 \frac{\partial}{\partial x_i} \left[ T_{,i}(\textbf{x}) - T_{i}^*(\textbf{x}) \right] = 0 \quad \text{or equivalently} \quad -K^0 T_{,ii}(\textbf{x})= - K^0 T_{i,i}^*(\textbf{x}) 
    \label{eq:heat_gov}
\end{equation}
where $K^0$ is the thermal conductivity of the matrix, and $T_{i}^*$ refers to position-dependent eigen-temperature-gradient (ETG). Based on the concept of Green's function, the temperature $T(\textbf{x})$ at the field point can be obtained through domain integrals of Green's function multiplied by the ETG,

\begin{equation}
\begin{aligned}
    T(\textbf{x}) & = -K^0\sum_{I=1}^{N_I} \int_{\Omega_I} G_{,i}(\textbf{x}, \textbf{x}') T_{i}^{I*}(\textbf{x}') \thinspace d\textbf{x}' = -\frac{K^0}{4 \pi} \sum_{J=1}^{N^I} \int_{\Omega^I} \phi_{,n} T_{i}^{I*} \thinspace d\textbf{x}' 
\end{aligned}
    \label{eq:temp_disturb}
\end{equation}
where ETG is zero over the matrix, and $T^{I*}_{i}$ is the continuous ETG distributed within the $I^{th}$ subdomain $\Omega^I$; and $G(\textbf{x}, \textbf{x}')$ is Green's function for the steady-state heat conduction,

\begin{equation}
    G(\textbf{x}, \textbf{x}') = \frac{1}{4\pi K^0} \phi, \quad \phi
    = -\ln |\textbf{x} - \textbf{x}'|^2
    \label{eq:thermal_green}
\end{equation}

When ETGs are uniform over subdomains, ETGs factor out of the domain integrals in Eq. (\ref{eq:temp_disturb}), and thus the remaining geometry-dependent integral becomes the uniform Eshelby's tensor. As pointed by Rodin \cite{Rodin1996} on polygonal inclusions, a uniform eigen-field can lead to $\ln r$ singular behaviors for temperature gradients near vertices. However, the following derivations show that (i) the ETG cannot be uniform for angular inhomogeneities; and (ii) the ETG is singular in the vicinity of the vertex, which is substantially different from the singularities of polygonal inclusions with uniform eigen-fields \cite{Rodin1996, Trotta2017, Liu2013}. 

\subsection{Equivalent flux condition}
The ETGs are determined by the equivalent flux condition over each subdomain $\Omega^I$ \cite{Hatta1986}, 
\begin{equation}
    -K^0 \left( T_{,i} - T_i^{I*} \right) = -K^I T_{,i}
    \label{eq:equiv_flux}
\end{equation}

Note that the EIM on steady-state heat conduction is analogous to Eshelby's original problem in elasticity, which shares similar features of Eshelby's tensors. For instance, for an ellipsoidal inclusion with a uniform eigen-field, Eshelby's tensors are uniform for interior points. When an ellipsoidal inhomogeneity is subjected to a uniform temperature gradient, the uniform ETG induces uniform disturbed temperature gradients, which satisfies Eq. (\ref{eq:equiv_flux}) at every interior point. Although polynomial-form or piece-wise continuous ETG can provide acceptable predictions away from the vertex, they cannot satisfy the equivalent flux conditions exactly at the vertex. Following Chen and Huang \cite{Chen1992}, this section conducts an asymptotic analysis of heat flux and achieves flux equivalence with the analytical solution. 

For the completeness of the derivation, the essential steps are briefly provided below. With the separation of variables, the temperature can be expressed as functions of the distance $R = \sqrt{x_1^2 + x_2^2}$ and angle $\theta = \tan^{-1} \left[\frac{x_2}{x_1} \right]$. Based on the heat equation in Eq. (\ref{eq:heat_gov_ori}), the temperature $T^I$ in the $I^{th}$ subdomain can be written as,

\begin{equation}
    T^I(r, \theta)=  R^{1-m} \left[f_1^I \cos(1-m) \theta + f_2^I \sin(1-m)\theta \right]
    \label{eq:temp-field}
\end{equation}
where $f_1^I$ and $f_2^I$ are two coefficients for the $I^{th}$ subdomain; and $m$ refers to the singularity parameter. The temperature gradient in the Cartesian coordinate can be written as,

\begin{equation}
\begin{split}
    T^I_{,1} & = R^{-m} (1-m)\Big(f_1^I \cos m\theta - f_2^I \sin m\theta\Big) \\ T_{,2}^I &= R^{-m} (1-m)\Big(f_1^I \sin m\theta + f_2^I \cos m\theta \Big)
\end{split}
    \label{eq:tempgrad-field}
\end{equation}

Substituting Eq. (\ref{eq:tempgrad-field}) into Eq. (\ref{eq:equiv_flux}), the ETG within the $I^{th}$ subdomain can be expressed in terms of temperature gradients, 

\begin{equation}
\begin{split}
    T^{I*}_{1} = R^{-m}\frac{K^0-K^I}{K^0}(1-m)\Big(f_1^I \cos m\theta - f_2^I \sin m\theta\Big)\\ 
    T^{I*}_{2} = R^{-m}\frac{K^0-K^I}{K^0}(1-m)\Big(f_1^I \sin m\theta + f_2^I \cos m\theta \Big)
    \label{eq:ETG}
    \end{split}
\end{equation}
where the ETG to simulate thermal conductivity mismatch generally exhibits the same singularity level as the temperature gradient. Keep in mind that even under two extreme cases, the void $K^0 / K^1 = \infty$ or $0$, the ETG in Eq. (\ref{eq:ETG}) is non-zero. Therefore, the original problem is now converted into searching for a proper singularity parameter $m$, which enables the temperature gradient and ETG to share the same singularity in the vicinity of the vertex. 

The ETG can be expressed in terms of temperature gradients in Eq. (\ref{eq:equiv_flux}). Differentiating Eq. (\ref{eq:temp_disturb}), the temperature gradients can be written as follows 

\begin{equation}
    \thinspace T_{,i}(\textbf{x}) = -K^0 \sum_{J=1}^{N^I} \int_{\Omega^J} G_{,ni}(\textbf{x}, \textbf{x}') T^{J*}_{n}(\textbf{x}') \thinspace d\textbf{x}' = -\frac{1}{4\pi} \sum_{J=1}^{N^I} \int_{\Omega^J} \phi_{,ni} \thinspace T_{n}^{J*} \thinspace d\textbf{x}'
    \label{eq:flux_int}
\end{equation}
where the temperature gradients contain disturbances from each subdomain. Note that each subdomain contains two unknown coefficients and that equivalent flux conditions can provide two equations. Although the temperature gradient and ETG are given in Eq. (\ref{eq:tempgrad-field}) and Eq. (\ref{eq:ETG}), respectively, the singularity parameter depends on Eq. (\ref{eq:flux_int}). Collecting equivalent flux conditions and unknown coefficients from each subdomain, a global system of linear equations ($2 \times N^I$ unknowns) can be constructed formally,

\begin{equation}
    R^{-m} \begin{bmatrix} A^1_{1} & A^1_{2} & \ldots & A^1_{2N^I-1} & A^1_{2 N^I} \\ A^2_{1} & A^2_{2} & \ldots & A^2_{2N^I-1} & A^2_{2 N^I} \\ \vdots & \vdots & \vdots & \vdots & \vdots \\ A^{2 N^I - 1}_{1} & A^{2 N^I - 1}_{2} & \ldots & A^{2 N^I -1}_{2 N^I - 1} & A^{2 N^I - 1}_{2 N^I} \\ A^{2 N^I}_{1} & A^{2 N^I}_{2} & \ldots & A^{2 N^I}_{2 N^I - 1} & A^{2 N^I}_{2 N^I} \end{bmatrix} \begin{bmatrix} f_1^1 \\ f_2^1 \\ \vdots \\ f_1^{N^I}  \\ f_2^{N^I} \end{bmatrix} = \begin{bmatrix} 0 \\ 0 \\ \vdots \\ 0 \\ 0 \end{bmatrix}    
    \label{eq:global_thermal}
\end{equation}
where the matrix is composed of $2 N^I \times 2 N^I$ coefficients $A^{i}_{j}$, and the superscript $i$ and subscript $j$ refers to the $i^{th}$ row and $j^{th}$ column, respectively. For the ease of expressions, the singular part $R^{-m}$ is separated. Keep in mind that coefficients $A^i_j$ are symbolic expressions here, which contain the singularity parameter $m$. The evaluations of $A^i_j$ through domain integrals of Green's function and ETG are conducted in the following subsection.

\subsection{Disturbances of the power-form eigen-temperature-gradient}
Based on the singularity of the second-order partial derivative of Green's function, the evaluation of disturbances should be classified into two categories: (i) the disturbances in the $I^{th}$ subdomain caused by its own ETG, namely the "diagonal coefficient"; and (ii) the disturbances in the $I^{th}$ subdomain caused by ETG over the $J^{th}$ subdomain ($I \neq J$), known as "off-diagonal coefficients". The two types of components will be evaluated in turn. 

Note that the domain integrals are dependent on the positions of the field point and coordinates, as well as the opening angles of the subdomain. Without the loss of any generality, set the local coordinate of each triangular subdomain. For instance, Fig. \ref{fig:problem} plots the local coordinate of the first subdomain, (i) the origin of the local coordinate is set as $(0, 0)$ (same as the vertex); and (ii) the local basis vector $e_1'$ is aligned with the symmetric line (from the origin to the base), and the other component $e_2'$ is perpendicular to $e_1'$ following the right-hand rule. Following Moschovidis and Mura's scheme on differently oriented ellipsoidal inhomogeneity\cite{Moschovidis1975}, for each subdomain $\Omega^I$, equations of row ($2 I - 1$ and $2 I$) are grouped and constructed in the local coordinate. Consequently, this setting avoids the need for repeated evaluations of diagonal coefficients. 

\subsubsection{Evaluation of diagonal coefficients}
The diagonal coefficients should be interpreted as $2 \times 2$ block, for instance, $A_{2I-1}^{2I-1}, A_{2I-1}^{2I}, A_{2I}^{2I-1}, A_{2I}^{2I}$. As mentioned above, since equations are constructed per local coordinate of each subdomain, only a one-time evaluation is required. Without the loss of any generality, the following derivation is based on the first subdomain $\Omega^1$ with the opening angle $2 \beta^1$. 

Since the equivalent flux condition is constructed at the vertex, it is natural to evaluate domain integrals for an interior field point along the symmetric line near the vertex, i.e., ($0^+, 0$). However, it is challenging to evaluate such domain integrals due to contributions from the Dirac Delta function of Eshelby's tensors. Recently, Wu and Yin \cite{Wu2024} provided an alternative integral strategy, (i) evaluate domain integrals at the exterior field point ($0^-$, 0), as no Dirac delta function is involved; and (ii) superpose the results with the difference between Eshelby's tensor of the interior/exterior parts of the vertex. \textit{In analogy with the elastic problem, a complete derivation using the transformed coordinate is given in \ref{sec:dif_eshelby}.} 

Note that the following domain integrals are conducted along the symmetric line, $\textbf{x} = (x_1, 0)$ ($x_1 \to 0^-$), which do not contain the contribution by the Dirac Delta function in Eq. (\ref{eq:Eshelby tensor diff-final}). For field points along the symmetric line, the second-order partial derivatives of Green's function are provided below:

\begin{small}
\begin{equation}
   G_{,11} = \frac{r^2\cos{2\theta} - 2r\cos{\theta}x_1 + x_1^2}{2\pi K^0 \left(r^2 - 2r\cos{\theta}x_1 + x_1^2\right)^2}, \quad
   G_{,12}=  \frac{r\sin{\theta}\left(r\cos{\theta} - x_1\right)}{\pi  K^0 \left(r^2 - 2r\cos{\theta}x_1 + x_1^2\right)^2}, \quad
   G_{,22} = -\frac{r^2\cos{2\theta} - 2r\cos{\theta}x_1 + x_1^2}{2\pi  K^0 \left(r^2 - 2r\cos{\theta}x_1 + x_1^2\right)^2}
    \label{eq:harmonic potentials}
\end{equation}
\end{small}
where $r = \sqrt{x_i' x_i'}$ refers to the distance between the vertex (origin) and the source point $\textbf{x}' $. Let $\Omega^{1p}$ denote the triangular polar integral domain of the first subdomain. The disturbed temperature gradients within the first subdomain induced by its own ETG ($T_{i}^{1*}$) can be written as:

\begin{small}
\begin{equation}
\begin{split}
    T_{,1}^{1 \prime} = -\frac{1}{4 \pi}\int_{\Omega^{1p}} \phi_{,1j} T^{1*}_{j}\thinspace d\textbf{x}' & = -\int_{-\beta^1}^{\beta^1} \int_0^1 \left( \frac{r^2\cos{2\theta} - 2r\cos{\theta}x_1 + x_1^2}{2\pi\left(r^2 - 2r\cos{\theta}x_1 + x_1^2\right)^2} T_{1}^{1*} + \frac{r\sin{\theta}\left(r\cos{\theta} - x_1\right)}{\pi\left(r^2 - 2r\cos{\theta}x_1 + x_1^2\right)^2} T_{2}^{1*} \right) r \thinspace dr \thinspace d\theta \\ 
    T_{,2}^{1 \prime} = -\frac{1}{4 \pi}\int_{\Omega^{1p}} \phi_{,2j} T^{1*}_{j}\thinspace d\textbf{x}' & = -\int_{-\beta^1}^{\beta^1} \int_0^1 \left( \frac{r\sin{\theta}\left(r\cos{\theta} - x_1\right)}{\pi\left(r^2 - 2r\cos{\theta}x_1 + x_1^2\right)^2} T_{1}^{1*} -\frac{r^2\cos{2\theta} - 2r\cos{\theta}x_1 + x_1^2}{2\pi\left(r^2 - 2r\cos{\theta}x_1 + x_1^2\right)^2} T_{2}^{1*} \right) r \thinspace dr \thinspace d\theta 
\end{split}
\label{eq:diagonal_form}
\end{equation}
\end{small}
where $T_{,i}^{I \prime}$ refers to the disturbed temperature gradient caused by the $I^{th}$ subdomain, and here $I = 1$ is applied for diagonal coefficients. The evaluation of the above domain integrals in Eq. (\ref{eq:diagonal_form}) can be conducted through three steps: (i) integrate the polar radius-related part, i.e., $r^{1-m}$; (ii) conduct series expansion around the vertex, $x_1 \to 0^-$; and (iii) complete the angle integral. 

Since the domain integrals for $T_{,1}^{1'}$ and $T_{,2}^{1'}$ share the similar derivation process, only details of $T_{,1}^{1'}$ is elaborated in the main text, and \textit{the derivation of $T_{,2}^{1'}$ is provided in Appendix B.1.} With some straightforward and lengthy derivation, the integral result in step (i) becomes: 

\begin{small}
\begin{align}
   \frac{1}{4 \pi}\int \phi_{,1j} & T^{1*}_{j}  rdr= -\frac{K^0-K^1}{4\pi K^0 x_1} r^{1 - m} \Bigg\{ \notag  \\&\Bigg(
   - \thinspace_2F_1\left[2,1-m,2-m,\frac{r(\cos{\theta} - i\sin{\theta})}{x_1}\right]\cos{(-1 + m)\theta} - _2F_1\left[2,1-m,2-m,\frac{r(\cos{\theta} + i\sin{\theta})}{x_1}\right]\cos{(-1 + m)\theta} \notag  \\
   &+ \thinspace _2F_1\left[1,1-m,2-m,\frac{r(\cos{\theta} - i\sin{\theta})}{x_1}\right]\Big(\cos{(-1 + m)\theta} - i\sin{(-1 + m)\theta}\Big) \notag  \\
   &+ \thinspace _2F_1\left[1,1-m,2-m,\frac{r(\cos{\theta} + i\sin{\theta})}{x_1}\right]\Big(\cos{(-1 + m)\theta} + i\sin{(-1 + m)\theta}\Big) \notag  \\
   &+ i \thinspace _2F_1\left[2,1-m,2-m,\frac{r(\cos{\theta} - i\sin{\theta})}{x_1}\right] \sin{(-1 + m)\theta} - i\thinspace _2F_1\left[2,1-m,2-m,\frac{r(\cos{\theta} + i\sin{\theta})}{x_1}\right] \sin{(-1 + m)\theta}
   \Bigg) f_1^{1} \notag  \\
   &+\Bigg(
     \thinspace i_2F_1\left[2,1-m,2-m,\frac{r(\cos{\theta} - i\sin{\theta})}{x_1}\right]\cos{(-1 + m)\theta} -  \thinspace i_2F_1\left[2,1-m,2-m,\frac{r(\cos{\theta} + i\sin{\theta})}{x_1}\right]\cos{(-1 + m)\theta} \notag  \\
   &+ \thinspace _2F_1\left[1,1-m,2-m,\frac{r(\cos{\theta} - i\sin{\theta})}{x_1}\right]\Big(-i\cos{(-1 + m)\theta} - \sin{(-1 + m)\theta}\Big) \notag  \\
   &+ i \thinspace _2F_1\left[1,1-m,2-m,\frac{r(\cos{\theta} + i\sin{\theta})}{x_1}\right]\Big(\cos{(-1 + m)\theta]} + i\sin{(-1 + m)\theta}\Big) \notag  \\
   &+ \thinspace _2F_1\left[2,1-m,2-m,\frac{r(\cos{\theta} - i\sin{\theta})}{x_1}\right] \sin{[(-1 + m)\theta]} + \thinspace _2F_1\left[2,1-m,2-m,\frac{r(\cos{\theta} + i\sin{\theta})}{x_1}\right] \sin{(-1 + m)\theta}
   \Bigg) f_2^{1} 
   \Bigg\}
    \label{eq:diagonal_step_1}
    \end{align}
\end{small}
where $_2F_1 \left[.\right]$ denotes the hypergeometric function\cite{Bailey1935} . Because the singularity parameter $m \in (0, 1)$, when $r \to 0$ the integral vanishes, while $r \to 1$ the term $r^{1-m}$ becomes 1. Note that the upper limit of $r$ does not affect on the singular pattern of the disturbed temperature, because $r^{1-m}$ remains finite for a bounded subdomain. And the singularity pattern is dependent on the position of the field point. Conduct the series expansion of Eq. (\ref{eq:diagonal_step_1}) with $x_1 \to 0^-$, the result of step (ii) becomes: 

\begin{small}
\begin{equation}
\begin{split}
   \frac{1}{4 \pi} \int_0^1 \phi_{,1j} & T^{1*}_{j} r dr  = \frac{K^0 - K^1}{2  \pi K^0 m}(m - 1) \Big(\cos{(m - 2) \theta} f_1^1 - \sin{(m - 2) \theta} f_2^1 \Big) + \mathcal{O}(x_1) \\
   &+\Big[ \frac{K^0 - K^1}{2 K^0}  (m - 1)^2\csc m \pi \thinspace  \Big(\sin [ 2 \theta +(\pi -2 \theta ) m_1 ]f_2^1 +  \cos [ 2 \theta +(\pi -2 \theta ) m ] f_1^1\Big)+ \mathcal{O}(x_1^2)\Big]x_1^{-m}\\
    \end{split}
    \label{eq:diagonal_step_2}
\end{equation}
\end{small}
where the integral result in step (ii) contains a non-singular part and a singular part ($x_1^{-m}$). As previously mentioned, the non-singular part is caused by the far-end contribution, i.e., the upper limit of $r$; whereas the singular part results from the close-end contributions, which is is highly related to the singularity parameter of the vertex. Note that the series expansion formulation in Eq. (\ref{eq:diagonal_step_2}) only presents exterior results, and the interior results can be obtained by superposing the difference $D_{ij}^d$ in Eq. (\ref{eq:Eshelby tensor diff-final}), which will be added for equivalent flux conditions. Finally, the angle integral of singular parts in Eq. (\ref{eq:diagonal_step_2}) and Eq. (\ref{eq:diagonal_step_2_ap}) with respect to step(iii) yield two singular disturbed temperature gradients: 

\begin{equation}
\begin{aligned}
   T_{,1}^{1 \prime} &= x_1^{-m} \frac{(K^0 - K^1)}{K^0} (m - 1) \csc m \pi \thinspace \cos [m (\pi - \beta^1) + \beta^1] \sin[(m - 1)\beta^1] f_1^1
   \\
   T_{,2}^{1 \prime} & = -x_1^{-m} \frac{(K^0 - K^1) }{K^0}  (m - 1)\csc m \pi \thinspace \cos [m(\pi - \beta^1) + \beta^1] \sin [(m - 1)\beta^1] f_2^1 
\label{eq:diagonal_final}
\end{aligned}
\end{equation}

\subsubsection{Evaluation of off-diagonal coefficients}
As previously mentioned, the equivalent flux conditions for each subdomain are constructed in its local coordinate system. Accordingly, this subsection employs coordinate transformation to evaluate the disturbances in the $I^{th}$ subdomain caused by ETG in the $J^{th}$ subdomain. Without the loss of any generality, this subsection considers the case of two subdomains, $\Omega^1, \Omega^2$. 
The formulae derived in this subsection depend on the local coordinate of the $I^{th}$ subdomain and angles between two symmetric lines, which can be straightforwardly extended to the case of multiple subdomains. 

In the following, the ETG of the second subdomain is transformed into the local coordinate of the first subdomain. Specifically, the ETG of the second subdomain is written as: 

\begin{equation}
\begin{aligned}
    \overline{T}^{2*}_1 & = -\frac{K^0-K^2}{K^0} (m - 1) \Big(\cos [ (1 - m) \gamma + m \theta] f_2^1 - \sin [(1 - m) \gamma + m\theta] f_2^2 \Big) \\ 
    \overline{T}^{2*}_2 & = -\frac{K^0-K^2}{K^0} (m - 1) \Big(\sin [ (1-m) \gamma + m \theta] f_1^2 + \cos[(1-m) \gamma + m \theta] f_2^2\Big)
\end{aligned}
    \label{eq:ETG_second}
\end{equation}
where $\overline{T}^{2*}_{,i}$ stands for ETG of the second subdomain transformed into the local coordinate of the first subdomain. Substituting Eq. (\ref{eq:ETG_second}) into Eq. (\ref{eq:flux_int}), the disturbances caused by the ETG over the second subdomain can be evaluated as domain integrals: 

\begin{small}
\begin{equation}
\begin{split}
    T_{,1}^{2 \prime} = -\frac{1}{4\pi}\int_{\Omega^{2p}} \phi_{,1j} \overline{T}^{2*}_{j}\thinspace d\textbf{x}' & = -\int_{\gamma-\beta^2}^{\gamma+\beta^2} \int_0^1 \left( \frac{r^2\cos{2\theta} - 2r\cos{\theta}x_1 + x_1^2}{2\pi\left(r^2 - 2r\cos{\theta}x_1 + x_1^2\right)^2} \overline{T}_{1}^{2*} + \frac{r\sin{\theta}\left(r\cos{\theta} - x_1\right)}{\pi\left(r^2 - 2r\cos{\theta}x_1 + x_1^2\right)^2} \overline{T}_{2}^{2*} \right) r \thinspace dr \thinspace d\theta \\ 
    T_{,2}^{2 \prime} = -\frac{1}{4\pi}\int_{\Omega^{2p}} \phi_{,2j} \overline{T}^{2*}_{j}\thinspace d\textbf{x}' & = -\int_{\gamma-\beta^2}^{\gamma + \beta^2} \int_0^1 \left( \frac{r\sin{\theta}\left(r\cos{\theta} - x_1\right)}{\pi\left(r^2 - 2r\cos{\theta}x_1 + x_1^2\right)^2} \overline{T}_{1}^{2*} -\frac{r^2\cos{2\theta} - 2r\cos{\theta}x_1 + x_1^2}{2\pi\left(r^2 - 2r\cos{\theta}x_1 + x_1^2\right)^2} \overline{T}_{2}^{2*} \right) r \thinspace dr \thinspace d\theta 
\end{split}
\label{eq:off_diagonal_form}
\end{equation}
\end{small}

Because the equivalent flux condition is constructed based on the local coordinate of the first subdomain, partial derivatives of Green's function remain unchanged, but the angular limits change as $\theta \in [\gamma - \beta^2, \gamma + \beta^2]$. Note that the domain integrals in Eq. (\ref{eq:off_diagonal_form}) do not involve contributions from the Dirac Delta function, because the field point and source points do not coincide with each other. \textit{As the domain integral of $T_{,1}^{2 \prime}$ and $T_{,2}^{2 \prime}$ share the similar procedure, only the derivation of $T_{,1}^{2 \prime}$ is kept in the main text, and details of $T_{,2}^{2\prime}$ is elaborated in Appendix B.2.}  

Following the same fashion in the preceding subsection, the domain integrals are conducted in three steps. The only difference lies in the second step, the series expansion is conducted with respect to $x_1$ ($x_1 \to 0^+$). With some straightforward and lengthy derivation, the result of step (i) becomes: 

\begin{align}
   -\frac{1}{4\pi}\int_{0}^{1} \phi_{,1j} &\overline{T}_{j}^{2*} r dr =  \frac{K^0-K^2}{4\pi K^0 x_1} \Bigg\{ \Big(
   -\cos{[(\gamma - \theta)(-1 + m)]}_2F_1\left[2,1-m,2-m,\frac{(\cos{\theta} - i\sin{\theta})}{x_1}\right] \notag \\
   &-\cos{[(\gamma - \theta)(-1 + m)]}_2F_1\left[2,1-m,2-m,\frac{(\cos{\theta} + i\sin{\theta})}{x_1}\right] \notag \\
   &+ _2F_1\left[1,1-m,2-m,\frac{(\cos{\theta} + i\sin{\theta})}{x_1}\right]\left(\cos{[(\gamma - \theta)(-1 + m)]} - i\sin{[(\gamma - \theta)(-1 + m)]}\right) \notag \\
   &+ _2F_1\left[1,1-m,2-m,\frac{(\cos{\theta} - i\sin{\theta})}{x_1}\right]\left(\cos{[(\gamma - \theta)(-1 + m)]} + i\sin{[(\gamma - \theta)(-1 + m)]}\right) \notag \\
   &- i_2F_1\left[2,1-m,2-m,\frac{(\cos{\theta} - i\sin{\theta})}{x_1}\right] \sin{[(\gamma - \theta)(-1 + m)]} \notag \\
   &+ i_2F_1\left[2,1-m,2-m,\frac{(\cos{\theta} + i\sin{\theta})}{x_1}\right] \sin{[(\gamma - \theta)(-1 + m)]}
   \Big) f_1^2 \notag \\
   &+\Big(
    i\cos{[(\gamma - \theta)(-1 + m)]}_2F_1\left[2,1-m,2-m,\frac{(\cos{\theta} - i\sin{\theta})}{x_1}\right] \notag \\
   &- i\cos{[(\gamma - \theta)(-1 + m)]}_2F_1\left[2,1-m,2-m,\frac{(\cos{\theta} + i\sin{\theta})}{x_1}\right]  \notag \\
   &+ _2F_1\left[1,1-m,2-m,\frac{(\cos{\theta} - i\sin{\theta})}{x_1}\right]\left(-i\cos{[(\gamma - \theta)(-1 + m)]} + \sin{[(\gamma - \theta)(-1 + m)]}\right) \notag \\
   &+ i_2F_1\left[1,1-m,2-m,\frac{(\cos{\theta} + i\sin{\theta})}{x_1}\right]\left(i\cos{[(\gamma - \theta)(-1 + m)]} + \sin{[(\gamma - \theta)(-1 + m)]}\right) \notag \\
   &- _2F_1\left[2,1-m,2-m,\frac{(\cos{\theta} - i\sin{\theta})}{x_1}\right] \sin{[(\gamma - \theta)(-1 + m)]} \notag \\
   &- _2F_1\left[2,1-m,2-m,\frac{(\cos{\theta} + i\sin{\theta})}{x_1}\right] \sin{[(\gamma - \theta)(-1 + m)]}
   \Big) f_2^2 
   \Bigg\}
    \label{eq:off_diagonal_step_1}
\end{align}
Conduct the series expansion with $x_1$ at the point $x_1 \to 0^+$, and the result of step (ii) can be obtained: 

\begin{small}
\begin{align}
   -\frac{1}{4\pi}\int_{0}^{1} \phi_{,1j}  &\overline{T}_{j}^{2*} r dr = -\frac{K^0 - K^2}{2 \pi K^0 m} (m - 1) \Big(\cos{[(m-1)(\gamma - \theta) + \theta]} f_1^2 + \sin{[(m - 1)(\gamma - \theta) + \theta]} f_2^2 \Big) + \mathcal{O}(x_1) \notag \\
   &+ \Big[ \frac{
    \left(m-1\right)^2 (K^0 - K^2) 
    \csc m \pi
}{
    2 K^0
}
\Big(
    \sin [ (1 - m) (\gamma -2 \theta) - m \pi ]f_2^2  -
    \cos [ (1 - m) (\gamma -2 \theta) - m \pi ] f_1^2
\Big) + \mathcal{O}(x_1^2) \Big]x_1^{-m}
   \label{eq:off_diagonal_step_2}
    \end{align} 
\end{small}

Conduct the angle integrals of Eq. (\ref{eq:off_diagonal_step_2}) and Eq. (\ref{eq:off_diagonal_step_2_ap}), and the results in step (iii) can be obtained: 

\begin{equation}
\begin{aligned}
   T_{,1}^{2 \prime} &= - x_1^{-m} \frac{K^0 - K^2}{2 K^0} (m - 1) \csc m \pi \thinspace \sin[2 (m - 1) \beta^2] \Big( \cos [m (\pi - \gamma) + \gamma] f_1^2 + \sin [m (\pi - \gamma) + \gamma] f_2^2 \Big) \\
   T_{,2}^{2 \prime} &= - x_1^{-m} \frac{K^0 - K^2}{2 K^0} (m - 1) \csc m \pi \sin [2 (m - 1) \beta^2] \Big( \sin [m (\pi - \gamma) + \gamma] f_1^2 + \cos [m (\pi - \gamma) + \gamma] f_2^2 \Big)
\label{eq:off_diagonal_final}
\end{aligned}
\end{equation}

Note that Eq. (\ref{eq:off_diagonal_final}) provides disturbances by the second subdomain at the field point ($0^+, 0$) with respect to the local coordinate of the first subdomain. Following the same fashion, the reciprocal disturbances, by the first subdomain $\Omega^1$ evaluated in the local coordinate of the second subdomain $\Omega^2$, can be derived through domain integrals again. Alternatively, such disturbances can be obtained through the replacement of material properties and geometric properties. Specifically, replace $K^2, \beta^2, \gamma, f_1^2, f_2^2$ with $K^1, \beta^1, 2\pi - \gamma, f_1^1, f_2^1$, respectively, one can obtain the reciprocal disturbances: 

\begin{equation}
\begin{aligned}
   T_{,1}^{1 \prime} &= -\frac{K^0 - K^1}{2 K^0} (m - 1) \csc m \pi \thinspace \sin[2 (m - 1) \beta^1] \Big( \cos [m (-\pi + \gamma) + 2 \pi - \gamma] f_1^1 + \sin [m (-\pi + \gamma) + 2 \pi - \gamma] f_2^1 \Big) \\
   T_{,2}^{1 \prime} &= -\frac{K^0 - K^1}{2 K^0} (m - 1) \csc m \pi \sin [2 (m - 1) \beta^1] \Big( \sin [m (-\pi + \gamma) + 2 \pi - \gamma] f_1^1 + \cos [m (-\pi + \gamma) + 2 \pi - \gamma] f_2^1 \Big)
\label{eq:off_diagonal_final_2}
\end{aligned}
\end{equation}

Therefore, Eq. (\ref{eq:diagonal_final}) provides closed-form diagonal coefficients, and Eq. (\ref{eq:off_diagonal_final}) and Eq. (\ref{eq:off_diagonal_final_2}) together handle interactions between inhomogeneities, which can be easily extended for multiple inhomogeneities. In the following, these coefficients are filled in Eq. (\ref{eq:global_thermal}). Discussions for the single and two-inhomogeneity cases are provided below. 

\subsection{Case 1: a single inhomogeneity}
When the single inhomogeneity exists, the expressions in Eq. (\ref{eq:global_thermal}) can be significantly simplified. The global system of linear equations becomes (ignoring $x_1^{-m}$ as the factor): 

\begin{equation}
    \begin{split}
	\begin{bmatrix} A_{1}^1 & A_2^1 \\ A_1^2 & A_{2}^2 \end{bmatrix} \begin{bmatrix} f_1^1 \\ f_2^1 \end{bmatrix} = \begin{bmatrix} 0 \\ 0 \end{bmatrix}
    \end{split}
\label{eq:single_thermal}
\end{equation}
where 
\begin{align}
     A_{1}^1 &= \frac{K^0 - K^1}{2K^0} (m - 1) \left\{ K^0 + K^1 + (K^0 - K^1) \csc m \pi \thinspace \sin \left[ 2 (1 - m) \beta^1 + m \pi \right] \right\} \notag \\
     A_1^2& = A_2^1 = 0 \notag \\
     A_{2}^2 &= \frac{K^0 - K^1}{2K^0} (m - 1) \thinspace  \left\{ K + K^0 - (K^0 - K^I) \sin m \pi \thinspace \thinspace  \sin \left[ 2 (1- m) \beta^1  + m \pi\right] \right\}
    \label{eq:single_coeff}
\end{align}

As Eq. (\ref{eq:single_thermal}) indicates, the non-trivial solution of coefficients $f_1^1$ and $f_2^1$ requires the determinant of the matrix to be zero. Hence, the singularity parameter $m$ is governed by the determinant of Eq. (\ref{eq:single_thermal}), which can be explicitly written as: 

\begin{equation}
    \text{Det} =  -\frac{1}{4 (K^0) ^2} \left(K^0-K^1 \right)^2 (m - 1)^2 \left\{ - \left(K^0 + K^1 \right)^2 + \left(K^0 - K^1 \right)^2 \csc^2[m \pi]  \thinspace \sin^2 \left[2 (1 - m) \beta^1 + m \pi \right] \right\}
    \label{eq:Det_single}
\end{equation}

When the inhomogeneity is a void ($K^1 = 0$), Eq. (\ref{eq:Det_single}) reduces as: 

\begin{equation}
    \text{Det} = \frac{1}{4} (m - 1)^2 (K^0)^2 \csc^2[m \pi]  \thinspace \cos^2 \left[2 (1 - m) \beta^1 + m \pi  \right]
    \label{eq:Det_single_void}
\end{equation}

\subsection{Case 2: two inhomogeneities}
As Eq. (\ref{eq:single_thermal}) and Eq. (\ref{eq:single_coeff}) indicate, the diagonal block ($2 \times 2$) only contains two non-zero components. Hence, for two inhomogeneities, the global system of linear equations can be written as: 

\begin{equation}
\begin{split}
	\begin{bmatrix} A_{1}^1 & 0 & A_{3}^1 & A_{4}^1 \\ 0 & A_{2}^2 & A_{3}^2  & A_{4}^2 \\ A^3_{1}& A^3_{2} & A_{3}^3 & 0\\ A_1^4 & A_{2}^4 & 0 & A_{4}^4 \end{bmatrix} \begin{bmatrix} f_1^1 \\ f_2^1 \\ f_1^2 \\ f_2^2 \end{bmatrix} = \begin{bmatrix} 0 \\ 0 \\ 0 \\ 0 \end{bmatrix}
    \label{eg:bmatrix_multiple}
    \end{split}
\end{equation}
where $A_1^{1}, A_2^2$ are provided in Eq. (\ref{eq:single_coeff}), and the other two diagonal coefficients can be obtained through replacing properties of the first subdomain with those of the second subdomain. For the completeness of the derivation, all coefficients are listed below: 

\begin{align}
     A_{1}^1 &= \frac{K^0 - K^1}{2K^0} (m - 1) \left\{ K^0 + K^1 + (K^0 - K^1) \csc m \pi \thinspace \sin \left[2 (1 - m) \beta^1 m \pi \right] 
     \right\} \notag \\
     A_{3}^1 &= -\frac{(K^0 - K^1)(K^0 - K^2)}{2K^0}(m - 1) \csc m \pi \thinspace \cos \left[ (1 - m) \gamma + m \pi \right] \sin \left[2 \left(m-1\right)\beta^2\right] \notag \\
     A_{4}^1 &= -\frac{(K^0 - K^1)(K^0 - K^2)}{2K^0}(m - 1) \csc m \pi \thinspace \sin \left[ (1 - m) \gamma  + m \pi \right] \sin \left[2 \left(m-1\right)\beta^2\right] \notag \\
     A_{2}^2 &= \frac{(K^0 - K^1)}{2K^0}(m - 1) \thinspace \left\{K^0 + K^1 + (K^1-K^0) \csc m \pi \thinspace \sin \left[ 2 (1 - m) \beta^1  + m \pi  \right]\thinspace 
     \right\} \notag \\
     A_{3}^2 &= -\frac{(K^0 - K^1)(K^0 - K^2)}{2K^0}(m - 1) \csc m \pi \thinspace \sin \left[ (1 - m) \gamma  + m \pi  \right] \sin \left[2 \left(m-1\right)\beta^2\right] \notag\\
     A_{4}^2 &= \frac{(K^0 - K^1)(K^0 - K^2)}{2K^0}(m - 1) \csc m \pi \thinspace \cos \left[ (1 - m) \gamma  + m \pi  \right] \sin \left[2 \left(m-1\right)\beta^2\right] \notag\\
     A_{1}^3 &= -\frac{(K^0 - K^1)(K^0 - K^2)}{2K^0}(m - 1) \csc m \pi \thinspace \cos \left[ (1 - m) \gamma  + m \pi  \right] \sin \left[2 \left(m-1\right)\beta^1\right] \notag \\
     A_{2}^3 &= \frac{(K^0 - K^1)(K^0 - K^2)}{2K^0}(m - 1) \csc m \pi \thinspace \sin \left[ (1 - m) \gamma  + m \pi  \right] \sin \left[2 \left(m-1\right)\beta^1\right] \notag \\
     A_{3}^3 &= \frac{(K^0 - K^2)}{2K^0}(m - 1)\left\{ K^0 + K^2 + (K^0-K^2) \csc m \pi \thinspace \sin \left[ 2 (1 - m) \beta^2  + m \pi  \right] \right\} \notag \\
     A_{1}^4 &= \frac{(K^0 - K^1)(K^0 - K^2)}{2K^0}(m - 1) \csc m \pi \thinspace \sin \left[ (1 - m) \gamma  + m \pi \right] \sin \left[2 \left(m-1\right)\beta^1\right]  \notag \\
     A_{2}^4 &= \frac{(K^0 - K^1)(K^0 - K^2)}{2K^0}(m - 1) \csc m \pi \thinspace \cos \left[ (1 - m) \gamma  + m \pi  \right] \sin \left[2 \left(m-1\right)\beta^1\right] \notag \\
     A_{4}^4 &= \frac{(K^0 - K^2)}{2K^0}(m - 1) \thinspace \left\{K^0 + K^2 + (K^2-K^0) \csc m \pi \thinspace \sin \left[ 2 (1 - m) \beta^2  + m \pi  \right]\thinspace 
     \right\}
    \label{eg:two_coeff}
\end{align}

Following the same fashion as the single inhomogeneity, the non-trivial solutions of coefficients require the determinant of the matrix to be zero:

\begin{small}
\begin{align}
\text{Det}=\;&
          \frac{(m-1)^3 (K^0 - K^1)^2 (K^0 - K^2)^2 \csc^2 m \pi}{16 (K^0)^4} \Bigg\{
          (1-m) (K^0 - K^1) (K^0 - K^2) \sin[2 \beta^1 (m-1)] \sin[2 \beta^2 (m-1)] \notag\\
          &\times \Big[ - (K^0 - K^1) (K^0 - K^2) \csc^2 m \pi \sin^2[ \gamma + (\pi - \gamma) m ] \sin[2 \beta^1 (m-1)] \sin[2 \beta^2 (m-1)]  - \csc m \pi\notag\\
          &\times \sin^2[ \gamma + (\pi - \gamma) m ] \big( (K^1 - K^0) \sin[ m \pi - 2 \beta^1 (m-1) ] + (K^0 + K^1) \sin m \pi \thinspace \big) \big( (K^0 - K^2) \csc m \pi \notag\\
          &\times \sin[ m \pi - 2 \beta^2 (m-1) ] + K^0 + K^2 \big) - (K^0 - K^1) (K^0 - K^2) \csc^2 [m \pi] \cos^2[ \gamma + (\pi - \gamma) m ] \sin[2 \beta^1 (m-1)] \notag\\
          &\times \sin[2 \beta^2 (m-1)]  + \cos^2[ \gamma + (\pi - \gamma) m ] \big( (K^0 - K^1) \csc m \pi \sin[ m \pi - 2 \beta^1 (m-1) ] + K^0 + K^1 \big)  \big( (K^0 - K^2) \notag\\
          &\times \csc m \pi \sin[ m \pi - 2 \beta^2 (m-1) ] + K^0 + K^2 \big) \Big]  + (m-1) \big( (K^2 - K^0) \sin[ m \pi - 2 \beta^2 (m-1) ] + (K^0 + K^2)\notag\\
          &\times \sin m \pi \thinspace \big)  \Big[ (K^0 - K^1) (K^0 - K^2) \csc m \pi \sin^2[ \gamma + (\pi - \gamma) m ] \sin[2 \beta^1 (m-1)] \sin[2 \beta^2 (m-1)]  \big( (K^0 - K^1)\notag\\
          &\times \csc m \pi \sin[ m \pi - 2 \beta^1 (m-1) ] + K^0 + K^1 \big) + \big( (K^1 - K^0) \sin[ m \pi - 2 \beta^1 (m-1) ] + (K^0 + K^1) \sin m \pi \thinspace \big) \notag\\
          &\times \Big( \big( (K^0 - K^1) \csc m \pi \sin[ m \pi - 2 \beta^1 (m-1) ] + K^0 + K^1 \big)  \big( (K^0 - K^2) \csc m \pi \sin[ m \pi - 2 \beta^2 (m-1) ] + K^0 + K^2 \big)  \notag\\
          &- (K^0 - K^1) (K^0 - K^2) \csc^2 [m \pi] \cos^2[ \gamma + (\pi - \gamma) m ] \sin[2 \beta^1 (m-1)] \sin[2 \beta^2 (m-1)] \Big) \Big]
          \Bigg\} 
\label{eq:det_thermal_multiple}
\end{align}
\end{small}

\textit{The numerical implementation of one- and two-inhomogeneity cases in Mathematica script are provided as the supplemental file "Thermal-singularity.nb".} 

\section{Formulation of stress singularity}

\subsection{Governing equation and fundamental solution}
Navier's equation governs stress transfer in the composite domain $\mathcal{D}$ (in the absence of body force),  

\begin{equation}
    \frac{\partial}{\partial x_j} \left[ C_{ijkl}(\textbf{x})  \varepsilon_{kl}(\textbf{x}) \right]= 0
    \label{eq:elastic_govern}
\end{equation}
where $C_{ijkl}(\textbf{x})$ refers to position-dependent stiffness tensor. Because this paper uses eigenstrain to simulate mismatch of stiffness, Eq. (\ref{eq:elastic_govern}) can be rewritten with matrix stiffness and together with eigenstrain, 

\begin{equation}
   C^0_{ijkl} \frac{\partial}{\partial x_j} \left[ \varepsilon_{kl}(\textbf{x}) - \varepsilon_{kl}^*(\textbf{x}) \right] = 0 \quad \text{or equivalently} \quad C^0_{ijkl} \varepsilon_{kl,j} = C^0_{ijkl} \varepsilon^*_{kl,j} (\textbf{x})
    \label{eq:elastic_govern_2}
\end{equation}
where $C^0_{ijkl}$ is the stiffness of the matrix, and $\varepsilon^*_{ij}$ is the eigenstrain. Based on the concept of Green's function, the displacement gradient and strain at a field point can be obtained through a domain integral of the modified Green's functions multiplied by the eigenstrain, 

\begin{align}
    u_{i,j}(\textbf{x}) & = \int_{\mathcal{D}} g_{ikl,j}(\textbf{x}, \textbf{x}') \varepsilon^*_{kl} (\textbf{x}') \thinspace d\textbf{x}' = \sum_{I=1}^{N^I} \int_{\Omega^I} g_{ikl,j}(\textbf{x}, \textbf{x}') \varepsilon^{I*}_{kl}(\textbf{x}') \thinspace d\textbf{x}' \\
    \varepsilon_{ij}(\textbf{x}) & = \int_{\mathcal{D}} s_{ijkl}(\textbf{x}, \textbf{x}') \varepsilon^*_{kl} (\textbf{x}') \thinspace d\textbf{x}' = \sum_{I=1}^{N^I} \int_{\Omega^I} s_{ijkl}(\textbf{x}, \textbf{x}') \varepsilon_{kl}^{I*}(\textbf{x}') \thinspace d\textbf{x}'
    \label{eq:disturbed_elastic}
\end{align}
where $\varepsilon_{ij}^{I*}$ represents eigenstrain over the $I^{th}$ subdomain; and $g_{ikl}$ and $s_{ijkl}$ are modified Green's functions relating eigenstrain to displacement and strain, respectively. The two modified Green's functions were summarized in \cite{Mura1987} as: 

\begin{equation}
\begin{aligned}
    g_{ikl}(\textbf{x}, \textbf{x}') & = \frac{1}{8 \pi (1 - \nu^0)} \left[ \psi_{,ikl} - 2 \nu^0 \phi_{,i} \delta_{kl} - 2 (1 - \nu^0) \left( \phi_{,k} \delta_{il} +  \phi_{,l} \delta_{ik}\right) \right] \\
    s_{ijkl}(\textbf{x}, \textbf{x}') & = \frac{1}{8 \pi (1-\nu^0)} \left[ \psi_{,klij} - 2 \nu^0 \phi_{,ij} \delta_{kl} - (1-\nu^0) (\phi_{,kj} \delta_{il} + \phi_{,ki} \delta_{jl} + \phi_{,lj} \delta_{ik} + \phi_{,li} \delta_{jk})  \right]
\end{aligned}
    \label{eq:modify_green}
\end{equation}
where $\phi$ is the harmonic potential function provided in Eq. (\ref{eq:thermal_green}), and $\psi = \frac{\phi - 1}{2} |\textbf{x} - \textbf{x}'|^2$ is the biharmonic potential function. 

\subsection{Equivalent stress condition}
The eigenstrain in Eq. (\ref{eq:elastic_govern_2}) can be determined through solving the equivalent stress condition \cite{Eshelby1957} over each subdomain $\Omega^I$:

\begin{equation}
    C_{ijkl}^0 (\varepsilon_{kl} - \varepsilon_{kl}^*) = C^I_{ijkl} \varepsilon_{kl}
\label{eq:equiv_stress}
\end{equation}
In parallel to the inhomogeneity problem in steady-state heat conduction, the equivalent stress condition needs to be satisfied at every interior point of each subdomain. Recently, using Airy's stress function and separation of variables by \cite{Williams1952, Dempsey1979}, Wu and Yin \cite{Wu2024} conducted an asymptotic analysis of the singular eigenstrain for a single triangular inhomogeneity. Building upon that result, this paper extends the framework to conduct an asymptotic analysis of multiple inhomogeneities. 

Following Dempsey and Sinclair's work \cite{Dempsey1979}, the stress within the $I^{th}$ subdomain in the polar coordinate can be expressed in terms of the distance and the angle: 

\begin{equation}
\begin{aligned}
    \sigma_{rr}^{I} & = \frac{\partial}{\partial \lambda} \Big\{ R^{-\lambda} (\lambda-1) \Big[ c_1^I (2 - \lambda) \sin(2- \lambda) \theta + c_2^I (2 - \lambda) \cos(2- \lambda) \theta
    \\ & - c_3^I (\lambda + 2) \sin\lambda \theta - c_4^I (\lambda + 2) \cos\lambda \theta \Big] \Big\}
     \\ 
    \sigma_{\theta \theta}^I & = \frac{\partial}{\partial \lambda} \Big\{ R^{-\lambda} (1 - \lambda) (2 - \lambda) \Big[ c_1^I \sin(2 - \lambda) \theta + c_2^I \cos(2- \lambda) \theta  + c_3^I \sin\lambda \theta + c_4^I \cos\lambda \theta \Big] \Big\} \\
    \sigma_{r \theta}^I &= \frac{\partial}{\partial \lambda} \Big\{ R^{-\lambda} (\lambda -1) \Big[ c_1^I (2 - \lambda) \cos (2 - \lambda) \theta - (2 - \lambda) c_2^I \sin (2 - \lambda) \theta
    + c_3^I \lambda \cos\lambda \theta - c_4^I \lambda \sin\lambda \theta
    \Big] \Big\}
\end{aligned}
\label{eq:stress_Polar}
\end{equation}
where $\bm{\sigma}^I$ refers to stress within the $I^{th}$ subdomain; and $c_1^I, c_2^I, c_3^I, c_4^I$ are four coefficients to describe stress distribution, which are functions of the singularity parameter $\lambda$. Based on the energy criteria, the singularity parameter $\lambda \in (0, 1)$. The partial derivative with respect to the singularity parameter $\lambda$ aims to handle potential logarithmic singularities \cite{Dempsey1979}, which may not exist for every case. 

Using the equivalent stress condition, the eigenstrain over the $I^{th}$ subdomain can be expressed in terms of stress, which can be formally expressed as: 

\begin{equation}
    \varepsilon^{I*}_{ij} = \left( C^0_{ijkl} \right)^{-1} \left( C^0_{klmn} - C^I_{klmn} \right) \varepsilon_{mn}^{I} = \left( C^0_{ijkl} \right)^{-1} \left( C^0_{klmn} - C^I_{klmn} \right) \left(C^I_{mnpq} \right)^{-1} \sigma^{I}_{pq}
    \label{eq:eigenstrain}
\end{equation}
and the explicit eigenstrain can be obtained by substituting Eq. (\ref{eq:stress_Polar}) into Eq. (\ref{eq:eigenstrain}). For the convenience of domain integrals with Green's function, the eigenstrain (in polar coordinates) has been converted into Cartesian coordinates:

\begin{small}
\begin{align}
    \varepsilon_{11}^{I*} & = \frac{\partial}{\partial \lambda} \Big\{ \frac{(\lambda -1)}{2 \mu^0 \mu^I} R^{-\lambda} \Big( c_1^I (\lambda -2) (\mu^0-\mu^I) \sin \lambda \theta-c_2^I (\lambda -2) (\mu^0-\mu^I) \cos \lambda \theta-2 c_3^I \mu^0 \sin \lambda \theta \notag \\ & -c_3^I \lambda  \mu^0 \sin (\lambda +2) \theta -2 c_4^I \mu^0 \cos \lambda \theta-c_4^I \lambda  \mu^0 \cos (\lambda +2) \theta   +2 c_3^I \mu^I \sin \lambda \theta \notag \\ & +c_3^I \lambda  \mu^I \sin (\lambda +2) \theta +2 c_4^I \mu^I \cos \lambda \theta+c_4^I \lambda  \mu^I \cos (\lambda +2) \theta  -4 c_3^I \mu^I \nu^0\sin \lambda \theta \notag \\ & -4 c_4^I \mu^I \nu^0 \cos \lambda \theta+4 c_3^I \mu^0 \nu^I \sin \lambda \theta+4 c_4^I \mu^0 \nu^I \cos \lambda \theta \Big) \Big\} \notag \\
    \varepsilon_{12}^{I*} & = -\frac{\partial}{\partial \lambda} \Big\{ \frac{(\lambda -1) (\mu^0-\mu^I)}{2 \mu^0 \mu^I} R^{-\lambda} \Big( c_2^I (\lambda -2) \sin \lambda \theta+c_4^I \lambda  \sin (\lambda +2) \theta +c_1^I (\lambda -2) \cos \lambda \theta-c_3^I \lambda  \cos (\lambda +2) \theta  \Big) \Big\} \notag \\ 
    \varepsilon_{22}^{I*} & = -\frac{\partial}{\partial \lambda} \Big\{ \frac{(\lambda -1)}{2 \mu^0 \mu^I} R^{-\lambda} \Big( c_1^I (\lambda -2) (\mu^0-\mu^I) \sin \lambda \theta-c_2^I (\lambda -2) (\mu^0-\mu^I) \cos \lambda \theta+2 c_3^I \mu^0 \sin \lambda \theta \notag \\ & -c_3^I \lambda  \mu^0 \sin (\lambda +2) \theta +2 c_4^I \mu^0 \cos \lambda \theta-c_4^I \lambda  \mu^0 \cos (\lambda +2) \theta -2 c_3^I \mu^I \sin \lambda \theta \notag \\ &+c_3^I \lambda  \mu^I \sin (\lambda +2) \theta  -2 c_4^I \mu^I \cos \lambda \theta+c_4^I \lambda  \mu^I \cos (\lambda +2) \theta +4 c_3^I \mu^I \nu^0\sin \lambda \theta \notag \\ & +4 c_4^I \mu^I \nu^0\cos \lambda \theta-4 c_3^I \mu^0 \nu^I \sin \lambda \theta-4 c_4^I \mu^0 \nu^I \cos \lambda \theta \Big) \Big\}
\label{eq:eigen_cartesian}
\end{align}
\end{small}

With Eq. (\ref{eq:eigen_cartesian}), the eigenstrain has been written in terms of stress. Substituting eigenstrain in Eq. (\ref{eq:disturbed_elastic}), the Fredholm integral equation of the second kind can be obtained similarly to the heat conduction problem. Some differences are emphasized that for the elastic problem, each subdomain requires four coefficients, i.e., $c_1^I - c_4^I$, to describe local fields. To form a well-posed global system of linear equations, each subdomain requires four linearly independent equations. However, the equivalence of stress/strain can only supply three, such as $\varepsilon_{11}, \varepsilon_{12}, \varepsilon_{22}$, because they compress four independent equations by the equivalences of displacement gradients into three. Instead of using the equivalence of $\varepsilon_{12}$, the equivalences of $u_{1,2}$ and $u_{2,1}$ are employed to provide two linearly independent equations. Therefore, for each subdomain, equations are built upon equivalences of $\varepsilon_{11}, \varepsilon_{22}, \varepsilon_{12}$, and $u_{2,1}$. Formally, a global system of linear equations can be constructed in analogy with the heat conduction, 

\begin{equation}
    R^{-\lambda} \begin{bmatrix} B^1_{1} & B^1_{2} & B^1_3 & B^1_4 & \ldots & B^1_{4N^I-3} & B^1_{4 N^I - 2} & B^1_{4 N^I - 1} & B^1_{4 N^I} \\ B^2_{1} & B^2_{2} & B^2_3 & B^2_4 & \ldots & B^2_{4N^I-3} & B^2_{4 N^I - 2} & B^2_{4 N^I - 1} & B^2_{4 N^I} \\ B^3_{1} & B^3_{2} & B^3_3 & B^3_4 & \ldots & B^3_{4N^I-3} & B^3_{4 N^I - 2} & B^3_{4 N^I - 1} & B^3_{4 N^I} \\ B^4_{1} & B^4_{2} & B^4_3 & B^4_4 & \ldots & B^4_{4N^I-3} & B^4_{4 N^I - 2} & B^4_{4 N^I - 1} & B^4_{4 N^I} \\ \vdots & \vdots & \vdots & \vdots & \vdots & \vdots& \vdots& \vdots& \vdots \\ B^{4 N^I - 3}_{1} & B^{4 N^I - 3}_{2} & B^{4 N^I - 3}_{3} & B^{4 N^I - 3}_{4} & \ldots & B^{4 N^I - 3}_{4 N^I - 3} & B^{4 N^I - 3}_{4 N^I - 2} & B^{4 N^I - 3}_{4 N^I - 1} & B^{4 N^I - 3}_{4 N^I} \\  
    B^{4 N^I - 2}_{1} & B^{4 N^I - 2}_{2} & B^{4 N^I - 2}_{3} & B^{4 N^I - 2}_{4} & \ldots & B^{4 N^I - 2}_{4 N^I - 3} & B^{4 N^I - 2}_{4 N^I - 2} & B^{4 N^I - 2}_{4 N^I - 1} & B^{4 N^I - 2}_{4 N^I} \\
    B^{4 N^I - 1}_{1} & B^{4 N^I - 1}_{2} & B^{4 N^I - 1}_{3} & B^{4 N^I - 1}_{4} & \ldots & B^{4 N^I - 1}_{4 N^I - 3} & B^{4 N^I - 1}_{4 N^I - 2} & B^{4 N^I - 1}_{4 N^I - 1} & B^{4 N^I - 1}_{4 N^I} \\
    B^{4 N^I}_{1} & B^{4 N^I}_{2} & B^{4 N^I}_{3} & B^{4 N^I}_{4} & \ldots & B^{4 N^I}_{4 N^I - 3} & B^{4 N^I}_{4 N^I - 2} & B^{4 N^I}_{4 N^I - 1} & B^{4 N^I}_{4 N^I} \\
    \end{bmatrix} \begin{bmatrix} c_1^1 \\ c_2^1 \\ c_3^1 \\ c_4^1 \\ \vdots \\ c_1^{N^I} \\ c_2^{N^I} \\ c_3^{N^I} \\ c_4^{N^I} \end{bmatrix} = \begin{bmatrix} 0 \\ 0 \\ 0 \\ 0 \\ \vdots \\ 0 \\ 0 \\ 0 \\ 0 \end{bmatrix}    
    \label{eq:global_elastic}
\end{equation}
where the matrix is composed of $4 N^I \times 4 N^I$ coefficients $B^{i}_{j}$, and the superscript $i$ and subscript $j$ refers to the $i^{th}$ row and $j^{th}$ column, respectively. Note that coefficients $B^i_j$ are symbolic expressions here, and they are related to the singularity parameter $\lambda$. The evaluations of $B^i_j$ through domain integrals of Green's function multiplied by the power-form eigenstrain are conducted in the following subsection. 

\subsection{Disturbances of the power-form eigenstrain}
The evaluation of disturbed displacement gradients should be divided into two parts: (i) the disturbances in the $I^{th}$ subdomain caused by its own eigenstrain, namely the "diagonal coefficient"; and (ii) the disturbances in the $I^{th}$ subdomain caused by eigenstrain over the $J^{th}$ subdomain, known as "off-diagonal coefficients". \textit{Note that Wu and Yin \cite{Wu2024} have derived diagonal coefficients ($4 \times 4$ block), and such a derivation process will not be repeated below. For the completeness of this work, diagonal coefficients will be specified per case.} 

\subsubsection{Evaluation of off-diagonal coefficients}
Without the loss of any generality, this subsection derives the disturbance in the first subdomain caused by eigenstrain within the second subdomain. Specifically, let $\varepsilon^{J \prime}_{ij}$ and $u_{i,j}^{J \prime}$ represent disturbed strain and displacement gradient caused by eigenstrain within the $J^{th}$ subdomain. Following the same fashion as the heat conduction problem, the reciprocal interactions can be considered by the replacement of material/geometric properties, and the formulation is general, which can be straightforwardly extended for multiple inhomogeneities. 

Since the eigenstrain over each subdomain are expressed in its local coordinate, i.e., Eq. (\ref{eq:eigen_cartesian}), the coordinate transformation is required. Hence, the eigenstrain over the second subdomain will be written in terms of the local coordinate of the first subdomain. As previously mentioned in Section 3.3.2, the coordinate transformation avoids the further change of modified Green's function (and their components). Specifically, the eigenstrain over the second subdomain can be written in the local coordinate of the first subdomain explicitly, 

\begin{small}
\begin{align}
    \overline{\varepsilon}_{11}^{2*} = & \frac{\partial}{\partial \lambda} \Bigg\{ \frac{(-1 + \lambda)}{2 \mu^0 \mu^2}  R^{-\lambda}
    \Bigg[ 2 \Big( -\mu^0 + 2 \nu^2 \mu^0 + \mu^2 - 2 \nu^0 \mu^2 \Big) \Big(  c_3^2\sin(-\gamma + \theta) \lambda + c_4^2\cos(-\gamma + \theta) \lambda  \Big) \notag \\
    & + (\mu^0 - \mu^2) (-2 + \lambda) \sin\Big(2 \gamma + (-\gamma + \theta) \lambda\Big) c_1^2  - (\mu^0 - \mu^2) (-2 + \lambda)\cos\Big(2 \gamma + (-\gamma + \theta) \lambda\Big) c_2^2 \notag \\
    &- (\mu^0 - \mu^2) \Bigg( \sin\Big(2 \theta + (-\gamma + \theta) \lambda\Big) c_3^2 + \cos\Big(2 \theta + (-\gamma + \theta) \lambda\Big) c_4^2 \Bigg) \lambda \Bigg] \Bigg\} \notag\\
    \overline{\varepsilon}_{12}^{2*} = & - \frac{\partial}{\partial \lambda} \Bigg\{ \frac{(\mu^0 - \mu^2) (-1 + \lambda)}{2 \mu^0 \mu^2} R^{-\lambda}
    \Bigg[  (-2 + \lambda)\cos\Big(2 \gamma + (-\gamma + \theta) \lambda\Big) c_1^2 + (-2 + \lambda) \sin \Big(2 \gamma + (-\gamma + \theta) \lambda\Big) c_2^2 \notag \\
    &+ \Bigg( -\cos \Big(2 \theta + (-\gamma + \theta) \lambda\Big) c_3^2 + \sin \Big(2 \theta + (-\gamma + \theta) \lambda \Big) c_4^2 \Bigg) \lambda \Bigg] \Bigg\} \notag\\ 
    \overline{\varepsilon}_{22}^{2*} = & \frac{\partial}{\partial \lambda} \Bigg\{ \frac{(-1 + \lambda)}{2 \mu^0 \mu^2} R^{-\lambda}
    \Bigg[ 2 \Big(-\mu^0 + 2 \nu^2 \mu^0 + \mu^2 - 2 \nu^0 \mu^2\Big) \Big( c_3^2\sin(-\gamma + \theta) \lambda  + c_4^2\cos(-\gamma + \theta) \lambda  \Big) \notag \\
    &- (\mu^0 - \mu^2) (-2 + \lambda)\sin\Big(2 \gamma + (-\gamma + \theta) \lambda\Big) c_1^2  + (\mu^0 - \mu^2) (-2 + \lambda) \cos\Big(2 \gamma + (-\gamma + \theta) \lambda\Big) c_2^2 \notag \\
    &+ (\mu^0 - \mu^2) \Bigg( \sin\Big(2 \theta + (-\gamma + \theta) \lambda\Big) c_3^2 + \cos\Big(2 \theta + (-\gamma + \theta) \lambda\Big) c_4^2 \Bigg) \lambda \Bigg] \Bigg\}
\label{eq:eigenstrain_Cartesian_1}
\end{align}
\end{small}
where $\overline{\varepsilon}^{2*}_{,kl}$ represents the eigenstrain of the second subdomain transformed into the local coordinate system of the first subdomain. Substituting Eq. (\ref{eq:eigenstrain_Cartesian_1}) into Eq. (\ref{eq:disturbed_elastic}), the disturbed strain/displacement gradient caused by the eigenstrain over the second subdomain can be evaluated as domain integrals: 

\begin{equation}
    \begin{aligned}
    u_{2,1}^{2'} & = \int_{\Omega^{2p}} g_{2kl,1} \overline{\varepsilon}_{kl}^{2*} \thinspace d\textbf{x}' = \int_{\gamma-\beta^2}^{\gamma+\beta^2} \int_0^1 g_{2kl,1} \overline{\varepsilon}_{kl}^{2*} r \thinspace dr \thinspace d\theta = \frac{\partial}{\partial \lambda} \left[ \int_{\gamma-\beta^2}^{\gamma+\beta^2} \int_0^1 g_{2kl,1} r^{1 - \lambda} H_{kl}^{2*}(\lambda, \gamma, \theta) \thinspace dr \thinspace d\theta \right] \\ 
    \varepsilon_{ij}^{2 \prime} & = \int_{\Omega^{2p}} s_{ijkl} \overline{\varepsilon}_{kl}^{2*}\thinspace d\textbf{x}'  = \int_{\gamma-\beta^2}^{\gamma+\beta^2} \int_0^1 s_{ijkl} \overline{\varepsilon}_{kl}^{2*} r \thinspace dr \thinspace d\theta = \frac{\partial}{\partial \lambda} \left[ \int_{\gamma-\beta^2}^{\gamma+\beta^2} \int_0^1 s_{ijkl} r^{1 - \lambda} H^{2*}_{kl}(\lambda, \gamma, \theta) \thinspace dr \thinspace d\theta \right]
    \end{aligned}
    \label{eq:off-diagonal_form_elastic}
\end{equation}
where $\Omega^{2p}$ presents the triangular polar coordinate integration domain of the second subdomain; the eigenstrain can be expressed as $\overline{\varepsilon}_{kl}^{2*} = \frac{\partial \left[ r^{-\lambda} H^{2*}(\lambda, \gamma, \theta) \right]}{\partial \lambda}$. Since the elastic Green's function is more complicated than the thermal Green's function, the domain integrals of the modified Green's function multiplied by the power-form eigenstrain are evaluated in several components. 

Since the elastic Green's function is composed of partial derivatives of the biharmonic and harmonic potential functions, \textit{Appendix C.1 provides the fourth-order partial derivatives of $\psi$ and second-order partial derivatives of $\phi$, when the field point is along the symmetric line ($x_1, 0$).} Because domain integrals share similar integral procedures, i.e., the heat conduction problem, \textit{the explicit integral results are provided in Appendix C.2 and Appendix C.3 for integrals associated with $\psi$ and $\phi$, respectively.} 

It should be emphasized that when there is only a single inhomogeneity, the disturbances (domain integrals) can be simplified by eliminating symmetric or asymmetric components. For instance, $c_1^1$ and $c_3^1$ can be omitted in the symmetric case. However, the second inhomogeneity (or more inhomogeneities) breaks the symmetric/anti-symmetric conditions. Consequently, the domain integrals in Appendix C include terms such as $c_1^2, c_2^2, c_3^2$, and $c_4^2$, which cannot be simply dropped.

\subsection{The case of two inhomogeneities}
For two inhomogeneities, the global system of linear equations can be written as, 
\begin{equation}
   \left[
    \begin{array}{cccc|cccc}
        0 & B^1_{2} & 0 & B^1_4 &  B^1_{5} & B^1_{6} & B^1_{7} & B^1_{8} \\ 
        0 & B^2_{2} & 0 & B^2_4  & B^2_{5} & B^2_{6} & B^2_{7} & B^2_{8} \\ 
        B^3_{1} & 0 & B^3_3 & 0 & B^3_{5} & B^3_{6} & B^3_{7} & B^3_{8} \\ 
        B^4_{1} & 0 & B^4_3 & 0 &  B^4_{5} & B^4_{6} & B^4_{7} & B^4_{8}  \\ \hline
        B^{5}_{1} & B^{5}_{2} & B^{5}_{3} & B^{5}_{4} & 0 & B^{5}_{6} & 0 & B^{5}_{8} \\ 
        B^{6}_{1} & B^{6}_{2} & B^{6}_{3} & B^{6}_{4} &  0 & B^{6}_{6} & 0 & B^{6}_{8} \\
        B^{7}_{1} & B^{7}_{2} & B^{7}_{3} & B^{7}_{4} & B^{7}_{5} & 0 & B^{7}_{7} & 0 \\
        B^{8}_{1} & B^{8}_{2} & B^{8}_{3} & B^{8}_{4} & B^{8}_{5} & 0 & B^{8}_{7} & 0 \\
    \end{array}
    \right]
    \begin{bmatrix} c_1^1 \\ c_2^1 \\ c_3^1 \\ c_4^1 \\  c_1^{2} \\ c_2^{2} \\ c_3^{2} \\ c_4^{2} \end{bmatrix} = 
    \begin{bmatrix} 0 \\ 0 \\ 0 \\ 0  \\ 0 \\ 0 \\ 0 \\ 0 \end{bmatrix}     \quad \text{or} \quad \begin{bmatrix} \bm{\mathcal{B}}^{UL} & \bm{\mathcal{B}}^{UR} \\ \bm{\mathcal{B}}^{LL} & \bm{\mathcal{B}}^{LR} \end{bmatrix} \begin{bmatrix} \bm{c}^1 \\ \bm{c}^2 \end{bmatrix} = \begin{bmatrix} \bm{0} \\ \bm{0} \end{bmatrix} 
    \label{eq:two_elastic} 
\end{equation}
where the coefficient matrix can be partitioned into four blocks ($4 \times 4$); $\bm{\mathcal{B}}^{UL}$, $\bm{\mathcal{B}}^{UR}$, $\bm{\mathcal{B}}^{LL}$, and $\bm{\mathcal{B}}^{LR}$ represents the upper left, upper right, lower left, and lower right blocks, respectively; $\bm{c}^1$ and $\bm{c}^{2}$ represents $4$ unknown coefficients of the first and second subdomain, respectively. Note that $\bm{\mathcal{B}}^{UL}$ and $\bm{\mathcal{B}}^{LR}$ contain diagonal coefficients, while $\bm{\mathcal{B}}^{UR}$ and $\bm{\mathcal{B}}^{LL}$ collect off-diagonal coefficients. The diagonal coefficients of the upper left block ($\bm{\mathcal{B}}^{UL}$) are given row by row, 

\noindent (i) the first row of $\bm{\mathcal{B}}^{UL}$
\begin{align}
    &B_2^1 = \frac{(2 - \lambda) (\lambda -1) }{8 \mu^0 \mu^1 (1 - \nu^0)} \Big[-\mu^0 +\csc \lambda \pi \thinspace (\mu^0 - \mu^2) \Big((\lambda -1) \sin [\lambda (\pi - 2\beta^1) + 4 \beta^1] \notag \\ & \quad\quad - (\lambda - 4 \nu^0- 2) \sin [\lambda (\pi - 2\beta^1) + 2 \beta^1] \Big) - \mu^1 (3 - 4 \nu^0) \Big] \notag \\ 
    & B_4^1 = \frac{(1 - \lambda) \csc \lambda \pi}{8 \mu^0 \mu^1 (1 - \nu^0)} \Big[   (\lambda -1) \Big( 2 \sin [\lambda \pi + 2\beta^1] \big( \mu^0 (1 - 2\nu^1) - \mu^1 (1 - 2\nu^0) \big) - (\mu^0 - \mu^1) (\lambda - 4 \nu^0 +2) \notag \\ & \quad \quad \times \sin [\lambda (\pi - 2 \beta^1)] \Big) + \sin [\lambda (\pi - 2 \beta^1) + 2 \beta^1]  \Big(\mu^0  \big[ (\lambda -2) \lambda + 4 \nu^1 - 2 \big]- \mu^1 \big[ (\lambda -2) \lambda +4 \nu^0 -2 \big] \Big) \notag \\ & \quad \quad - \sin \lambda \pi \thinspace (\lambda -4 \nu^0 +2) \big( \mu^0 (3 - 4 \nu^1) + \mu^1 \big)  \Big]
    \label{eq:first_line}
\end{align}

\noindent (ii) the second row of $\bm{\mathcal{B}}^{UL}$
\begin{align}
   & B_2^2 = \frac{(2 - \lambda) (\lambda -1)}{8 \pi \mu^0 \mu^1 (1 - \nu^0)} \Big[  \mu^0 + \csc \lambda \pi (\mu^0 -\mu^1) \Big( (\lambda +4 \nu^0-2) \sin \left[\lambda (\pi - 2\beta^1) + 2\beta^1 \right] \notag \\ & \quad \quad - (\lambda -1) \sin \left[\lambda (\pi - 2\beta^1) + 4\beta^1 \right] \Big)+ \mu^1 (3-4 \nu^0) \Big] \notag \\ 
    & B_4^2 = \frac{(1 - \lambda) \csc \lambda \pi }{8 \mu^0 \mu^1 (1 - \nu^0)}  \Big[ (\lambda -1) \Big( (\mu^0 -\mu^1) (\lambda +4 \nu^0 -2) \sin [\lambda (\pi - 2 \beta^1)] +2 \sin [\lambda \pi + 2\beta^1] \notag \\ & \quad \quad  \times \big[ \mu^1(1 - 2\nu^0) - \mu^0 (1 - 2\nu^1) \big]  \Big) + \sin \left[ \lambda (\pi - 2\beta^1) + 2 \beta^1 \right] \Big( \mu^1 [(\lambda -2) \lambda +4 \nu^0 -2] - \mu^0 [(\lambda -2) \lambda +4 \nu^1 -2] \Big) \notag \\ & \quad \quad + \sin
   \lambda \pi  \thinspace (\lambda +4 \nu^0 -2) (\mu^0  (3-4\nu^1) + \mu^1)  \Big] 
    \label{eq:second_line}
\end{align}

\noindent (iii) the third row of $\bm{\mathcal{B}}^{UL}$
\begin{small}
\begin{align}
    & B_1^3 = \frac{(\lambda -2) (\lambda -1)}{8 \mu^0 \mu^1 (1 - \nu^0)}   \Big[ \csc \lambda \pi \thinspace (\mu^0 - \mu^1) \Big( 2 \lambda  \sin \beta^1 \thinspace \cos \left[ \lambda (\pi - 2 \beta^1) + 3 \beta^1  \right] - \sin \left[ \lambda (\pi - 2 \beta^1) + 4 \beta^1  \right] \Big) +\mu^0 + \mu^1 (3-4 \nu^0) \Big] \notag \\ 
   & B_3^3 = \frac{(1 - \lambda) \csc \lambda \pi}{8 \mu^0 \mu^1 (1 - \nu^0)} \Big[ (\lambda -1) \Big( \lambda  (\mu^1-\mu^0) \sin [\lambda (\pi - 2 \beta^1)] +2 \sin [\lambda \pi + 2\beta^1] 
   \big[ \mu^1 (1 - 2\nu^0) - \mu^0 (1- 2\nu^1) \big] \Big) \notag \\ & \quad \quad +\sin \left[ \lambda (\pi - 2\beta^1) + 2\beta^1 \right] \Big(\mu^0  [(\lambda -2) \lambda
   +4 \nu^1 -2]- \mu^1 [(\lambda -2) \lambda +4 \nu^0 -2] \Big)+\lambda  \sin \lambda \pi \thinspace [ \mu^1 +\mu^0  (3-4 \nu^1) ] \Big] \notag \\ 
    \label{eq:third_line}
\end{align}
\end{small}

\noindent (iv) the fourth row of $\bm{\mathcal{B}}^{UL}$
\begin{align}
    & B_1^4 = \frac{(2 - \lambda) (\lambda -1)}{8 \mu^0 \mu^1 (1 - \nu^0)} \Big[ \csc \lambda \pi \thinspace (\mu^0 -\mu^1) \Big( (\lambda +4 \nu^0-4) \sin \left[ \lambda (\pi - 2\beta^1) + 2 \beta^1 \right] - (\lambda -1) \sin \left[ \lambda (\pi - 2\beta^1) + 4 \beta^1 \right] \Big) \notag \\ & \quad \quad -\mu^0 - \mu^1 (3 - 4\nu^0) \Big] \notag \\
    & B_3^4 = \frac{(1 - \lambda) \csc \lambda \pi}{8 \mu^0 \mu^1 (1 - \nu^0)} \Big[ (\lambda -1) \Big( 2 \sin [\lambda \pi + 2\beta^1] \big[ \mu^1 (1 - 2\nu^0) - \mu^0 (1- 2\nu^1) \big] -(\mu^0 - \mu^1) (\lambda +4 \nu^0-4) \notag \\ & \quad \quad  \times \sin [\lambda (\pi - 2\beta^1)] \Big) + \sin \left[ \lambda (\pi - 2\beta^1) + 2\beta^1 \right]  \Big( \mu^0 [ (\lambda -2) \lambda +4 \nu^1 -2 ]-\mu^1 [(\lambda -2) \lambda +4 \nu^0 -2] \Big) \notag \\ & \quad \quad+ \sin \lambda \pi (\lambda +4 \nu^0-4) ( \mu^0  (3 - 4\nu^1 ) +\mu^1 )  \Big]
    \label{eq:fourth_line}
\end{align}

The coefficients in the lower right block ($\bm{\mathcal{B}}^{LR}$) can be derived from Eqs. (\ref{eq:first_line} - \ref{eq:fourth_line}) by: (i) replacing the material properties $\mu^1, \nu^1$ with $\mu^2, \nu^2$, respectively; and (ii) substituting the opening angles $\beta^1$ with $\beta^2$. Additionally, the off-diagonal coefficients of the upper right block ($\bm{\mathcal{B}}^{UR}$) are provided below, as derived from formulae of domain integrals in Appendix C.2 and Appendix C.3. 

\noindent (i) the first row of $\bm{\mathcal{B}}^{UR}$
\begin{small}
\begin{align}
     B_5^1 = & \frac{(2 - \lambda) (\lambda -1) \csc \lambda \pi (\mu^0 - \mu^2)}{8 \mu^0 \mu^2 (1 - \nu^0)} \Big[ (\lambda -1) \sin [\lambda (\pi - \gamma) + 2\gamma] \sin [2 (2 - \lambda) \beta^2]  - (\lambda -4 \nu^0 +2) \sin [\lambda (\pi - \gamma)] \notag \\ 
    &\times \sin [2 (1 - \lambda) \beta^2] \Big] \notag \  \notag \\
    B_6^1 = & \frac{(2 - \lambda) (\lambda -1) \csc \lambda \pi  (\mu^0 - \mu^2)}{8 \mu^0 \mu^2 (1 - \nu^0)} \Big[ (\lambda -1) \cos [ \lambda (\pi - \gamma) + 2\gamma] \sin [ 2 (2 - \lambda) \beta^2] - (\lambda -4 \nu^0 +2) \cos [ \lambda (\pi - \gamma) ]  \notag \\ 
    &\times \sin [ 2 (1 - \lambda) \beta^2] \Big]   \notag\\
    B_7^1 = & \frac{(1 - \lambda) \csc \lambda \pi}{8 \mu^0 \mu^2 (1 - \nu^0)} \Big[ (\lambda -1) (\mu^0 - \mu^2) (-\lambda +4 \nu^0 -2) \sin [ \lambda (\pi - \gamma) ] \sin  2 \lambda \beta^2 \thinspace - \sin [ \lambda (\pi - \gamma) + 2\gamma ] \Big( \lambda^2 \mu^0 \sin [ 2 (1 - \lambda) \beta^2] \notag \\ 
    & + 2 (\lambda -1) \big[ \mu^2 (1 - 2\nu^0) - \mu^0 (1-2\nu^2) \big] \sin  2 \beta^2 \thinspace +\big( \mu^2 \big[ (\lambda -2) \lambda +4 \nu^0 -2 \big] + 2 \mu^0 (\lambda - 2 \nu^2 + 1) \big)  \sin [ 2 (\lambda -1) \beta^2  ] \Big) \Big] \notag  \notag\\
    B_8^1 = & \frac{(1 - \lambda) \csc \lambda \pi}{8 \mu^0 \mu^2 (1 - \nu^0)} \Big[  (\lambda -1) (\mu^0 - \mu^2) (\lambda -4 \nu^0 +2) \cos [ \lambda (\pi - \gamma) ] \sin  2 \lambda \beta^2  \thinspace+ \cos [ \lambda (\pi - \gamma) + 2\gamma ] \Big( \big[ \mu^2 ( (\lambda -2) \lambda  \notag \\ 
    &+4 \nu^0 -2 ) - \mu^0 ( (\lambda -2) \lambda +4 \nu^2 -2 ) \big]\sin [ 2 \beta^2 (\lambda -1) ] - 2 (\lambda -1)  \big[ \mu^2 (1 - 2\nu^0) - \mu^0 (1-2\nu^2) \big] \sin  2 \beta^2  \Big) \Big]
    \label{eq:off_first_line}
\end{align}
\end{small}

\noindent (ii) the second row of $\bm{\mathcal{B}}^{UR}$
\begin{small}
\begin{align}
    B_5^2 = & \frac{(2 - \lambda)(\lambda-1) \csc \lambda \pi (\mu^0 - \mu^2)}{8 \mu^0 \mu^2 (1 - \nu^0)} \Big[  \Big( (\lambda +4 \nu^0-2) \sin[\lambda (\pi - \gamma)] \sin[2 (1 - \lambda)\beta^2 ]  - (\lambda-1) \sin[\lambda (\pi - \gamma )+2\gamma] \notag \\
    &\times \sin[2 (2 - \lambda) \beta^2] \Big) \Big]   \notag\\
    B_6^2 = & \frac{(\lambda -2) (\lambda -1) \csc \lambda \pi \thinspace (\mu^0 - \mu^2)}{8 \mu^0 \mu^2 (1 - \nu^0)} \Big[ (\lambda -1) \cos[ \lambda (\pi - \gamma) + 2\gamma ] \sin[ 2 (2 - \lambda) \beta^2 ] + (\lambda + 4\nu^0 -2) \cos[ \lambda (\pi - \gamma) ]\notag \\
    &\times  \sin[ 2(\lambda -1) \beta^2 ] \Big]  \notag\\
    B_7^2 = & \frac{(1 - \lambda) \csc \lambda \pi}{8 \mu^0 \mu^2 (1 - \nu^0)} \Big[ (\lambda -1) (\mu^0 - \mu^2) (\lambda +4 \nu^0 -2) \sin[\lambda (\pi - \gamma)] \sin 2 \lambda \beta^2\thinspace + \sin[ \lambda (\pi - \gamma) + 2\gamma ] \Big( \lambda^2 \mu^0 \sin[2 (1 - \lambda) \beta^2 ] \notag \\
    & + 2 (\lambda -1) \big[ \mu^2 (1 - 2\nu^0) - \mu^0 (1-2\nu^2) \big] \sin 2 \beta^2 \thinspace  + \big( \mu^2 \big[ (\lambda-2)\lambda +4 \nu^0 -2 \big] + 2 \mu^0 (\lambda -2 \nu^2 +1) \big) \sin[2 (\lambda -1) \beta^2 ]  \Big) \Big]   \notag\\
    B_8^2 = & \frac{(1 - \lambda) \csc \lambda \pi}{8 \mu^0 \mu^2 (1 - \nu^0)} \Big[ \cos [ \gamma (\lambda -2) - \lambda \pi ] \Big( 2 (\lambda -1)  \big[ \mu^2 (1 - 2\nu^0) - \mu^0 (1-2\nu^2) \big] \sin 2 \beta^2 \thinspace + \big( \mu^0 \big[ (\lambda -2) \lambda +4 \nu^2 -2 \big]  \notag \\
    &- \mu^2 \big[ (\lambda -2) \lambda +4 \nu^0 -2 \big] \big) \sin [2 (\lambda -1) \beta^2 ] \Big) - (\lambda -1) (\mu^0 - \mu^2) (\lambda +4 \nu^0 -2) \cos [ \lambda (\pi - \gamma) ] \sin 2 \lambda \beta^2 \thinspace \Big] 
    \label{eq:off_second_line}
\end{align}
\end{small}

\noindent (iii) the third row of $\bm{\mathcal{B}}^{UR}$
\begin{small}
\begin{align}
   B_5^3 = & \frac{(\lambda -2) (\lambda -1) \csc \lambda \pi \thinspace (\mu^0 - \mu^2)}{8 \mu^0 \mu^2 (1 - \nu^0)} \Big[ \lambda \cos[ \lambda (\pi - \gamma) ] \sin[2  (\lambda -1) \beta^2]+ (\lambda -1) \cos[ \lambda (\pi - \gamma) + 2\gamma]\notag \\
    &\times \sin[2 (2 - \lambda) \beta^2] \Big]   \notag\\
    B_6^3 = & \frac{(2 - \lambda) (\lambda -1) \csc \lambda \pi \thinspace (\mu^0 - \mu^2)}{8 \mu^0 \mu^2 (1 - \nu^0)} \Big[ \lambda \sin[ \lambda (\pi - \gamma) ] \sin[ 2  (\lambda -1) \beta^2]+ (\lambda -1) \sin[ \lambda (\pi -\gamma) + 2\gamma ] \notag \\
    &\times \sin[ 2 (2 - \lambda) \beta^2 ] \Big]  \notag\\
    B_7^3 = & \frac{(1 - \lambda) \csc \lambda \pi}{8 \mu^0 \mu^2 (1 - \nu^0)} \Big[ \lambda (\lambda -1) (\mu^0 - \mu^2) \cos[ \lambda (\pi - \gamma) ] \sin 2 \lambda \beta^2 \thinspace + \cos[ \lambda (\pi - \gamma) + 2\gamma ] \Big( 2 (\lambda -1)  \big[ \mu^2 (1 - 2\nu^0) \notag \\ & - \mu^0 (1-2\nu^2) \big] \sin 2 \beta^2 \thinspace + \sin[2  (\lambda -1) \beta^2] \big( \mu^2 ( (\lambda -2) \lambda +4 \nu^0 -2 ) - \mu^0 \big[ (\lambda -2) \lambda +4 \nu^2 -2 \big] \big) \Big) \Big]   \notag\\
    B_8^3 = & \frac{(1 - \lambda) \csc \lambda \pi}{8 \mu^0 \mu^2 (1 - \nu^0)} \Big[  \lambda(\lambda -1) (\mu^0 - \mu^2) \sin[ \lambda (\pi - \gamma) ] \sin 2 \lambda \beta^2 \thinspace - \sin[ \lambda (\pi - \gamma) + 2 \gamma ] \Big( 2 (\lambda -1)  \big[ \mu^2 (1 - 2\nu^0) \notag \\ & - \mu^0 (1-2\nu^2) \big] \sin2 \beta^2\thinspace+  \big( \mu^0 \big[(\lambda -2) \lambda +4 \nu^2 -2\big] - \mu^2 \big[ (\lambda -2) \lambda +4 \nu^0 -2 \big] \big)\sin[2  (\lambda -1) \beta^2] \Big) \Big]  
    \label{eq:off_third_line}
\end{align}
\end{small}

\noindent (iv) the fourth row of $\bm{\mathcal{B}}^{UR}$
\begin{small}
\begin{align}
    B_5^4 = & \frac{(\lambda -2) (\lambda -1) \csc \lambda \pi \thinspace (\mu^0 - \mu^2)}{8 \mu^0 \mu^2 (1 - \nu^0)} \Big[ (\lambda -1) \cos [ \lambda (\pi - \gamma) + 2\gamma] \sin [2 (2 - \lambda) \beta^2] + (\lambda +4 \nu^0 -4) \cos [ \lambda (\pi - \gamma) ]\notag \\
    &\times \sin [2  (\lambda -1) \beta^2] \Big]  \notag\\
    B_6^4 = & \frac{(2 - \lambda) (\lambda -1) \csc \lambda \pi \thinspace (\mu^0 - \mu^2)}{8 \mu^0 \mu^2 (1 - \nu^0)} \Big[ (\lambda -1) \sin [ \lambda (\pi - \gamma) + 2\gamma ] \sin [2 (2 - \lambda) \beta^2] + (\lambda +4 \nu^0 -4) \sin [ \lambda (\pi - \gamma) ]\notag \\
    &\times \sin [2  (\lambda -1) \beta^2] \Big]  \notag\\
    B_7^4 = & \frac{(1 - \lambda) \csc \lambda \pi}{8 \mu^0 \mu^2 (1 - \nu^0)} \Big[ (\lambda -1) (\mu^0 - \mu^2) (\lambda +4 \nu^0 -4) \cos[ \lambda (\pi - \gamma) ] \sin 2 \lambda \beta^2 \thinspace + \cos[ \lambda (\pi - \gamma) + 2 \gamma ] \Big( 2 (\lambda -1) \big[ \mu^2 (1 - 2\nu^0) \notag \\ & - \mu^0 (1-2\nu^2) \big] \sin 2 \beta^2 \thinspace  +\big( \mu^2 \big[ (\lambda -2) \lambda +4 \nu^0 -2 \big] - \mu^0 \big[ (\lambda -2) \lambda +4 \nu^2 -2 \big] \big) \sin[2 (\lambda -1) \beta^2 ]  \Big) \Big]  \notag\\
    B_8^4 = & \frac{(1 - \lambda) \csc \lambda \pi}{8 \mu^0 \mu^2 (1 - \nu^0)} \Big[ (\lambda -1) (\mu^0 - \mu^2) (\lambda +4 \nu^0 -4) \sin[ \lambda (\pi - \gamma) ] \sin 2 \lambda \beta^2 \thinspace - \sin[ \lambda (\pi - \gamma) + 2\gamma ] \Big( 2 (\lambda -1)  \big[ \mu^2 (1 - 2\nu^0) \notag \\ & - \mu^0 (1-2\nu^2) \big] \sin 2 \beta^2 \thinspace + \big( \mu^0 \big[ (\lambda -2) \lambda +4 \nu^2 -2 \big] - \mu^2 \big[ (\lambda -2) \lambda +4 \nu^0 -2\big] \big) \sin[2 (\lambda -1) \beta^2 ] \Big) \Big]  \notag\\
    \label{eq:off_fourth_line}
\end{align}
\end{small}

The off-diagonal coefficients of the lower left block ($\bm{\mathcal{B}}^{LL}$) can be derived from Eqs. (\ref{eq:off_first_line} - \ref{eq:off_fourth_line}) by: (i) replacing the material properties $\mu^2, \nu^2$ with $\mu^1, \nu^1$, respectively; (ii) substituting the opening angles $\beta^2$ with $\beta^1$; and (iii) replacing the angle $\gamma$ with $2 \pi - \gamma$. The singularity parameter $\lambda$ is determined by the determinant of the matrix in Eq. (\ref{eq:two_elastic}). \textit{Due to the length of the determinant, the Mathematica script entitled "Elastic Singularity.nb" is provided as the supplemental material.}

\section{Verification and comparison of the EIM formulae}
Williams \cite{Williams1952,Williams1956} first investigated stress singularity of the angular single-material wedge, and Dempsey and Sinclair extended Williams' method on composite wedges. Following their solution scheme, the singular behaviors were successfully detected for composite wedges. As previously mentioned in Section 3 and Section 4, the singularity parameter $m$ (for heat conduction) and $\lambda$ (for elasticity) can be determined by solving the linear eigenvalue problem of global matrices. Based on the energy criteria, singularity parameters within the range [0, 1] are of our interests, despite there are numbers of roots satisfying the eigenvalue problem. Hence, this section compares dominant singularity parameters solved by Eshelby's equivalent inclusion method and these celebrated works, which serve as both a verification and tribute to pioneers. 

Prior to comparisons, some details are mentioned here: (i) the opening angle of each triangular subdomain $2 \beta^I \in [0, \pi]$. For ``wedge-like'' inhomogeneities with opening angles exceeding $\pi $, they can be represented by two triangular inhomogeneities of the same material. For instance, say the opening angle is $\frac{3}{2} \pi$, it can be considered as two inhomogeneities $2 \beta^1 = 2\beta^2 = \frac{3}{4} \pi$, and the angle $\gamma$ between their symmetric lines is $\frac{3}{4} \pi$, see Fig. \ref{fig:verification of elastic_JMPS_F} and Fig.\ref{fig:verification of elastic_JMPS_S}. (ii) The interfacial effects of bimaterial media are achieved by changing the symmetric line and opening angles of one inhomogeneity. Two representative configurations are shown in Fig. \ref{fig:bimaterial_thermal_verify} and Fig. \ref{fig:example of thermal_interface}.

\subsection{Heat flux singularity}
\begin{figure}
    \centering
    \includegraphics[width=0.5\linewidth]{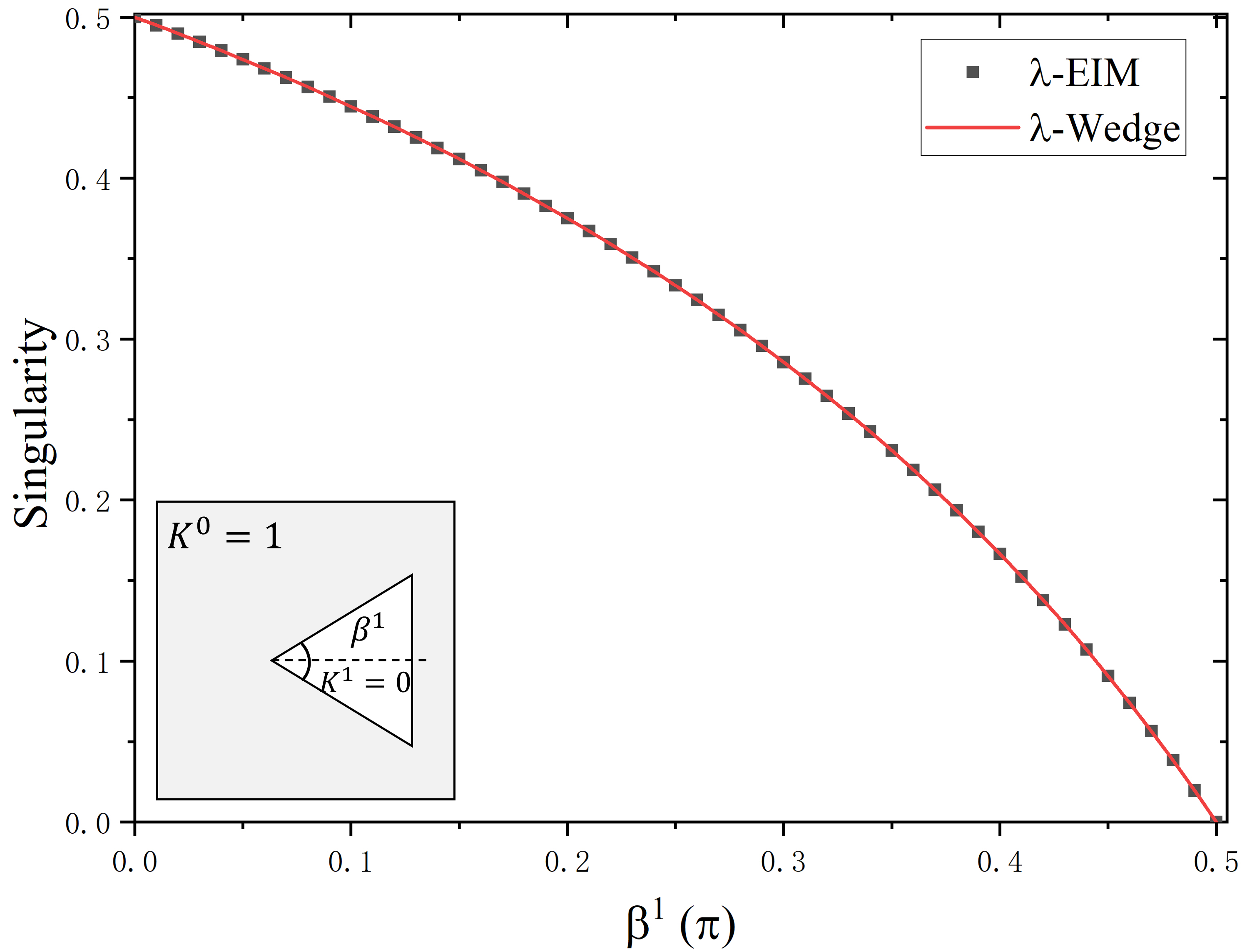}
    \caption{Comparison of the dominant singularity $m$ of a triangular void versus the opening angle $\beta^1$ by Eshelby's EIM with the classic wedge solution of the re-entrant corner.}
    \label{fig:verification of thetrmal_single void case}
\end{figure}

Solving Eq. (\ref{eq:Det_single_void}) with various of $\beta^1 \in [0, \pi/2]$, the singularity parameter $m$ for heat flux can be determined. Fig. \ref{fig:verification of thetrmal_single void case} plots the variation of $m$ versus the opening angle $\beta^1$. The singularity parameter decreases with the opening angle, which reaches the maxima and minima when the opening angle is $0$ and $0.5 \pi$, respectively. When the opening angle is $0$, the triangular void reduces to the classic slit-like crack problem, and the singularity level is $0.5$. The excellent agreement between two curves reveal that the singularity parameter solved by Eshelby's EIM is exact as the classic solution of the edge. 

\begin{figure}
\includegraphics[width = 1\textwidth,keepaspectratio]{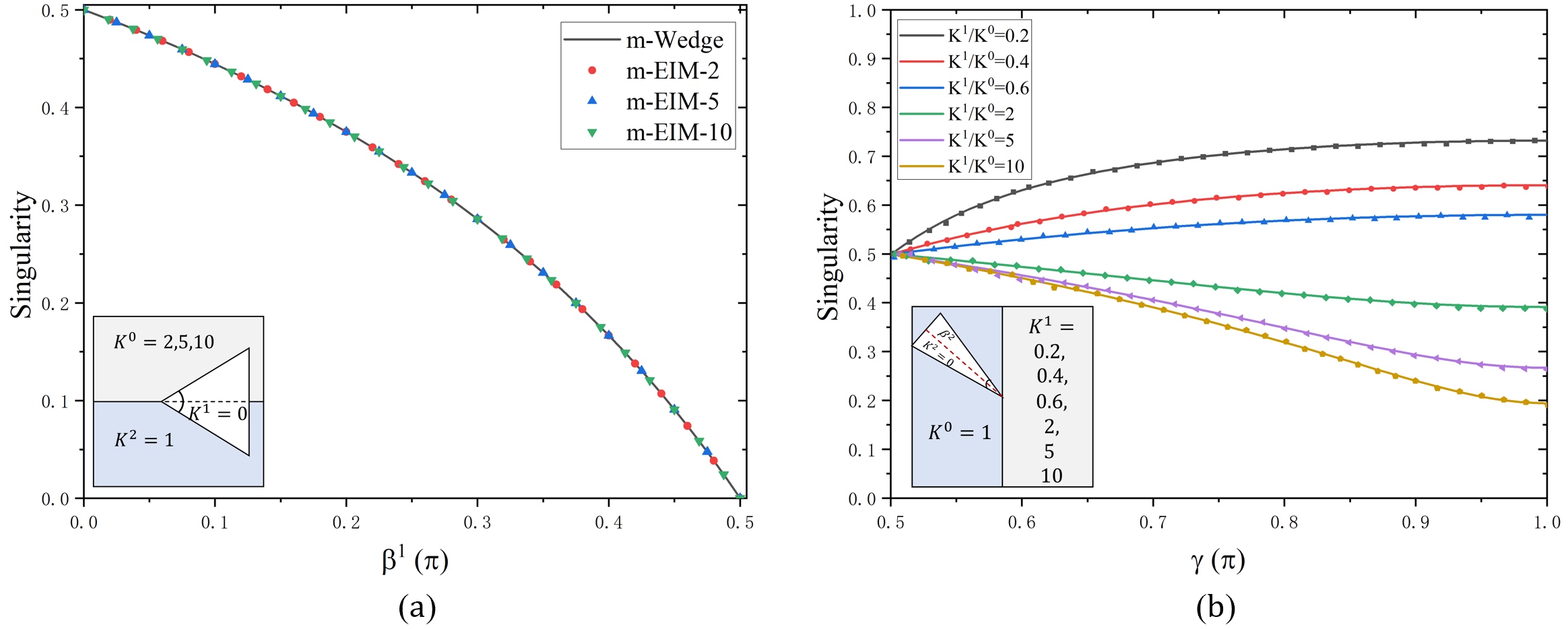}
\caption{Comparison and verification of heat flux singularity $m$ between Eshelby's EIM at the tip of a triangular inhomogeneity and the classic solutions. (a) When the triangular void (opening angle $\beta^1 \in [0, \frac{1}{2}] \pi$) is parallel with the bimaterial interface, and the thermal conductivity ratio $K^0 / K^2 = 2, 5$ and $10$; (b) When the triangular void (slit-like crack $\beta^1 \to 0^+$) is inclinedly intersected with the bimaterial interface, and the thermal conductivity ratio $K^1 / K^0 = 0.2, 0.4, 0.6, 2, 5$ and $10$. The straight lines denote EIM's predictions and colored circles represent Chen and Huang's solution \cite{Chen1992}}
\label{fig:bimaterial_thermal_verify}
\end{figure}

Subsequently, Eq. (\ref{eq:det_thermal_multiple}) governs the singularity parameter for two triangular inhomogeneities. The schematic diagram in Fig. \ref{fig:bimaterial_thermal_verify} (a) show that Eshelby's EIM with two inhomogeneities can be extended to simulate bimaterial interfacial effects. Specifically, two subdomains are arranged as: (i) the symmetric line of the triangular void is placed along the $x_1$ axis, with the opening angle $\beta^1$; (ii) the symmetric line of the second triangular subdomain $\gamma = \frac{1}{2} (\pi + \beta^1)$, with opening angles $\beta^2 = \frac{1}{2} (\pi - \beta^1)$. Fig. \ref{fig:bimaterial_thermal_verify} (a) reveals that the singularity parameter is completely independent from the thermal conductivity ratio of the upper and lower layers, and exact predictions can be obtained from Eshelby's EIM by solving Eq. (\ref{eq:det_thermal_multiple}). In addition, this interesting phenomenon can be well explained in mathematical expressions. Substituting $K^1 = 0, \gamma = \frac{1}{2} (\pi + \beta^1), \beta^2 = \frac{1}{2} (\pi - \beta^1)$ into Eq. (\ref{eq:det_thermal_multiple}) yields,

\begin{equation}
    \text{Det} = \frac{K^0 (K^0 + K^2) (K^0 - K^2)^2}{8 (K^0)^3 K^2} (1 - m)^4 \csc ^2 [m \pi] \thinspace \sin \left[2 (m-1) \beta^1\right] \sin \left[2 m (\pi - \beta^1) + 2\beta^1 \right] = 0
    \label{eq:det_bimaterial_special}
\end{equation}

Since $K^0$, $K^2$ are positive and $K^0 \neq K^2$, the root (singularity parameter) $m$ that makes the determinant zero is completely independent from $K^0$ and $K^2$. Note that when $K^0 = K^2$, the problem reduces to a single triangular void embedded in the infinite media. Based on Eq. (\ref{eq:Det_single_void}), the singularity parameter $m$ is still independent from $K^0$. 

Beyond the case of one inhomogeneity parallel with the bimaterial interface, Fig. \ref{fig:bimaterial_thermal_verify} (b) considers a slit-like crack, which is inclinedly intersected with the bimaterial interface. As the schematic diagram in Fig. \ref{fig:bimaterial_thermal_verify} (b) show, two subdomains are arranged as: (i) the symmetric line of the first subdomain is placed along the $x_1$ axis, with the opening angle $\beta^1 = \frac{\pi}{2}$; and (ii) the symmetric line of the second triangular void is placed along $\gamma \in [\frac{1}{2}, 1] \pi$, with the opening angle $\beta^2 = 0$. Fig. \ref{fig:bimaterial_thermal_verify} (b) compares singularity parameters evaluated by the EIM (lines) and Chen and Huang's solution \cite{Chen1992} (circles). When $\gamma = \frac{\pi}{2}$, all cases exhibit the same singularity $m = 0.5$, which has been demonstrated in Fig. \ref{fig:bimaterial_thermal_verify} (a). When the angle between the bimaterial interface and the slit-like crack increases, the singularity parameter increases when the thermal conductivity ratio $K^1 / K^0$ is less than 1, and vice versa. Such a phenomenon can be interpreted as the intensive bimaterial interfacial effect. The excellent agreement between lines and circles reveal that the EIM can provide the exact predictions. 

\subsection{Bimaterial stress singularity}
When a triangular inhomogeneity is embedded in a bimaterial medium, intensive bimaterial interfacial effects trigger more complicated singularities. In the following, let $\lambda_F$ and $\lambda_S$ denote the real part of the first and second singularity, and let $\xi$ denote the magnitude of the imaginary part. When the singularity is complex, two singularities occur as a conjugate pair, i.e., $\lambda = \lambda_F \pm i \xi$ ($\lambda_F = \lambda_S$). Because two singularities share the same real part and the same magnitude of the imaginary part $\xi$, it is unnecessary to distinguish the imaginary parts of the first and second singularities. For the special case of symmetric geometric settings, $\lambda_F$ and $\lambda_S$ reduces to the symmetric and anti-symmetric singularities, respectively.

\begin{figure}
    \includegraphics[width = 1\textwidth,keepaspectratio]{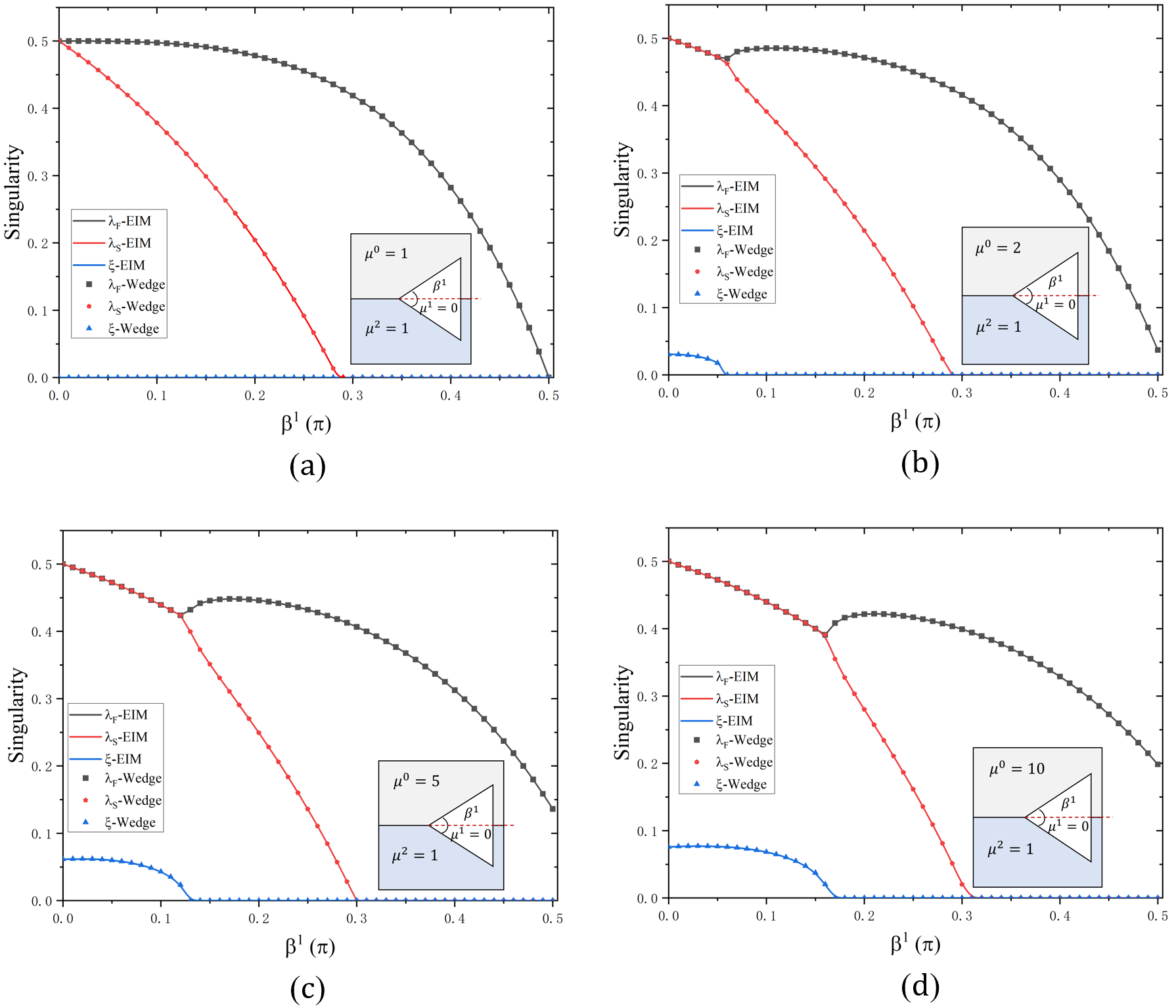}
    \caption{Comparison and verification of dominant complex stress singularity $\lambda$ between the EIM of singularity at the tip of a triangular inhomogeneity in infinite space and the classic Williams's composite wedges problems, $\lambda_F$ and $\lambda_s$ represents symmetric/anti-symmetric real stress singularity, respectively, and $\xi$ denotes virtual stress singularity, opening angle $\beta^1 \in (0, \frac{\pi}{2})$ and $\beta^2 = \frac{\pi}{2}$ with shear modulus $\mu^1 = 0$, and Poisson's ratio $\nu^0 = \nu^2 = 0.3.$ (a) ratio of shear modulus $\mu^0 / \mu^2 = 1$; (b) ratio of shear modulus $\mu^0 / \mu^2 = 2$; (c)ratio of shear modulus $\mu^0 / \mu^2 = 5$; (d) ratio of shear modulus $\mu^0 / \mu^2 = 10$;}
    \label{fig:verification of elastic_wedge}
\end{figure}

Figs. \ref{fig:verification of elastic_wedge} (a) verifies the EIM formulation against the classic Williams' solution of a single-material wedge \cite{Williams1952}. For a single triangular void, only the left upper part $\bm{\mathcal{B}}^{UL}$ in Eq. (\ref{eq:two_elastic}) is applied. Subsequently, the singularity parameters in Figs. \ref{fig:verification of elastic_wedge} (b-d) are based on the determinant of Eq. (\ref{eq:two_elastic}). When the triangular void is embedded in an infinitely large medium, the first (symmetric) and second (anti-symmetric) singularity parameters are real. For the bimaterial medium, Figs. \ref{fig:verification of elastic_wedge} (c-d) set the ratio of shear moduli ($\mu^0 / \mu^2$) between two layers as $2, 5$, and $10$, respectively. These singularities are solved following Dempsey and Sinclair's formulae \cite{Dempsey1979}. The mismatch of the upper and lower layers significantly changes singularities.

When the opening angle $\beta^1 = \frac{\pi}{2}$, the first singularity ($\lambda_F$) is non-zero. For instance, when the shear modulus ratio is $2, 5, 10$, the singularity parameters are $0.037$, $0.136$, $0.198$, respectively. The occurrence of the second singularity is delayed for larger opening angles. For example, when the shear modulus ratio is $2,5, 10$, the second singularity vanishes when the opening angles are $0.29\pi$, $0.30\pi$, $0.31\pi$, respectively. When the opening angle is below $0.06\pi$, $0.13\pi$, $0.17\pi$ for the shear modulus ratio is $ 2, 5, 10$, respectively, the first and second singularities merge, which becomes the complex singularity. While the real part of the singularity characterize the extent of singular behaviors, the imaginary part show the oscillatory features of elastic fields. The real and imaginary parts reaches maxima for $\beta^1 \to 0^+$ (slit-like cracks), and the maxima real part remain constant $0.5$. The good agreement among curves, including the real and imaginary parts of singularities, validate the EIM formulation in Eq. (\ref{eq:two_elastic}) is exact for elasticity.

\begin{figure}
    \centering
    \includegraphics[width=0.67\textwidth,keepaspectratio]{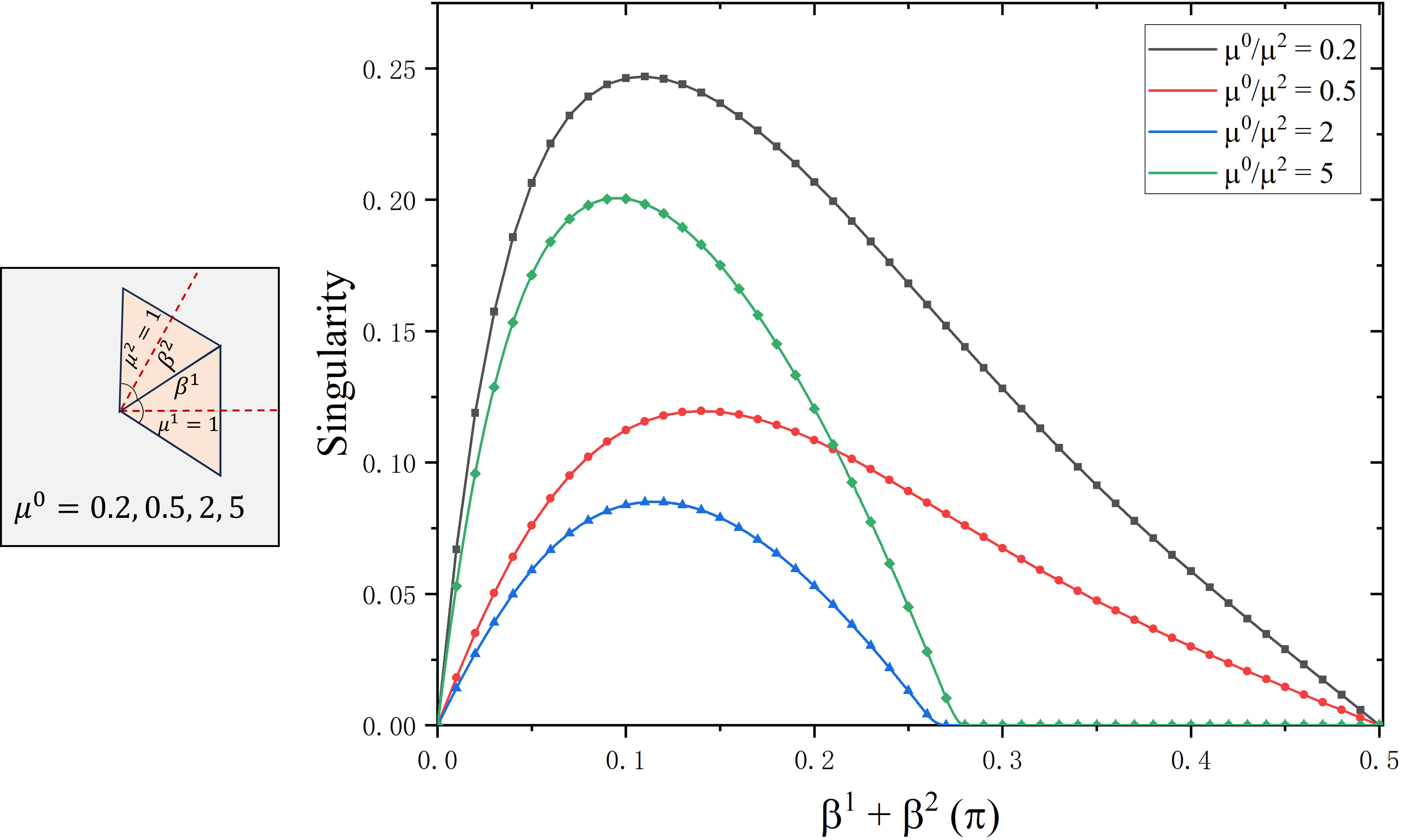}
    \caption{Comparison and verification of symmetric dominant stress singularity parameter $\lambda_F$ between the EIM of singularity at the tip of a single triangular inhomogeneity embedded in the infinite space.
    The straight lines denoted EIM's predictions is evaluated by placing two same triangular inhomogeneities bonded together, with opening angles $\beta^1 = \beta^2$, shear moduli $\mu^1 = \mu^2$, and Poisson's ratio $\nu^1 = \nu^2 = 0.3$. The colored symbols is evaluated by the EIM formulation Eq. (54) and Eq. (58) in \cite{Wu2024} of a single triangular inhomogeneity, with the opening angle $\beta^a = \beta^1 + \beta^2$, shear modulus $\mu^a = \mu^1$, and Poisson's ratio $\nu^a = \nu^1$. Four ratios of shear moduli are considered:  $\mu^0 / \mu^2 = 0.2, 0.5, 2, 5$.}
    \label{fig:verification of elastic_JMPS_F}
\end{figure}

\begin{figure}
    \centering
    \includegraphics[width=0.67\textwidth]{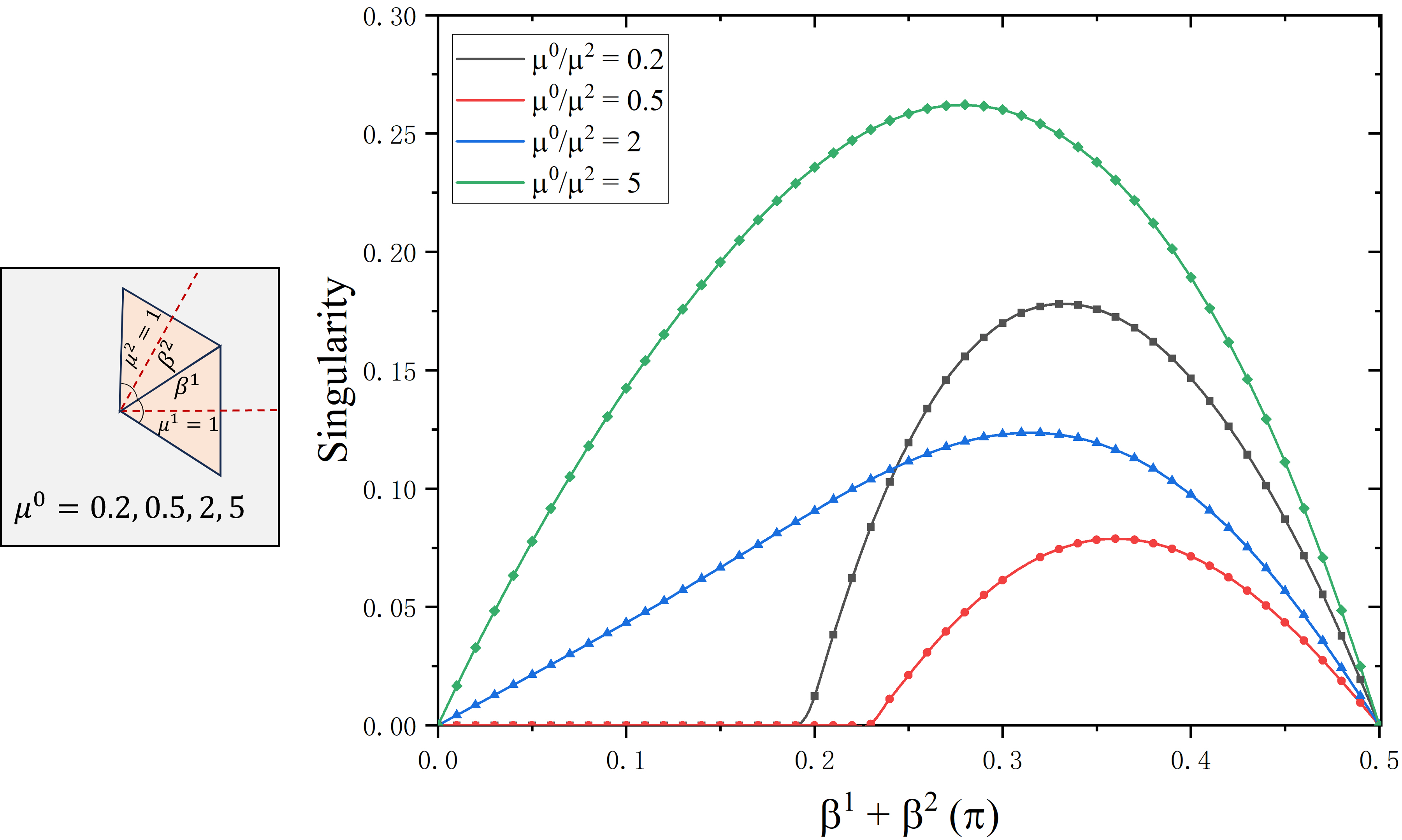}
    \caption{Comparison and verification of anti-symmetric dominant stress singularity parameter $\lambda_S$ between the EIM of singularity at the tip of a single triangular inhomogeneity embedded in the infinite space.
    The straight lines denoted EIM's predictions is evaluated by placing two same triangular inhomogeneities bonded together, with opening angles $\beta^1 = \beta^2$, shear moduli $\mu^1 = \mu^2$, and Poisson's ratio $\nu^1 = \nu^2 = 0.3$. The colored symbols is evaluated by the EIM formulation Eq. (54) and Eq. (58) in \cite{Wu2024} of a single triangular inhomogeneity, with the opening angle $\beta^a = \beta^1 + \beta^2$, shear modulus $\mu^a = \mu^1$, and Poisson's ratio $\nu^a = \nu^1$. Four ratios of shear moduli are considered: $\mu^0 / \mu^2 = 0.2, 0.5, 2, 5$.}
    \label{fig:verification of elastic_JMPS_S}
\end{figure}

Subsequently, it is essential to illustrate that when two identical triangular inhomogeneities are bonded together, whether its singular behaviors can be captured by the EIM formulation for the single triangular inhomogeneity. The Poisson's ratios are $\nu^0 = \nu^1 = \nu^2 = 0.3$, and four ratios of shear moduli $\mu^0 / \mu^1 = \mu^0 / \mu^2 = 0.2, 0.5, 2, 5$ are considered. Because the case study is dependent on the total opening angle, $\beta^1 = \beta^2$ is adopted for convenience. As the schematic diagrams in Fig. \ref{fig:verification of elastic_JMPS_F} and Fig. \ref{fig:verification of elastic_JMPS_S} indicate, (i) the symmetric line of the first triangular subdomain is along $x_1$ axis; and (ii) the symmetric line of the second subdomain is along $\gamma = 2 \beta^1$.  

The growth rate of the symmetric singular parameter $\lambda_F$ increases more rapidly than that of the anti-symmetric singular parameter $\lambda_S$ as the opening angle $\beta^1$ approaches $0.15\pi$. For the symmetric case shown in Fig. \ref{fig:verification of elastic_JMPS_F}, the symmetric singularity vanishes more quickly when the shear modulus ratio "$ \mu^0  > \mu^2$".  For instance, when the shear modulus ratio are $2,5$, the symmetric singularity vanishes when the opening angles are $0.27\pi$, $0.28\pi$, respectively. On the contrary, when the shear modulus ratio are $0.2, 0.5$, it did not disappear until the opening angle is $0.5 \pi$. At the same time, the anti-symmetric singularity does not exist until the opening angle is $0.2\pi$, $0.23\pi$ for the shear modulus ratio $ \mu^0 / \mu^2 = 0.2, 0.5$, respectively. These preceding analysis suggests that this approach provides a straightforward and robust method for reproducing the solution proposed by Wu and Yin.

This section compares the dominant heat flux and stress singularity parameters obtained by Eshelby's EIM with classical analytical solutions. Specifically, (i) EIM reproduces the heat flux singularity by Williams-type solutions and Chen and Huang's solution for bimaterial cracks; (ii) EIM reproduces the stress singularity for the single void embedded in the bimaterial medium, and recovers the stress singularity of one inhomogeneity in our recent work \cite{Wu2024} by two identical inhomogeneities. The above verifications validate the accuracy of the EIM framework on singularities in both heat conduction and elasticity.

\section{Results and discussion}
Section 6 applies the formulae to analyze singularities for inhomogeneities embedded in the bimaterial medium. Particularly, this section focuses on two special cases where the inhomogeneity is oriented parallel and perpendicular to the bimaterial interface. Note that other inclined cases can be straightforwardly implemented through adjusting symmetric lines of inhomogeneities, and one example has been plotted in Figs. \ref{fig:bimaterial_thermal_verify} in Section 5.  The two orientations of inhomogeneities are selected due to the reasons: (i) the parallel case is usually observed as debonding of the interface, i.e., laminate and coatings; and (ii) the perpendicular case can be found in through-interface cracks and defects. 

\subsection{Heat flux singularity} 
\begin{figure}
\centering
\includegraphics[width = 0.67\textwidth,keepaspectratio]{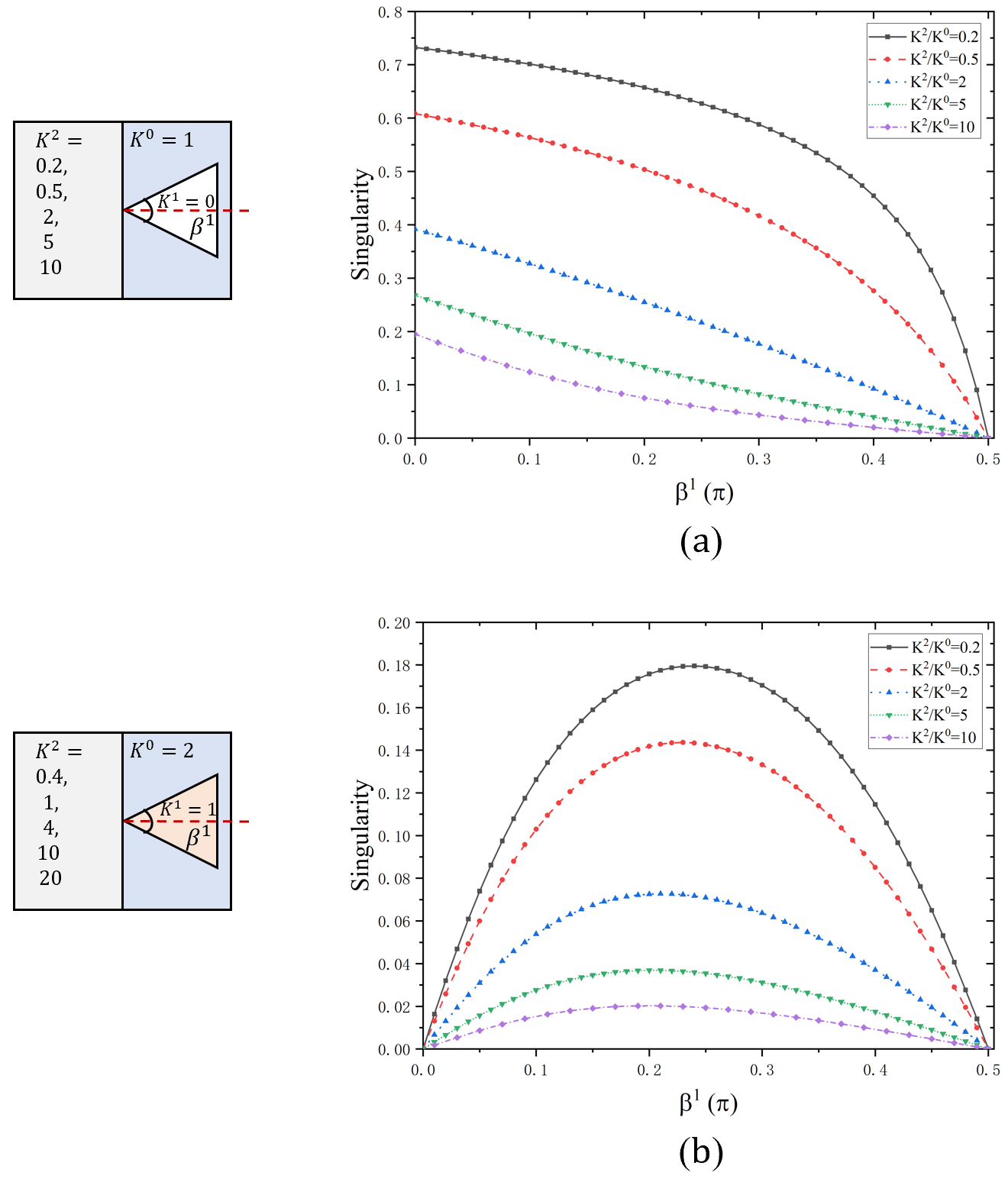}
\caption{Variation of heat flux singularity parameter $m$ versus the opening angle $\beta^1 \in [0, \frac{\pi}{2}]$, when a triangular (a) void ($K^1 / K^0 = 0$) and (b) triangular inhomogeneity ($K^1 / K^0 = \frac{1}{2}$) is oriented perpendicular to the bimaterial interface. Five ratios of thermal conductivity are considered: $K^2 / K^0 = 0.2, 0.5, 2, 5, 10$.}
\label{fig:example of thermal_interface}
\end{figure}

Fig. \ref{fig:example of thermal_interface} (a) displays the variation of the heat flux singularity parameter $m$ when a void is oriented perpendicular to the bimaterial interface. Two subdomains are arranged as: (i) the symmetric line of the first domain is placed along the $x_1$ axis, with the opening angle $\beta^1$; and (ii) the symmetric line of the second subdomain $\gamma = \pi$, with the opening angle $\beta^2 = \frac{\pi}{2}$. As Fig. \ref{fig:example of thermal_interface} (a) indicates, the thermal conductivity ratios significantly alter the singularity of the composite system. The singularity parameter increases with the thermal conductivity ratio ($ K^2/K^0$). For instance, when the void becomes a crack ($\beta^1 \to 0^+$), the singularity parameters are $0.732, 0.608, 0.392, 0.268$, and $0.195$, for thermal conductivity ratios $0.2, 0.5, 2, 5, 10$, respectively. Because the bimaterial interface allows for heat flux continuity, a smaller thermal conductivity ratio ($K^2 / K^0$) results in larger temperature gradients to achieve the same heat flux, which strengthens the singularity. In contrast, when the thermal conductivity ratio ($K^2 / K^0$) decreases, smaller temperature gradients weaken the singularity. Such a phenomenon should be interpreted as intensive bimaterial interfacial effects. For two extreme cases, the thermal conductivity ratio is infinite or approaching zero, the original bimaterial problem reduces to the half-space problem with Dirichlet and Neumann boundary conditions, respectively. For the Dirichlet case $K^2 / K^0 \to \infty$, the singularity parameter equals 1, which coincides with the singularity of the bimaterial fundamental solution \cite{Wu2023_JMPS}. Regarding the Neumann case $K^2 / K^0 = 0$, since the heat flux is zero across the interface, the singularity parameter becomes 0. Moreover, the singularity parameter decreases with the opening angle $\beta^1$, which is 0 when $\beta^1 \to \frac{\pi}{2}$.  

Fig. \ref{fig:example of thermal_interface} (a) displays the variation of the heat flux singularity parameter $m$ when an inhomogeneity is oriented perpendicular to the bimaterial interface. When the thermal conductivity ratio $K^1 / K^0$ is non-zero, the variation of the singularity parameter is significantly different from the void case. The singularity parameter is not a monotonic function of the opening angle, and there exists no singularities when $\beta^1 \to 0^+, \frac{\pi}{2}$. Compared to Fig. \ref{fig:example of thermal_interface} (a), when the inhomogeneity exhibits zero thermal conductivity, it yields the Neumann boundary condition, and the heat flux at the tip on the bimaterial interface suddenly drops to zero due to continuity conditions, which leads to a heat flux singularity. However, when the inhomogeneity exhibits finite thermal conductivity, it can be interpreted as a slit-like thin-layer with different thermal conductivity, which is similar to an imperfect bonding interface. Moreover, the maximum singularity occurs at larger opening angles; for instance, the maximum singularity parameter occurs at the opening angles of $\beta^1 = 0.24, 0.23, 0.21, 0.2 \pi$ for the thermal conductivity ratios $K^2 / K^0 = 0.2, 0.5, 2, 5, 10$, respectively. The maximum singularity parameter decreases with the increasing thermal conductivity ratios, which has been observed in Fig. \ref{fig:example of thermal_interface} (a) and interpreted as bimaterial interfacial effects.

\begin{figure}
    \centering
    \includegraphics[width=0.67\linewidth]{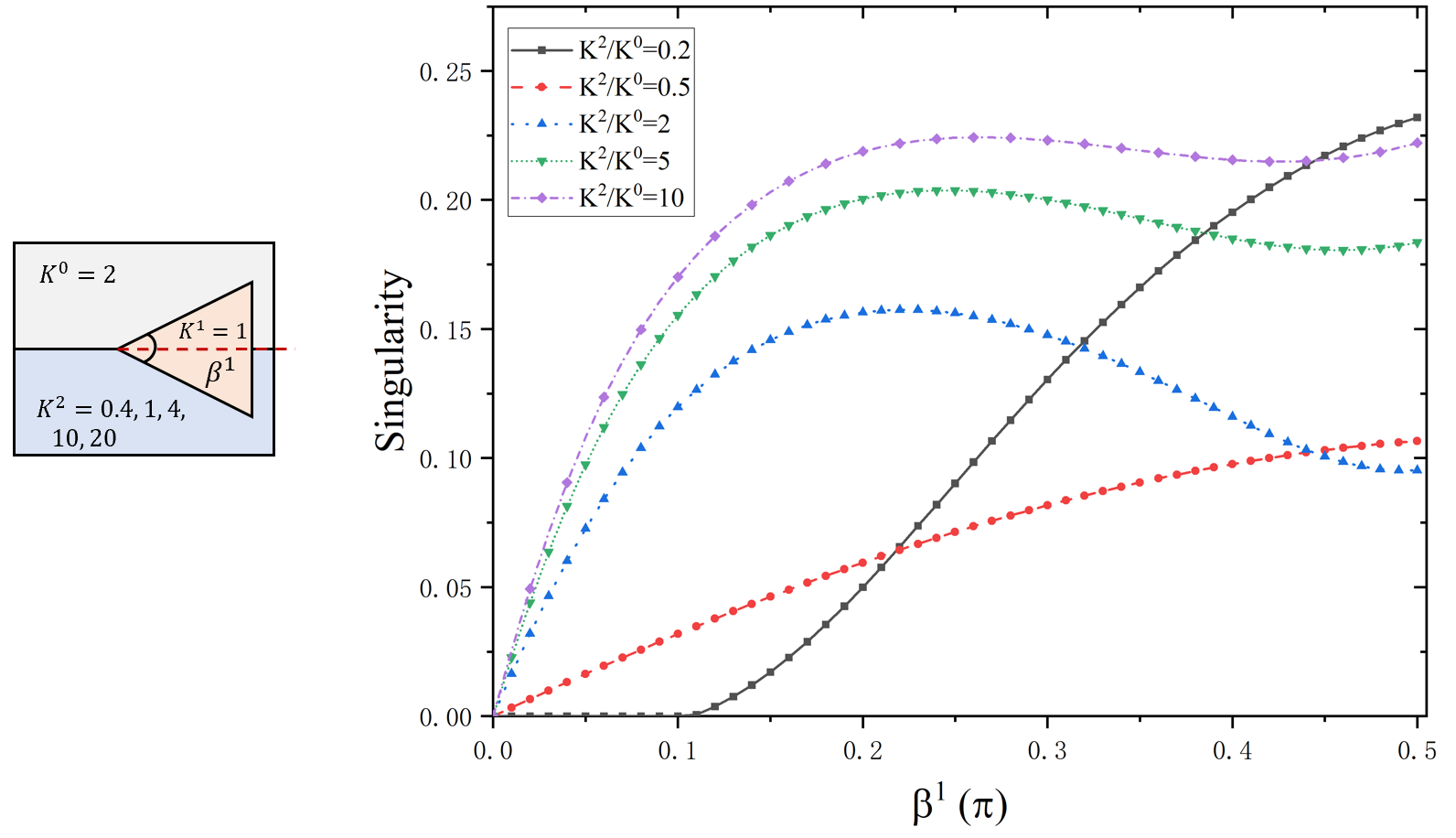}
    \caption{Variation of heat flux singularity parameter $m$ versus the opening angle $\beta^1 \in [0, \frac{\pi}{2}]$, when a triangular inhomogeneity is oriented parallel to the bimaterial interface, and the ratio of thermal conductivity between the inhomogeneity and the matrix is $K^1 / K^0 = \frac{1}{2}$. Five ratios of thermal conductivity are considered: $K^2 / K^0 = 0.2, 0.5, 2, 5, 10$.}
    \label{fig:example of thermal_crack}
\end{figure}

Fig. \ref{fig:example of thermal_crack} shows the singularity parameter $m$ versus the opening angle $\beta^1 \in [0, \frac{\pi}{2}]$, when the inhomogeneity is oriented parallel with the bimaterial interface. Note that Fig. \ref{fig:bimaterial_thermal_verify} (a) shows that when the inhomogeneity exhibits zero thermal conductivity, the singularity is independent of the thermal conductivity ratio of the two layers. However, as shown in Fig. \ref{fig:example of thermal_crack}, the singularity parameter is highly dependent on the thermal conductivity ratio $K^2 / K^0$, (i) when $K^2 > K^1$, the singularity parameter decreases with the increase of $K^2 / K^0$; i.e., when $\beta^1 \to \frac{\pi}{2}$, the singularity parameter is $0.0953$, $0.1835$, $0.2221$ when $K^2 / K^0 = 2, 5, 10$, respectively; (ii) when $K^2 = K^1$, the subdomain $\Omega^1$ and $\Omega^2$ merged as one ``wedge-like'' inhomogeneity with the angle $\pi + \beta^1$. The singularity parameter $m$ increases with the opening angle $\beta^1$, and the maxima occurs when $\beta^1 \to \frac{\pi}{2}$. When $K^2 < K^1$, $K^2 / K^0 = 0.2$, the singularity parameter exhibits thermal conductivity ratio-dependent delayed growth, which remains zero at smaller opening angles and starts to exist after a certain angle. 

\subsection{Stress singularity}
\begin{figure}
    \centering
    \includegraphics[width=1\linewidth]{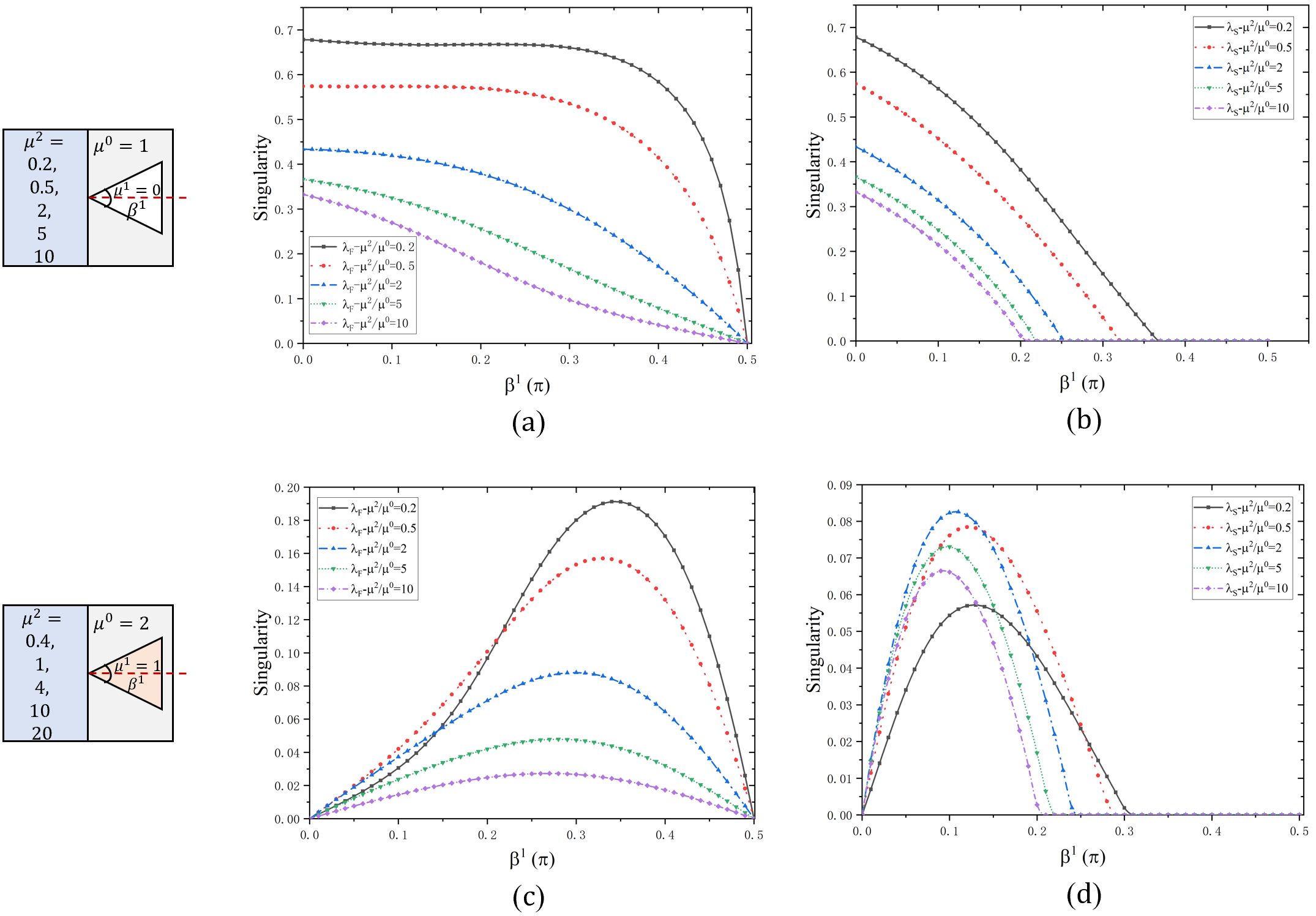}
\caption{Variation of the (a) symmetric ($\lambda_F$) and (b) anti-symmetric ($\lambda_S$) stress singularity regarding a triangular inhomogeneity of bimaterial problem perpendicular to the interface. The opening angle $\beta^1 \in (0,\frac{\pi}{2})$ and $\beta^2 = \frac{\pi}{2}$, Poisson's ratio $\nu^0 = \nu^2 = 0.3.$, shear modulus $\mu^1 = 0$, ratio of shear modulus $\mu^2 / \mu^0 = 0.2, 0.5, 2, 5, 10$; (c) and (d) $\mu^1 / \mu^0 = \frac{1}{2}$, five ratios of shear modulus $\mu^2 / \mu^0 = 0.2, 0.5, 2, 5, 10$.}
\label{fig:example of elastic_interface}
\end{figure}

When a void is oriented perpendicular to the bimaterial interface, Figs. \ref{fig:example of elastic_interface} (a) and (b) plot the variation of the stress singularity parameters $\lambda_F, \lambda_S$, respectively. The Poisson's ratios for the matrix and the second subdomain are 0.3, and the shear modulus ratios are $\mu^2 / \mu^0 = 0.1, 0.5, 2, 5$, and $10$, respectively. Note that the first subdomain is the void, which exhibits zero shear modulus and Poisson's ratio. As the diagram in Fig. \ref{fig:example of elastic_interface}(a) shows, the geometric setting is symmetric, and thus two singularity parameters $\lambda_F$ and $\lambda_S$ can be regarded as the symmetric and anti-symmetric singularity parameters, respectively. Recall the symmetric and anti-symmetric singularities in Fig. \ref{fig:verification of elastic_wedge} (a), when the shear modulus ratio $\mu^2 / \mu^0 = 1$, both symmetric and anti-symmetric cases exhibit the maximum singularity $0.5$ when $\beta^1 \to 0^+$. As Figs. \ref{fig:example of elastic_interface} (a-b) demonstrate, (i) the maximum symmetric and anti-symmetric singularity exist when $\beta^1 \to 0^+$ and the values are the same (machine precision has 0.1\% difference); (ii) the anti-symmetric singularity vanishes after the critical angle $\beta^c$, and $\beta^c$ decreases with the increase of the shear modulus ratio. When the second subdomain becomes stiffer, it prevent the growth of stress singularity. Particularly, when $\mu^2 \to 0$ or $\infty$, the original bimaterial problem reduces to the Neumann and Dirichlet half-space problems, respectively. The former boundary condition leads to $\frac{1}{r}$ singularity, which coincides with the singularity level of the fundamental solution; while the later boundary condition results in non-singular stress, as two subdomains exhibit the same material properties.  

When the void is filled in the elastic material ($\mu^1, \nu^1$), Figs. \ref{fig:example of elastic_interface} (c) and (d) plot the variation of stress singularity parameters $\lambda_F, \lambda_S$, respectively, and the Poisson's ratio $\nu^1 = 0.3$ and $\mu^1 / \mu^0 = \frac{1}{2}$ are utilized. Based on the shear modulus ratio $\mu^2 / \mu^0$, two cases are discussed that $\mu^2 > \mu^0$ and $\mu^2 < \mu^0$. When $\mu^2 > \mu^0$, the singularity parameter decreases as the shear modulus ratio gradually increases. For instance, Fig. \ref{fig:example of elastic_interface} (c) shows that the curves for the symmetric singularity $\lambda_F$ reach the maximum singularity when the opening angles are $0.30, 0.28, 0.27 \pi$ for $\mu^2/\mu^0 = 2, 5, 10$, respectively. The smaller shear modulus ratios postpone the opening angle for the maximum symmetric singularity. However, reverse trends can be observed for anti-symmetric singularities in Fig. \ref{fig:example of elastic_interface} (d). The curves for the anti-symmetric singularity $\lambda_S$ can reach $ 0.08253$, $ 0.0730$, $ 0.0665$ when the opening angle $\beta^1 = 0.11\pi, 0.10\pi, 0.09\pi$, respectively. When $\mu^2 < \mu^0$, the symmetric singularity parameter gradually increases and the anti-symmetric singularity parameter decreases with the shear modulus ratio $\mu^2 / \mu^0$. For the symmetric singularity shown in Fig. \ref{fig:example of elastic_interface} (c), the opening angle to reach the maxima value is even delayed. The maximum singularity for the curve ``$\mu^2 / \mu^0 = 0.2$'' is over two times the maximum value of the curve ``$\mu^2 / \mu^0 = 2$''. For the anti-symmetric singularity in Fig. \ref{fig:example of elastic_interface} (d), the anti-symmetric singularity can exist for larger opening angles. For instance, the singularity parameter did not vanish until the opening angle $\beta^1$ is approximately $0.31\pi$ for $\mu^2/\mu^0 = 0.2$, instead of $0.25\pi$ for $\mu^2/\mu^0 = 2$.

\begin{figure}
    \centering
    \includegraphics[width=1\linewidth]{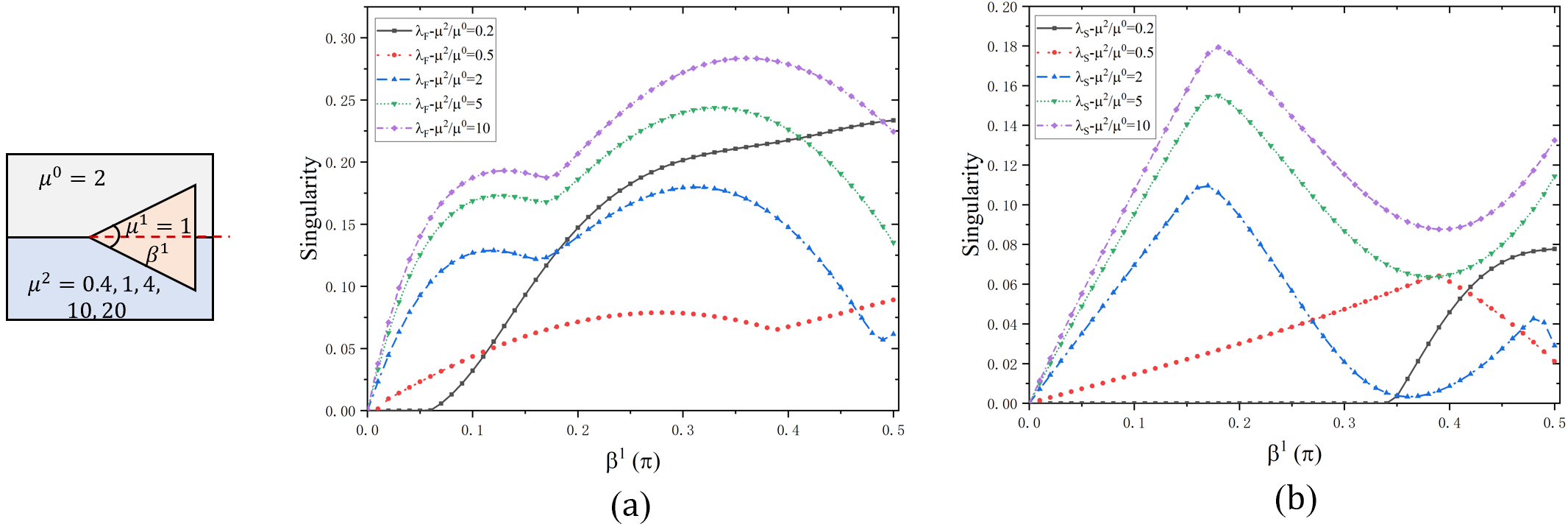}
    \caption{Variation of the (a) first ($\lambda_F$) and (b) second ($\lambda_S$) stress singularity regarding a triangular inhomogeneity of bimaterial problem oriented parallel to the interface. The opening angle $\beta^1 \in (0,\frac{\pi}{2})$ and $\beta^2 = \frac{\pi}{2}$, with Poisson's ratio $\nu^0 = \nu^1 = \nu^2 = 0.3.$, $\mu^1 / \mu^0 = \frac{1}{2}$, five ratios of shear modulus $\mu^2 / \mu^0 = 0.2, 0.5, 2, 5, 10$.}
    \label{fig:example of elastic_crack}
\end{figure}

Fig.\ref{fig:example of elastic_crack} (a) and (b) plot the variation of first ($\lambda_F$) and second ($\lambda_S$) singularity parameters of a triangular inhomogeneity oriented parallel to the bimaterial interface, when five shear modulus ratios $\mu^2 / \mu^0 = 0.2, 0.5, 2, 5, 10$ are considered. In Fig.\ref{fig:example of elastic_crack} (a), when the upper layer is stiffer, $\mu^2 / \mu^0 = 2, 5, 10$, there exists a dip around the opening angle $\beta^1 \approx 0.17 \pi$, which implies the competitive region of two eigenvalues. When the upper layer is softer, $\mu^2 / \mu^0 = 0.2$, an obvious singularity delay is observed, such that the singularity parameters do not occur until the opening angle is $0.06 \pi$ and $0.35 \pi$. In such a case, the bimaterial interfacial effects perform as a shield mechanism, which leads to the singularity delay. A similar phenomenon can be found in Figs. \ref{fig:verification of elastic_wedge} (a-d). The stress singularities depend on the shear modulus ratios of the inhomogeneity and the two layers, as well as the opening angles of the inhomogeneity. Although the above examples demonstrate some trends of stress singularities, it is necessary to analyze on a case-by-case basis.

\section{Conclusions}
This paper utilizes Eshelby's equivalent inclusion method to investigate the singularities at the vertex of multiply connected triangular inhomogeneities for heat conduction and elastic deformation. Using the separation of variables, the actual thermal and elastic fields can be expressed in terms of distance to the tip and angles. Based on the equivalent flux/stress conditions, the eigen-temperature-gradient (ETG) and eigenstrain are derived with the same singularity level as the actual fields. Since the actual thermal/elastic fields can be obtained by domain integrals of second-order partial derivatives of Green's function multiplied by the corresponding eigen-fields, the original boundary value problem has been converted into the Fredholm integral equation of the second kind. The dominant terms of domain integrals can be analytically captured through Taylor's series expansion. By employing the coordinate mapping and transformation, interactions between two triangular inhomogeneities have been derived, and the general formulae can be straightforwardly extended for the case of multiple inhomogeneities through switching of material and geometric properties. The present Eshelby's equivalent inclusion method compresses numbers of equations, and the extensions to half-space and bimaterial cases are implemented by modifying opening angles. The general formulae is validated against several classic solutions, including Williams', Dempsey and Sinclair's, and Chen and Huang's solution. For triangular inhomogeneities embedded in the infinite space, the EIM compresses the number of equations from $8$ to $4$ and from $16$ to $8$ for heat conduction and elasticity, respectively. With the systematic understanding of singularity pattern of eigen-fields, high fidelity calculation and modeling of the local field caused by angular inhomogeneities can be implemented with the EIM-based numerical methods. 

\section*{CRediT Author Statement}
\textbf{Yuangpeng Yang}: Methodology, Visualization, Writing-original draft; \textbf{Huiming Yin}: Writing Review \& Editing; \textbf{Chunlin Wu}: Conceptualization, Methodology, Supervision, Writing-original draft, Writing Review \& Editing. 

\section*{Acknowledgment}
CW's work is supported by the Chenguang Program No. 23CGA50 of Shanghai Education Development Foundation and Shanghai Municipal Education Commission, National Natural Science Foundation of China Grant No. 12302086. HY's work is sponsored by U.S. Department of Agriculture NIFA \#2021-67021-34201, and NIFA SBIR \#20233353039686, and National Science Foundation (NSF) grants IIP \#1738802 and IIP \#1941244. Those supports are gratefully acknowledged.

\appendix
\section{Difference between thermal Eshelby's tensors of interior/exterior parts of the vertex} \label{sec:dif_eshelby}
Fig. \ref{fig:discontinuous} plots the transformed coordinate \cite{Wu2021a} to evaluate domain integrals of the harmonic potential function. For the steady-state heat conduction, Eshelby's tensor for a uniformly distributed ETG can be written as:

\begin{figure}
    \centering
    \includegraphics[width=0.6\linewidth]{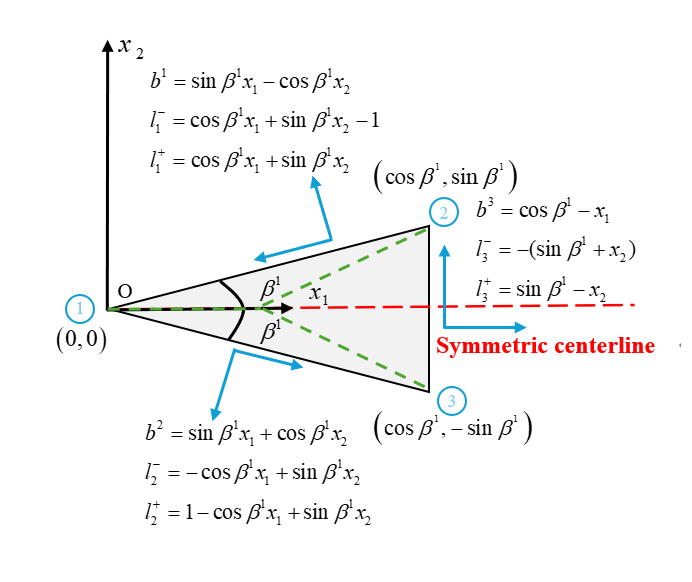}
    \caption{Schematic plot of a triangular subdomain $\Omega^I$ with the opening angle $2 \beta^I$, which is discretized into three sub-triangular domains characterized by the field point $\textbf{x}$ and three sides. The transformed coordinate is constructed to evaluate the domain integral of the harmonic potential function, including directional/normal vectors and $b_f$, $l_f^\pm$.}
    \label{fig:discontinuous}
\end{figure}

\begin{equation}
    D_{ij} = -\frac{1}{4\pi} \int_{\Omega^1} \phi_{,ij} \thinspace d\textbf{x}'
    \label{eq:thermal_Eshelby_tensor}
\end{equation}

The discontinuity term can be derived by finding the difference between Eshelby's tensors at two field points, ($0^+, 0$) and ($0^-, 0$) \cite{Wu2024}. Specifically, components of Eshelby's tensors, $D_{11}, D_{12}$, and $D_{22}$ are provided below: (the logarithmic parts are dropped)

\noindent ($\mathnormal{\mathrm{I}}$) Interior Cases: $x_1 \xrightarrow{}0^+$
\begin{equation}
\begin{split}
   D_{11} = &   \frac{1}{4 \pi} \left( 4\beta^1  + 2 \pi \sin^2\beta^1 \thinspace  + 4 \sin^2\beta^1 \thinspace \tan^{-1} \left[ \cot\beta^1 \right]\thinspace  \right)\\
   D_{12} = & \thinspace D_{21} = 0\\
   D_{22} = &  \frac{1}{4 \pi}\cos^2 \beta^1 \thinspace  \left( 2 \pi + 4 \tan^{-1} \left[ \cot \beta^1 \right] \right)
\end{split}
    \label{eq:Eshelby tensor_1}
\end{equation}

\noindent ($\mathnormal{\mathrm{II}}$) Exterior Cases: $x_1 \xrightarrow{}0^-$
\begin{equation}
\begin{split}
    D_{11} = & \frac{1}{4 \pi} \left( 4\beta^1  - 2 \pi \sin^2\beta^1 \thinspace  + 4 \sin^2 \beta^1 \tan^{-1}  \thinspace \left[\cot\beta^1 \right] \right)\\
   D_{12} = & \thinspace D_{21} = 0 \\
   D_{22} = &  -\frac{1}{4 \pi} \cos^2 \beta^1 \thinspace  \left(  2 \pi - 4 \tan^{-1} \left[ \cot\beta^1 \right] \right)
\end{split}
    \label{eq:Eshelby tensor_2}
\end{equation}

Therefore, the difference between Eshelby's tensors at the interior/exterior parts of the vertex along the symmetric line can be obtained as: 

\begin{equation}
    D^d_{11} = \sin^2 \beta^1, \quad D^d_{12} = D^d_{21} = 0, \quad D_{22}^d = \cos^2 \beta^1
    \label{eq:Eshelby tensor diff-final}
\end{equation}

\section{Domain integrals of Green's function and eigen-temperature-gradient} 

\subsection{\texorpdfstring{Diagonal coefficients: $T_{,2}^{1 \prime}$}{}}
After the radial integral, the result in step (i) becomes: 
\begin{small}
\begin{align}
   \int_0^1 \phi_{,2j}  &T^{1*}_{j} rdr = \frac{K^0-K^1}{4\pi K^0 x_1} \Bigg\{ \notag \\
   &\Bigg(
   -i_2F_1\left[2,1-m,2-m,\frac{r(\cos{\theta} - i\sin{\theta})}{x_1}\right]\cos(-1 + m)\theta
   + i_2F_1\left[2,1-m,2-m,\frac{r(\cos{\theta} + i\sin{\theta})}{x_1}\right]\cos(-1 + m)\theta \notag \\
   &- _2F_1\left[2,1-m,2-m,\frac{r(\cos{\theta} - i\sin{\theta})}{x_1}\right] \sin(-1 + m)\theta - _2F_1\left[2,1-m,2-m,\frac{r(\cos{\theta} + i\sin{\theta})}{x_1}\right] \sin(-1 + m)\theta \notag \\
   &+ _2F_1\left[1,1-m,2-m,\frac{r(\cos{\theta} + i\sin{\theta})}{x_1}\right]\Big(-i\cos(-1 + m)\theta + \sin(-1 + m)\theta\Big) \notag \\
   &+ _2F_1\left[1,1-m,2-m,\frac{r(\cos \theta - i\sin \theta)}{x_1}\right]\Big(i\cos(-1 + m)\theta + \sin(-1 + m)\theta\Big)\Bigg) f_1^1 \notag \\
   &+\Bigg(
    -_2F_1\left[2,1-m,2-m,\frac{r(\cos{\theta} - i\sin{\theta})}{x_1}\right]\cos(-1 + m)\theta - _2F_1\left[2,1-m,2-m,\frac{r(\cos{\theta} + i\sin{\theta})}{x_1}\right]\cos(-1 + m)\theta \notag \\
   &+ _2F_1\left[1,1-m,2-m,\frac{r(\cos{\theta} - i\sin{\theta})}{x_1}\right]\Big(\cos(-1 + m)\theta - i\sin(-1 + m)\theta\Big) \notag \\
   &+ i_2F_1\left[1,1-m,2-m,\frac{r(\cos{\theta} + i\sin{\theta})}{x_1}\right]\Big(\cos(-1 + m)\theta + i\sin(-1 + m)\theta\Big) \notag \\
   &+ i_2F_1\left[2,1-m,2-m,\frac{r(\cos{\theta} - i\sin{\theta})}{x_1}\right] \sin(-1 + m)\theta - i_2F_1\left[2,1-m,2-m,\frac{r(\cos{\theta} + i\sin{\theta})}{x_1}\right] \sin(-1 + m)\theta
   \Bigg) f_2^1 
   \Bigg\}
    \label{eq:diagonal_step_1_ap}
\end{align}
\end{small}
Conduct the series expansion of $x_1$ at $0^-$, the result in step (ii) becomes: 

\begin{align}
    \int_0^1 \phi_{,2j} &T^{1*}_{j} r dr = -\frac{K^0 - K^1}{2 \pi K^0 m}(m-1) \Big(\sin(m-2)\theta f_1^1 + \cos(m-2)\theta f_2^1 \Big)  + \mathcal{O}(x_1) \notag \\
   & + \Bigg\{\frac{K^0 - K^1}{2K^0 }(m-1)^2 (-1)^m \csc m \pi \thinspace \Big(\sin[2\theta(m-1)]f_1^1  + \cos[2\theta(m-1)]f_2^1  \Big) + \mathcal{O}(x_1^2) \Bigg\} x_1^{-m}
    \label{eq:diagonal_step_2_ap}
\end{align}

\subsection{\texorpdfstring{Off-diagonal coefficients: $T_{,2}^{2 \prime}$}{}}
After the radial integral, the result in step (i) becomes: 

\begin{align}
   \int_0^1 \phi_{,2j}  &\overline{T}^{2*}_{j} r dr=  \frac{K^0-K^2}{4\pi K^0 x_1} \Bigg\{ \Big(
   -i\cos{[(\gamma - \theta)(-1 + m)]}_2F_1\left[2,1-m,2-m,\frac{(\cos{\theta} - i\sin{\theta})}{x_1}\right] \notag \\
   &+ i\cos{[(\gamma - \theta)(-1 + m)]}_2F_1\left[2,1-m,2-m,\frac{(\cos{\theta} + i\sin{\theta})}{x_1}\right]\\
   &+ _2F_1\left[1,1-m,2-m,\frac{(\cos{\theta} + i\sin{\theta})}{x_1}\right]\left(-i\cos{[(\gamma - \theta)(-1 + m)]} - \sin{[(\gamma - \theta)(-1 + m)]}\right) \notag \\
   &+ _2F_1\left[1,1-m,2-m,\frac{(\cos{\theta} - i\sin{\theta})}{x_1}\right]\left(\cos{[(\gamma - \theta)(-1 + m)]} + i\sin{[(\gamma - \theta)(-1 + m)]}\right) \notag \\
   &+ _2F_1\left[2,1-m,2-m,\frac{(\cos{\theta} - i\sin{\theta})}{x_1}\right] \sin{[(\gamma - \theta)(-1 + m)]} \notag \\
   &+ _2F_1\left[2,1-m,2-m,\frac{(\cos{\theta} + i\sin{\theta})}{x_1}\right] \sin{[(\gamma - \theta)(-1 + m)]}
   \Big) f_1^2 + \Big(\notag \\
   &-\cos{[(\gamma - \theta)(-1 + m)]}_2F_1\left[2,1-m,2-m,\frac{(\cos{\theta} - i\sin{\theta})}{x_1}\right] \notag \\
   &- \cos{[(\gamma - \theta)(-1 + m)]}_2F_1\left[2,1-m,2-m,\frac{(\cos{\theta} + i\sin{\theta})}{x_1}\right] \notag \\
   &+ _2F_1\left[1,1-m,2-m,\frac{(\cos{\theta} + i\sin{\theta})}{x_1}\right]\left(\cos{[(\gamma - \theta)(-1 + m)]} - i\sin{[(\gamma - \theta)(-1 + m)]}\right) \notag \\
   &+ i_2F_1\left[1,1-m,2-m,\frac{(\cos{\theta} - i\sin{\theta})}{x_1}\right]\left(\cos{[(\gamma - \theta)(-1 + m)]} + i\sin{[(\gamma - \theta)(-1 + m)]}\right) \notag \\
   &- i_2F_1\left[2,1-m,2-m,\frac{(\cos{\theta} - i\sin{\theta})}{x_1}\right] \sin{[(\gamma - \theta)(-1 + m)]} \notag \\
   &+ i_2F_1\left[2,1-m,2-m,\frac{(\cos{\theta} + i\sin{\theta})}{x_1}\right] \sin{[(\gamma - \theta)(-1 + m)]}
   \Big) f_2^2 
   \Bigg\}
    \label{eq:off_diagonal_step_1_ap}
    \end{align}
Conduct the series expansion with respect to $x_1$ ($x_1 \to 0^+$), and the result of step (ii) becomes: 

\begin{align}
   \int_0^1 \phi_{,2j} & \overline{T}^{2*}_{j} r dr = -\frac{K^0 - K^2}{2 \pi K^0 m} (m - 1) \Big[\sin{[(-1 + m)(\gamma - \theta) + \theta]} f_1^2 - \cos{[(-1 + m)(\gamma - \theta) + \theta]} f_2^2 \Big] + \mathcal{O}(x_1) \notag \\
   & + \Big[ \frac{K^0 - K^2}{2K^0} (m - 1)^2 \csc m \pi \Big(\sin \left[ (\gamma -2 \theta) (1 - m) - m \pi \right] f_1^2  + \cos [ (\gamma -2 \theta) (1 - m) - m \pi) ] f_2^2\Big) \notag \\
   &+ \mathcal{O}(x_1^2) \Big] x_1^{-m}
    \label{eq:off_diagonal_step_2_ap}
    \end{align}

\section{Domain integrals of Green's function multiplied by the power-form eigenstrain} 
This subsection provides explicit formulation for domain integrals of Green's function multiplied by the power-form eigenstrain. These domain integrals form the global coefficient matrix in Eq. (\ref{eq:global_elastic}). Note that results presented in Appendix C.2 and Appendix C.3 handle disturbances from other subdomain, which are different from results in \cite{Wu2024}. 

\subsection{Partial derivatives of biharmonic and harmonic potential functions} 
For field points along the symmetric line $(x_1, 0)$, the fourth-order partial derivatives of $\psi$ and second-order partial derivatives of $\phi$ are provided below, 

\textit{Biharmonic potential function $\psi_{,ijkl}$}
\begin{equation}
\begin{aligned}
    & \psi_{,1111} = \frac{2 \left(-r^4 (\cos (4 \theta )-2 \cos 2\theta)+4 r^3 x_1 \cos \theta (\cos 2\theta-2)+6 r^2 x_1^2-4 r x_1^3 \cos \theta+x_1^4\right)}{\left(r^2-2 r x_1 \cos \theta+x_1^2\right){}^3} \\ 
    & \psi_{,1112} = -\frac{4 r \sin \theta \left(r \cos \theta-x_1\right) \left(r^2 (2 \cos 2\theta-1)-2 r x_1 \cos \theta+x_1^2\right)}{\left(r^2-2 r x_1 \cos \theta+x_1^2\right){}^3} \\ 
    & \psi_{,1122} = \frac{2 \left(r^4 \cos (4 \theta )-4 r^3 x_1 \cos (3 \theta )+6 r^2 x_1^2 \cos 2\theta-4 r x_1^3 \cos \theta+x_1^4\right)}{\left(r^2-2 r x_1 \cos \theta+x_1^2\right){}^3} \\ 
    & \psi_{,1212} = -\frac{r^4 \cos (4 \theta )-4 r^3 x_1 \cos (3 \theta )+6 r^2 x_1^2 \cos 2\theta-4 r x_1^3 \cos \theta+x_1^4}{2 \pi  (v-1) \left(r^2-2 r x_1 \cos \theta+x_1^2\right){}^3} \\
    & \psi_{,2212} = \frac{4 r \sin \theta \left(r \cos \theta-x_1\right) \left(r^2 (2 \cos 2\theta+1)+3 x_1 \left(x_1-2 r \cos \theta\right)\right)}{\left(r^2-2 r x_1 \cos \theta+x_1^2\right){}^3} \\ 
    & \psi_{,2222} = \frac{6 x_1 \left(r^3 \sin (4 \theta ) \csc \theta-x_1 \left(2 r^2 (2 \cos 2\theta+1)-4 r x_1 \cos \theta+x_1^2\right)\right)-2 r^4 (2 \cos 2\theta+\cos 4 \theta )}{\left(r^2-2 r x_1 \cos \theta+x_1^2\right){}^3} 
\end{aligned}
\end{equation}

\textit{Harmonic potential function $\phi_{,ij}$}
\begin{equation}
\begin{aligned}
    & \phi_{,11}  = \frac{2 \left(r^2 \cos 2\theta-2 r x_1 \cos \theta+x_1^2\right)}{\left(r^2-2 r x_1 \cos \theta+x_1^2\right){}^2} \\ & \phi_{,12} = \frac{4 r \sin \theta \left(r \cos \theta-x_1\right)}{\left(r^2-2 r x_1 \cos \theta+x_1^2\right){}^2} \\ &  \phi_{,22} = -\frac{2 \left(r^2 \cos 2\theta-2 r x_1 \cos \theta+x_1^2\right)}{\left(r^2-2 r x_1 \cos \theta+x_1^2\right){}^2}
\end{aligned}
\end{equation}

\subsection{Biharmonic potential related domain integrals}

\begin{small}
\begin{align}
    &\frac{1}{8 \pi (1 - \nu^0)}\int_{\Omega^{2p}} \psi_{,1111} \overline{H}_{11}^{2*}\thinspace r^{-\lambda} d\textbf{x}' = -|x_1|^{-\lambda} \frac{(\lambda - 1) \csc \lambda \pi}{192 \mu^0 \mu^2 (1- \nu^0)} \Bigg\{ 3 c_1^2  (\lambda -2) \thinspace (\mu^0 - \mu^2) \Big[ 2 (\lambda -4) \sin [\lambda (\pi - \gamma)] \notag \\ & \times \sin [2 (\lambda - 1) \beta^2 ] + (\lambda -1) \Big(-2 \sin \left[ \lambda (\pi - \gamma) + 2\gamma \right] \sin [2 (\lambda - 2)\beta^2] - 2 (\lambda -4) \sin 2 \beta^2 \thinspace \sin \left[ \lambda (\pi - \gamma) + 4 \gamma \right] \notag \\ & + (\lambda -2) \sin 4 \beta^2 \thinspace \sin \left[ \lambda (\pi - \gamma) + 6 \gamma \right] \Big) \Big]
    + 3 c_2^2 (\lambda -2) \thinspace (\mu^0 - \mu^2) \Big[ (\lambda -1) \notag \\ & \times \Big(2 (\lambda -4) \sin 2 \beta^2 \thinspace \cos \left[ \lambda (\pi - \gamma) + 4 \gamma \right] - (\lambda -2) \sin 4 \beta^2 \cos \left[ \lambda (\pi - \gamma) + 6 \gamma \right] - 2 \cos \left[ \lambda (\pi - \gamma) + 2 \gamma \right] \sin [2 (\lambda - 2) \beta^2] \Big) \notag \\ & + 2 (\lambda -4) \cos [\lambda (\pi - \gamma)] \sin [2 (\lambda - 1) \beta^2]  \Big] 
    + c_3^2  \Big[ -6 (\lambda -4) (\lambda -1) (\mu^0 -\mu^2) \notag \\ & \times \sin [\lambda (\pi - \gamma)] \sin 2 \lambda \beta^2  + 2 (\lambda -2) (\lambda -1) \lambda  \sin 6 \beta^2 \thinspace (\mu^0 - \mu^2) \sin [\lambda (\pi - \gamma) + 6 \gamma ] + 6 (\lambda -2) \lambda  (\mu^0 - \mu^2) \notag \\ & \times \sin \left[ \lambda (\pi - \gamma) + 2\gamma \right]  \sin [2 (\lambda -1) \beta^2] + 3 (\lambda -4) (\lambda -1) \lambda  \sin 4 \beta^2 \thinspace (\mu^0 - \mu^2) \sin \left[ \lambda (\pi - \gamma) + 4 \gamma\right] \notag \\ & - 12 (\lambda -4) \sin \left[ \lambda (\pi - \gamma) + 2 \gamma\right] \sin [2 (\lambda - 1) \beta^2] \big( \mu^0 (1 - 2\nu^2) - \mu^2 (1 - 2\nu^0) \big) \notag \\ & + 6 (\lambda -1) \Big(2 (\lambda -4) \sin 2 \beta^2 \thinspace \sin \left[ \lambda (\pi - \gamma) + 2\gamma \right] -\sin \left[ \lambda (\pi - \gamma) + 4\gamma \right] \big( (\lambda -2) \sin 4 \beta^2 \thinspace +2 \sin [2 (2 - \lambda) \beta^2] \big) \Big) \notag \\ & \times \big( \mu^0 (1 - 2\nu^2) - \mu^2 (1 - 2\nu^0) \big) \Big]
    + c_4^2  \Bigg[ 6 \cos [\lambda (\pi - \gamma) + 2\gamma]  \Big( \sin [2 (\lambda - 1) \beta^2] \big( \mu^2 [(\lambda - 4) (\lambda - 4\nu^0) + 8]  \notag \\ & -\mu^0 [(\lambda - 4) (\lambda - 4\nu^2) + 8]  \big) -2 (\lambda -4) (\lambda -1) \sin 2 \beta^2 \thinspace \big( \mu^2 (1 - 2\nu^0) - \mu^0 (1 - 2\nu^2) \big) \Big) \notag \\ & +(\lambda -1) \Bigg( 2 (\mu^0 - \mu^2) \big[ 3 (\lambda -4) \cos [\lambda (\pi - \gamma)] \sin (2 \beta^2  \lambda )-(\lambda -2) \lambda  \sin 6 \beta^2 \cos \left[ \lambda (\pi - \gamma) + 6\gamma \right] \big] \notag \\ & + 3 \cos \left[ \lambda (\pi - \gamma) + 4\gamma \right] \Big( \sin 4 \beta^2 \thinspace \big(\mu^0  ( (\lambda -6) \lambda +4 (\lambda -2) \nu^2 +4 ) - \mu^2 \big( (\lambda -6) \lambda +4 (\lambda -2) \nu^0+4 ) \big) \notag \\ &+4 \sin \left[ 2 (\lambda - 2) \beta^2 \right] \big( \mu^2 (1 - 2\nu^0) - \mu^0 (1 - 2\nu^2) \big)  \Big) \Bigg) \Bigg]
    \Bigg\}
\end{align}
\end{small}

\begin{small}
\begin{align}
    &\frac{1}{8 \pi (1 - \nu^0)}\int_{\Omega^{2p}} \psi_{,1112} \overline{H}_{12}^{2*}\thinspace r^{-\lambda} d\textbf{x}' = -|x_1|^{-\lambda} \frac{\csc \lambda \pi}{32 \mu^0 \mu^2 (1 - \nu^0)} \Bigg\{ c_1^2 (\mu^0 -\mu^2) \Big[ \left(\lambda ^2-3 \lambda + 2 \right)^2 \sin 2 \beta^2 \thinspace \Big(\sin \left[ \lambda (\pi - \gamma) + 4 \gamma \right] \notag \\ & - \cos 2 \beta^2 \thinspace \sin \left[ \lambda (\pi - \gamma) + 6 \gamma \right] \Big)+(\lambda -1) (\lambda -2)^2 \sin [\lambda (\pi - \gamma)] \sin [2 (\lambda - 1) \beta^2] - (\lambda -1)^2 (\lambda -2) \sin \left[ \lambda(\pi - \gamma) + 2\gamma \right] \notag \\ & \times \sin [2 (\lambda -2) \beta^2] \Big]
    + \frac{c_2^2}{2} (\mu^0 - \mu^2) \Big[ (\lambda -2) (\lambda -1) \Big( (\lambda -1) \big[ (\lambda -2) \sin 4 \beta^2 \thinspace \cos \left[ \lambda (\pi - \gamma) + 6 \gamma \right] - 2 \cos \left[ \lambda (\pi - \gamma) + 2 \gamma \right] \notag \\ & \times \sin [2 (\lambda -2) \beta^2] \big] + 2 (\lambda -2) \cos [\lambda (\pi - \gamma)] \sin [2 (\lambda -1) \beta^2] \Big) - 2 \left(\lambda ^2-3 \lambda +2\right)^2 \sin 2 \beta^2 \thinspace \cos \left[ \lambda (\pi - \gamma) + 4 \gamma \right] \Big] \notag \\ & 
    + \frac{c_3^2}{6} (\lambda -2) (\lambda -1) \csc \lambda \pi \thinspace (\mu^0 - \mu^2) \Big[-6 (\lambda -1) \sin [\lambda (\pi - \gamma)] \sin 2 \lambda \beta^2 - 3 (\lambda -1) \lambda  \sin 4 \beta^2 \thinspace \sin \left[ \lambda (\pi - \gamma) + 4 \gamma \right]  \notag \\ & + 6 \lambda  \sin \left[ \lambda (\pi - \gamma) + 2 \gamma \right] \sin [2 (\lambda -1) \beta^2] + 2 (\lambda -1) \lambda  \sin 6 \beta^2 \thinspace \sin \left[ \lambda (\pi - \gamma) + 6\gamma\right] \Big] \notag \\ &
    + \frac{c_4^2}{6} (\lambda -2) (\lambda -1) \csc \lambda \pi \thinspace (\mu^0 - \mu^2) \Big[ -\Big( (\lambda -1) \lambda  \big(3 \sin 4 \beta^2 \thinspace \cos \left[ \lambda (\pi - \gamma) + 4 \gamma \right] -2 \sin 6 \beta^2 \thinspace \cos \left[ \lambda (\pi - \gamma) + 6 \gamma \right] \big) \Big) \notag \\ & - 6 \lambda  \cos \left[ \lambda (\pi - \gamma) + 2 \gamma \right] \sin [2 (\lambda -1 ) \beta^2] + 6 (\lambda -1) \cos [\lambda (\pi - \gamma)]  \sin 2 \lambda \beta^2 \Big]
    \Bigg\}
\end{align}
\end{small}

\begin{small}
\begin{align}
    &\frac{1}{8 \pi (1 - \nu^0)} \int_{\Omega^{2p}} \psi_{,1122} \overline{H}_{22}^{2*} r^{-\lambda} d\textbf{x}' = |x_1|^{-\lambda} \frac{\csc \lambda \pi}{64 \mu^0 \mu^2 (1 - \nu^0)} (\mu^0 - \mu^2) \Bigg\{ 
    c_1^2 (\lambda -2) (\lambda -1) \bigg[ 2\lambda \sin[\lambda (\pi - \gamma)] \sin[2 (\lambda -1) \beta^2] \notag \\
    &- 2\lambda (\lambda -1) \sin 2 \beta^2 \thinspace \sin[\lambda (\pi - \gamma) + 4\gamma] + (\lambda -1)(\lambda -2) \sin 4 \beta^2 \thinspace \sin[\lambda (\pi - \gamma) + 6\gamma] \notag \\
    & + 2(\lambda -1) \sin[\lambda (\pi - \gamma) + 2\gamma] \sin[2 (\lambda -2) \beta^2] \bigg] 
    + c_2^2 \bigg[ 2(\lambda -2)\lambda (\lambda -1) \cos[\lambda (\pi - \gamma)] \sin[2 (\lambda -1) \beta^2] \notag \\
    & + 2(\lambda -2)(\lambda -1)^2 \lambda \sin 2 \beta^2 \thinspace \cos[\lambda (\pi - \gamma) + 4\gamma] + 2(\lambda -2)(\lambda -1)^2 \cos[\lambda (\pi - \gamma) + 2\gamma] \sin[2 (\lambda -2) \beta^2] \notag \\
    & - (\lambda -1)^2 (\lambda -2)^2 \sin 4 \beta^2 \thinspace \cos[\lambda (\pi - \gamma) + 6\gamma] \bigg]
    + \frac{c_3^2}{6} (\lambda -1) \bigg[ 12\lambda \sin[\lambda (\pi - \gamma) + 2\gamma] \Big( \sin[2 (\lambda -1) \beta^2] \notag \\ & \times \big[ \mu^2 (\lambda - 4 \nu^0) - \mu^0 (\lambda - 4 \nu^2) \big] - 2(\lambda -1) \sin 2 \beta^2 \thinspace (\mu^2 - 2 \mu^2 \nu^0 + \mu^0 (2 \nu^2 -1)) \Big) \notag \\
    &+ 6(\lambda -1) \sin 4 \beta^2 \thinspace \sin[\lambda (\pi - \gamma) + 4\gamma] \Big( \mu^2 (-\lambda (\lambda +2) + 4(\lambda -2) \nu^0 + 4) + \mu^0 (\lambda (\lambda +2) - 4(\lambda -2) \nu^2 -4) \Big) \notag \\
    & + (\lambda -1) \Big( \lambda (\mu^0 - \mu^2) (-12 \sin[\lambda (\pi - \gamma)] \sin 2 \lambda \beta^2 \thinspace - 4(\lambda -2) \sin 6 \beta^2 \thinspace \sin[\lambda (\pi - \gamma) + 6\gamma]) \notag \\
    & + 24 \sin[\lambda (\pi - \gamma) + 4\gamma] \sin[2 (\lambda -2) \beta^2] \big[ \mu^2 (1 - 2\nu^0) - \mu^0 (1 - 2\nu^2) \big] \Big) \bigg]
    + \frac{c_4^2}{3} (\lambda -1) \Bigg[ (\lambda -1) \Bigg( 2(\mu^0 - \mu^2) \notag \\ & \times \Big(3\lambda \cos[\lambda (\pi - \gamma)] \sin 2 \lambda \beta^2 \thinspace - (\lambda -2)\lambda \sin 6 \beta^2 \thinspace \cos[\lambda (\pi - \gamma) + 6\gamma] \Big) + 3 \cos[\lambda (\pi - \gamma) + 4\gamma] \notag \\
    &  \times \Big( \sin 4 \beta^2 \thinspace \big[\mu^2 (-\lambda (\lambda +2) + 4(\lambda -2) \nu^0 +4) + \mu^0 (\lambda (\lambda +2) - 4(\lambda -2) \nu^2 -4) \big] + 4 \sin[2 (\lambda -2) \beta^2] \notag \\ & \times \big[ \mu^0 (1 - 2\nu^2) - \mu^2 (1 - 2\nu^0) \big] \Big) \Bigg) + 6\lambda \cos[\lambda (\pi - \gamma) + 2\gamma] \Big( \lambda \mu^0 \sin[2 (1-\lambda) \beta^2] + 2(\lambda -1) \sin 2 \beta^2 \thinspace \notag \\
    & \times \big[ \mu^2 (1 - 2\nu^0) - \mu^0 (1 - 2\nu^2) \big] + \sin[2 (\lambda -1) \beta^2] (\lambda \mu^2 - 4 \mu^2 \nu^0 + 4 \mu^0 \nu^2) \Big) \Bigg] \Bigg\} 
\end{align}
\end{small}

\begin{small}
\begin{align}
    &\frac{1}{8 \pi (1 - \nu^0)} \int_{\Omega^{2p}} \psi_{,1211} \overline{H}_{11}^{2*} r^{-\lambda} d\textbf{x}' = |x_1|^{-\lambda} \frac{\csc \lambda \pi}{32 \mu^0 \mu^2 (1 - \nu^0)} \Bigg\{ 
    c_1^2 (\lambda -2) (\lambda -1) (\mu^0 - \mu^2) \Big[ -2 \sin 2 \beta^2 \thinspace \cos \left[ \lambda (\pi - \gamma) + 4 \gamma \right] \notag \\ & + \sin 4 \beta^2 \thinspace \cos \left[ \lambda (\pi - \gamma) + 6 \gamma \right] - 2 \cos \left[ \lambda (\pi - \gamma) + 2 \gamma \right] \sin [2 (\lambda -2 ) \beta^2] + (\lambda -2) \cos [\lambda (\pi - \gamma)] \sin [2 (\lambda -1) \beta^2] \Big] \notag \\ & 
    - c_2^2 (\lambda -2) (\lambda -1) (\mu^0 - \mu^2) \Big[ (\lambda -2) \sin [\lambda (\pi - \gamma)] \sin [2 (\lambda - 1) \beta^2] - 2 \sin \left[\lambda (\pi - \gamma ) + 2\gamma \right] \sin [2(\lambda -2) \beta^2] \notag \\ & + 2 \sin 2 \beta^2 \thinspace \sin \left[\lambda (\pi - \gamma ) + 4\gamma \right] - \sin 4 \beta^2 \thinspace \sin \left[\lambda (\pi - \gamma ) + 6 \gamma \right] \Big]  
    - \frac{c_3^2}{3} (\lambda -1) \Bigg[ (\lambda -2) (\mu^0 - \mu^2) \Big(2 \lambda  \sin 6 \beta^2 \thinspace \notag \\ & \times \cos \left[ \lambda (\pi - \gamma) + 6\gamma \right] + 3 (\lambda -1) \cos [\lambda (\pi - \gamma)] \sin 2 \lambda \beta^2 \Big) +3 \cos \left[ \lambda (\pi - \gamma) + 4 \gamma \right] \Big( 4 \sin [2 (\lambda -2 ) \beta^2] \notag \\ & \times \big( \mu^2 (1 - 2\nu^0) - \mu^0 (1 - 2\nu^2) \big) -(\lambda -2) \sin 4 \beta^2 \thinspace \big(\mu^0  [\lambda +4 \nu^2 -2]- \mu^2 [\lambda +4 \nu^0 -2] \big) \Big) \notag \\ & + 3 (\lambda -2) \cos \left[ \lambda (\pi - \gamma) + 2\gamma \right] \Big( \sin [2 (1 - \lambda) \beta^2] \big[\mu^0 (\lambda +4 \nu^2-2) - \mu^2 (\lambda +4 \nu^0-2)\big] \notag \\ & + 4 \sin 2 \beta^2 \thinspace \big( \mu^2 (1 - 2\nu^0) - \mu^0 (1 - 2\nu^2) \big) \Big) \Bigg]
    - \frac{c_4^2}{3} (\lambda -1) \Bigg[ 3 (\lambda -2) (\lambda -1) (\mu^0 - \mu^2) \sin [\lambda (\pi - \gamma)] \sin 2 \lambda \beta^2 \thinspace \notag \\ & - 2 (\lambda -2) \lambda  \sin 6 \beta^2 \thinspace (\mu^0 - \mu^2) \sin \left[ \lambda (\pi - \gamma) + 6\gamma \right] -3 (\lambda -2) \sin \left[ \lambda (\pi - \gamma) + 2 \gamma \right]  \Big(\sin [2 (\lambda -1) \beta^2] \notag \\ & \times \big( \mu^0  [\lambda +4 \nu^2 -2]- \mu^2 [\lambda +4 \nu^0-2] \big)+4 \sin 2 \beta^2 \thinspace \big( \mu^2 (1 - 2\nu^0) - \mu^0 (1 - 2\nu^2) \big) \Big) \notag \\ & + 3 \sin \left[ \lambda (\pi - \gamma) + 4 \gamma \right] \Big( (\lambda -2) \sin 4 \beta^2 \thinspace \big(\mu^0  [\lambda +4 \nu^2 -2] - \mu^2 [\lambda +4 \nu^0 -2] \big) \notag \\ & + 4 \sin [2 (\lambda -2) \beta^2] \big( \mu^2 (1 - 2\nu^0) - \mu^0 (1 - 2\nu^2) \big) \Big) \Bigg]
    \Bigg\} 
\end{align}
\end{small}

\begin{small}
\begin{align}
    &\frac{1}{8 \pi (1 - \nu^0)} \int_{\Omega^{2p}} \psi_{,1212} \overline{H}_{12}^{2*} r^{-\lambda} d\textbf{x}' = -|x_1|^{-\lambda} \frac{\csc \lambda \pi}{64 \mu^0 \mu^2 (1 - \nu^0)} \Bigg\{
    c_1^2 (\lambda -2) (\lambda -1) (\mu^0 - \mu^2) \Big[-2 (\lambda -1) \lambda  \sin 2 \theta \thinspace \notag \\ & \times \cos \left[ \lambda (\pi - \gamma) + 4 \gamma \right] + (\lambda -2) (\lambda -1) \sin 4 \theta \thinspace \cos \left[ \lambda (\pi - \gamma) + 6 \gamma \right] + 2 \lambda  \cos [\lambda (\pi - \gamma)] \sin [2 (1 - \lambda) \theta] \notag \\ & -2 (\lambda -1) \cos \left[ \lambda (\pi - \gamma) + 2 \gamma \right] \sin [2 (2 - \lambda) \theta] \Big]
    + c_2^2 (\mu^0 - \mu^2) \Big[ \left(\lambda ^2-3 \lambda +2\right)^2 \sin 4 \theta \thinspace \sin \left[ \lambda (\pi - \gamma) + 6 \gamma \right] \notag \\ & - 2 (\lambda -2) (\lambda -1)^2 \Big( \sin \left[ \lambda (\pi - \gamma) + 2 \gamma \right] \sin [2 (\lambda - 2) \theta] + \lambda \sin 2 \theta \thinspace \sin \left[ \lambda (\pi - \gamma) + 4 \gamma \right] \Big) \notag \\ & + 2 (\lambda -2) \lambda  (\lambda -1) \sin [\lambda (\pi - \gamma)] \sin [2 (\lambda - 1) \theta] \Big] 
    + \frac{c_3^2}{3} (\lambda -1) \lambda (\mu^0 - \mu^2) \Big[ 3 (\lambda -1) \lambda  \sin 4 \theta \thinspace \cos \left[ \lambda (\pi - \gamma) + 4 \gamma \right] \notag \\ & + 2 (\lambda -2) \Big( -\big[ (\lambda -1) \sin 6 \theta \thinspace \cos \left[ \lambda (\pi - \gamma) + 6 \gamma \right] \big] - 3 \cos \left[ \lambda (\pi - \gamma) + 2 \gamma \right] \sin [2 (\lambda - 1)\theta] \Big) \notag \\ &+ 6 (\lambda -1) \cos [\lambda (\pi - \gamma)] \sin 2 \lambda \theta \Big] 
    + \frac{c_4^2}{3} (\lambda -1) \lambda  \csc (\pi  \lambda ) (\mu -\text{$\mu $1}) \Big[ 6 (\lambda -1) \sin [\lambda (\pi - \gamma)] \sin 2 \lambda \theta \thinspace \notag \\ & - 3 (\lambda -1) \lambda  \sin 4 \theta \thinspace \sin \left[ \lambda (\pi - \gamma) + 4 \gamma \right] + 2 (\lambda -2) \Big(-3 \sin \left[ \lambda (\pi - \gamma) + 2 \gamma \right] \sin [2 (\lambda -1) \theta] \notag \\ & + (\lambda -1) \sin 6 \theta \thinspace \sin \left[ \lambda (\pi - \gamma) + 6 \gamma \right] \Big) \Big]
    \Bigg\}
\end{align}
\end{small}

\begin{small}
\begin{align}
    &\frac{1}{8 \pi (1 - \nu^0)} \int_{\Omega^{2p}} \psi_{,1222} \overline{H}_{22}^{2*} r^{-\lambda} d\textbf{x}' = |x_1|^{-\lambda} 
    \frac{\csc \lambda \pi}{192 \mu^0 \mu^2 (1 - \nu^0)} \Bigg\{ 
    3c_1^2  (\lambda -1) (\mu^0 - \mu^2) \Big[ 2 (\lambda -1) (\lambda^2 - 4) \sin 2 \beta^2 \thinspace \cos [\lambda (\gamma - 4) - \pi \lambda] \notag \\ 
    & + (\lambda -2) \Big( -(\lambda -2)(\lambda -1) \sin 4 \beta^2 \thinspace \cos [\lambda (\gamma - 6) - \pi \lambda] - 2 (\lambda +2) \cos [\lambda (\pi - \gamma)] \sin [2 (\lambda -1) \beta^2]  - 2 (\lambda -1) \notag \\
    &\times\cos [\lambda (\pi - \gamma) + 2 \gamma] \sin [4 \beta^2 - 2 \lambda \beta^2] \Big) \Big] 
    +3 c_2^2  (\lambda -2)(\lambda -1)(\mu^0 - \mu^2) \Big[ 2 (\lambda +2) \sin [\lambda (\pi - \gamma)] \sin [2 (\lambda -1) \beta^2] \notag \\
    & + (\lambda -1) \Big( 2 (\lambda +2) \sin 2 \beta^2  \thinspace \sin [\lambda (\pi - \gamma) + 4 \gamma] - (\lambda -2) \sin 4 \beta^2 \thinspace \sin [\lambda (\pi - \gamma) + 6 \gamma]  + 2 \sin [\lambda (\pi - \gamma) + 2 \gamma]\notag \\
    &\times \sin [4 \beta^2 - 2 \lambda \beta^2] \Big) \Big]
    + c_3^2 (\lambda -1) \Bigg[ (\lambda -1)(\mu^0 - \mu^2) \Big( 3 (\lambda +2) \big( \sin [\lambda (-\gamma + 2 \beta^2 + \pi)] - \sin [\lambda (-2\gamma - 2\beta^2 + \gamma + \pi)] \big) \notag \\
    & + 2 (\lambda -2)\lambda \sin 6 \beta^2 \thinspace \cos [\lambda (\pi - \gamma) + 6 \gamma] \Big) + 3 \cos [\lambda (\gamma - 4) - \pi \lambda] \Big( \sin 4 \beta^2 \thinspace \big( \mu^2 [\lambda (\lambda +4) - 4 (\lambda -2)\nu^0 - 4] \notag \\
    & + \mu^0 [-\lambda (\lambda +4) + 4 (\lambda -2)\nu^2 + 4] \big) + 4 \sin [2 (\lambda -2)\beta^2] \big( \mu^0 + \mu^2 (2\nu^0 - 1) - 2 \mu^0 \nu^2 \big) \Big) - 6 \cos [\lambda (\gamma - 2) - \pi \lambda] \notag \\
    & \times \Big( \lambda^2 \mu^2 \sin [2 (1 - \lambda)\beta^2] + 2 (\lambda^2 + \lambda - 2) \sin 2 \beta^2 \thinspace \big( \mu^2 - 2\mu^2\nu^0 + \mu^0 (2\nu^2 - 1) \big)  + \sin [2 (\lambda -1)\beta^2] \big( 4\mu^2 [(\lambda +2)\nu^0 - 1] \notag \\
    &+ \mu^0 [\lambda^2 - 4(\lambda +2)\nu^2 + 4] \big) \Big) \Bigg]
    + c_4^2 (\lambda -1) \Bigg[ 6 \sin [\lambda (\pi - \gamma) + 2 \gamma] \sin [2 (\lambda -1)\beta^2] \big( \mu^2 [\lambda^2 - 4(\lambda +2)\nu^0 + 4] \notag \\
    &+ \mu^0 [-\lambda^2 + 4(\lambda +2)\nu^2 - 4] \big)  + 3 (\lambda -1) \sin 4 \beta^2 \sin [\lambda (\pi - \gamma) + 4 \gamma] \big( \mu^2 [-\lambda (\lambda +4) + 4(\lambda -2)\nu^0 + 4] + \mu^0 [\lambda (\lambda +4)\notag \\
    & - 4(\lambda -2)\nu^2 - 4] \big) + 2 (\lambda -1) \Big( 3 (\lambda +2)(\mu^0 - \mu^2) \sin [\lambda (\pi - \gamma)] \sin 2 \lambda \beta^2 \thinspace - (\lambda -2)\lambda \sin 6 \beta^2 \thinspace (\mu^0 - \mu^2) \sin [\lambda (\pi - \gamma)  \notag \\
    &+ 6 \gamma] + 6 \big( \sin [\lambda (\gamma - 4) - \pi \lambda] \sin [2 (\lambda -2)\beta^2] + (\lambda +2) \sin 2 \beta^2 \thinspace \sin [\lambda (\pi - \gamma) + 2 \gamma] \big) \big( \mu^2 - 2\mu^2\nu^0 + \mu^0 (2\nu^2 - 1) \big) \Big) \Bigg]
    \Bigg\} 
\end{align}
\end{small}

\begin{small}
\begin{align}
    &\frac{1}{8 \pi (1 - \nu^0)} \int_{\Omega^{2p}} \psi_{,2211} \overline{H}_{11}^{2*} r^{-\lambda} d\textbf{x}' = -|x_1|^{-\lambda} \frac{\csc \lambda \pi}{64 \mu^0 \mu^2 (1 - \nu^0)} \Bigg\{ 
    c_1^2 (\lambda -2) (\lambda -1) (\mu^0 - \mu^2) \Big[ (\lambda-1) \big( 2\lambda \sin 2\beta^2 \thinspace  \sin[\lambda(\pi-\gamma)+4\gamma]\notag \\
    & - (\lambda-2) \sin 4\beta^2 \thinspace  \sin[\lambda(\pi-\gamma)+6\gamma] + 2 \sin[\lambda(\pi-\gamma)+2\gamma] \sin[2(\lambda-2)\beta^2] \big) - 2\lambda \sin[\lambda(\pi-\gamma)] \sin[2(\lambda-1)\beta^2] \Big] \notag \\
    &+ c_2^2 (\lambda -2) (\lambda -1) (\mu^0 - \mu^2) \Big[ (\lambda-1) \big( 2\lambda \sin 2\beta^2 \thinspace \cos[\lambda(\pi-\gamma)+4\gamma] - (\lambda-2) \sin 4\beta^2 \thinspace  \cos[\lambda(\pi-\gamma)+6\gamma] - 2 \cos[\lambda(\pi-\gamma)+2\gamma]\notag \\
    &\times  \sin[2(\lambda-2)\beta^2] \big) + 2\lambda \cos[\lambda(\pi-\gamma)] \sin[2(\lambda-1)\beta^2] \Big] 
    + \frac{c_3^2}{6} (\lambda-1) \Bigg[ (\lambda-1) \lambda (\mu^0 - \mu^2) \big( 2(\lambda-2) \cos[\lambda(\pi-\gamma)+6\gamma-6\beta^2] \notag \\
    &- 3\lambda \cos[\lambda(\pi-\gamma)+4\gamma-4\beta^2] \big) + 3(\lambda-1) \lambda (\mu^0 - \mu^2) \big( 4 \sin[\lambda(\pi-\gamma)] \sin 2\lambda\beta^2 \thinspace + \lambda \cos[4\gamma+4\beta^2+\lambda(\pi-\gamma)] \big) + 24(\lambda-1) \notag \\
    &\times \sin[\lambda(\pi-\gamma)+4\gamma] \sin[2(\lambda-2)\beta^2] \big( \mu^0 + \mu^2 (2\nu^0-1) - 2\mu^0 \nu^2 \big) + 24(\lambda-1) \lambda \sin 2\beta^2 \thinspace  \sin[\lambda(\pi-\gamma)+2\gamma] \big( \mu^0 + \mu^2 (2\nu^0-1) \notag \\
    &- 2\mu^0 \nu^2 \big) + 12\lambda \sin[\lambda(\pi-\gamma)+2\gamma] \sin[2(\lambda-1)\beta^2] \big( \mu^0 (\lambda+4\nu^2-4) - \mu^2 (\lambda+4\nu^0-4) \big) + 2(\lambda-2)(\lambda-1) \big( \lambda (\mu^2 - \mu^0) \notag \\
    &\times \cos[6\gamma+6\beta^2+\lambda(\pi-\gamma)] + 6 \sin 4\beta^2 \thinspace  \sin[\lambda(\pi-\gamma)+4\gamma] (\mu^2 - 2\mu^2\nu^0 + \mu^0 (2\nu^2-1)) \big) \Bigg] 
    + \frac{c_4^2}{3} (\lambda-1) \Bigg[ 6\lambda \cos[\lambda(\pi-\gamma)+2\gamma] \notag \\
    &\times \big( \lambda \mu^2 \sin[2(1-\lambda)\beta^2] + 2(\lambda-1) \sin 2\beta^2 \thinspace  (\mu^2 (1-2\nu^0) - \mu^0 (1-2\nu^2)) + \sin[2(\lambda-1)\beta^2] (\mu^0 (\lambda+4\nu^2-4) - 4\mu^2 (\nu^0-1)) \big) \notag \\
    &+ (\lambda-1) \big( 2(\mu^0 - \mu^2) ( (\lambda-2)\lambda \sin 6\beta^2 \thinspace  \cos[\lambda(\pi-\gamma)+6\gamma] - 3\lambda \cos[\lambda(\pi-\gamma)] \sin 2\lambda\beta^2 \thinspace  ) + 3 \cos[\lambda(\pi-\gamma)+4\gamma]\notag \\
    & \times ( \sin 4\beta^2 \thinspace  (\mu^2 (\lambda(\lambda-2)+4(\lambda-2)\nu^0+4) - \mu^0 (\lambda(\lambda-2)+4(\lambda-2)\nu^2+4)) + 4 \sin[4\beta^2-2\lambda\beta^2] (\mu^2 (1-2\nu^0) - \mu^0 (1-2\nu^2)) ) \big) \Bigg] \Bigg\}     
    \Bigg\}
\end{align}
\end{small}

\begin{small}
\begin{align}
    &\frac{1}{8 \pi (1 - \nu^0)} \int_{\Omega^{2p}} \psi_{,2212} \overline{H}_{12}^{2*} r^{-\lambda} d\textbf{x}' = -|x_1|^{-\lambda} \frac{\csc \lambda \pi}{64 \mu^0 \mu^2 (1 - \nu^0)} \Bigg\{ c_1^2
    (\lambda -2) (\lambda -1) (\mu^0 - \mu^2) \Bigg[-2 (\lambda +2) \cos 2 \beta^2 \thinspace \sin [\lambda (\pi - \gamma)] \notag \\ & \times \sin 2 \lambda \beta^2 \thinspace + 2 \cos 2 \lambda \beta^2 \thinspace \Big( (\lambda +2) \sin 2 \beta^2 \thinspace \sin [\lambda (\pi - \gamma)] - (\lambda -1) \sin 4 \beta^2 \thinspace \sin \left[ \lambda (\pi - \gamma) + 2 \gamma \right] \Big) \notag +(\lambda -1) \\ & \times \Big( -2 (\lambda +2) \sin 2 \beta^2 \thinspace \sin \left[ \lambda (\pi - \gamma) + 4 \gamma \right] + (\lambda -2) \sin 4 \beta^2 \thinspace \sin \left[ \lambda (\pi - \gamma) + 6 \gamma \right] + 2 \cos 4 \beta^2 \thinspace \sin \left[ \lambda (\pi - \gamma) + 2 \gamma \right] \sin 2 \lambda \beta^2 \Big)  \Bigg]
    \notag \\ & + c_2^2 (\lambda -2) (\lambda -1) (\mu^0 -\mu^2) \Bigg[ 2 \left(\lambda ^2+\lambda -2\right) \sin 2 \beta^2 \thinspace \cos \left[ \lambda (\pi - \gamma) + 4\gamma \right] - (\lambda -2) (\lambda -1) \sin 4 \beta^2 \thinspace \cos \left[ \lambda (\pi - \gamma) + 6 \gamma \right] \notag \\ & + 2 (\lambda +2) \cos [\lambda (\pi - \gamma)] \sin [2 (1 - \lambda) \beta^2] -2 (\lambda -1) \cos \left[ \lambda (\pi - \gamma) + 2 \gamma \right] \sin [2 (2 - \lambda) \beta^2] \Bigg]
    + \frac{c_3^2}{3} (\lambda -1) (\mu^0 - \mu^2) \notag \\ & \times \Big[ 6 (\lambda -1) (\lambda +2) \sin [\lambda (\pi - \gamma)] \sin 2 \lambda \beta^2 - 2 (\lambda -2) (\lambda -1) \lambda  \sin 6 \beta^2 \thinspace \sin \left[ \lambda (\pi - \gamma) + 6 \gamma \right] - 6 (\lambda -2) \lambda  \notag \\ & \times \sin \left[ \lambda (\pi - \gamma) + 2 \gamma \right] \sin [2(\lambda - 1) \beta^2]  + 3 (\lambda -1) \lambda  (\lambda +2) \sin 4 \beta^2 \thinspace \sin\left[ \lambda (\pi - \gamma) + 4 \gamma \right] \Big] \notag \\ &
    + \frac{c_4^2}{3} (\lambda -1) (\mu^0 - \mu^2) \Big[ 3 (\lambda -1) \lambda  (\lambda +2) \sin 4 \beta^2 \thinspace \cos \left[ \lambda (\pi - \gamma) + 4 \gamma \right] - 2 (\lambda -2) (\lambda -1) \lambda  \sin 6 \beta^2 \thinspace \cos \left[ \lambda (\pi - \gamma) + 6 \gamma \right] \notag \\ & + 6 (\lambda -2) \lambda  \cos \left[ \lambda (\pi - \gamma) + 2 \gamma \right] \sin [2(\lambda - 1) \beta^2]  - 6 (\lambda -1) (\lambda +2) \cos [\lambda (\pi - \gamma)] \sin 2 \lambda \beta^2 \Big]
    \Bigg\}
\end{align}
\end{small}

\begin{small}
\begin{align}
    &\frac{1}{8 \pi (1 - \nu^0)} \int_{\Omega^{2p}} \psi_{,2222} \overline{H}_{22}^{2*} r^{-\lambda} d\textbf{x}' = -|x_1|^{-\lambda} \frac{\csc \lambda \pi}{64 \mu^0 \mu^2 (1 - \nu^0)} \Bigg\{ 
    c_1^2 (\lambda -2) (\lambda -1) (\mu^0 - \mu^2) \Big( (\lambda -1) \Big( 2 \sin \left[ \lambda (\pi - \gamma) + 2\gamma \right] \notag \\ & \times \sin [2(\lambda -2) \beta^2] + 2 (\lambda +4) \sin 2 \beta^2 \thinspace \sin \left[ \lambda (\pi - \gamma) + 4\gamma \right] - (\lambda -2) \sin 4 \beta^2 \thinspace \sin \left[ \lambda (\pi - \gamma) + 6 \gamma \right] \Big) \notag \\ & - 2 (\lambda +4) \sin [\lambda (\pi - \gamma)] \sin [2 (\lambda -1) \beta^2]  \Big) 
    + c_2^2 (\lambda -2) (\lambda -1) (\mu^0 - \mu^2) \Big[ -2 (\lambda -1) (\lambda +4 ) \sin 2 \beta^2 \thinspace \notag \\ & \times \cos \left[ \lambda (\pi - \gamma) + 4 \gamma \right]  +(\lambda -1) \Big( (\lambda -2) \sin 4 \beta^2 \thinspace \cos \left[ \lambda (\pi - \gamma) + 6 \gamma \right] + 2 \cos \left[ \lambda (\pi - \gamma) + 2 \gamma \right] \sin [2 (\lambda -2) \beta^2] \Big) \notag \\ & +2 (\lambda +4) \cos [\lambda (\pi - \gamma)] \sin [2 (1 - \lambda) \beta^2] \Big]
    + \frac{c_3^2}{3} (\lambda -1) \Bigg[ 3 (\lambda -1) (\lambda +4) (\mu^0 - \mu^2) \Big(2 \sin [\lambda (\pi - \gamma)] \sin 2 \beta^2 \lambda \lambda \thinspace \notag \\ & - \lambda  \sin 4 \beta^2 \thinspace \sin \left[ \lambda (\pi - \gamma) + 4\gamma \right] \Big) + 2 (\lambda -2) \lambda  (\mu^0 - \mu^2) \Big(-3 \sin \left[ \lambda (\pi - \gamma) + 2\gamma \right] \sin [2(\lambda - 1) \beta^2] \notag \\ & +(\lambda -1) \sin 6 \beta^2 \thinspace \sin \left[ \lambda (\pi - \gamma) + 6 \gamma \right] \Big) - 12 (\lambda -1) \sin \left[ \lambda (\pi - \gamma) + 4\gamma \right] \sin [2 (\lambda - 2) \beta^2] \big[ \mu^2 (1 - 2\nu^0) - \mu^0 (1 - 2\nu^2)  \big]  \notag \\ & + 12 (\lambda +4) \sin \left[ \lambda (\pi - \gamma) + 2 \gamma \right] \sin [2 (\lambda - 1) \beta^2] \big[ \mu^2 (1 - 2\nu^0) - \mu^0 (1 - 2\nu^2)  \big] + 6 (\lambda -1) \Big( (\lambda -2) \sin 4 \beta^2 \thinspace \sin \left[ \lambda (\pi - \gamma) + 4\gamma \right] \notag \\ & - 2 (\lambda +4) \sin 2 \beta^2 \thinspace \sin \left[ \lambda (\pi - \gamma) + 2 \gamma \right] \Big) \big[ \mu^2 (1 - 2\nu^0) - \mu^0 (1 - 2\nu^2)  \big] \Bigg]
    + c_4^2 (\lambda -1) \Bigg[ 6 \cos \left[ \lambda (\pi - \gamma) + 2 \gamma \right]  \Big[ \sin [2 (\lambda -1 ) \beta^2] \notag \\ & \times \Big( \mu^0  \left(\lambda ^2-4 (\lambda +4) \nu^2 + 8\right) - \mu^2 \left(\lambda ^2 -4 (\lambda +4) \nu^0 - 8\right)  \Big) -2 (\lambda -1) (\lambda +4) \sin 2 \beta^2 \thinspace \big[ \mu^2 (1 - 2\nu^0) - \mu^0 (1 - 2\nu^2)  \big] \Big] \notag \\ & +(\lambda -1) \Bigg( 2 (\mu^0 - \mu^2) \Big[ (\lambda -2) \lambda  \sin 6 \beta^2 \thinspace \cos \left[ \lambda (\pi - \gamma) + 6 \gamma \right] -3 (\lambda +4) \cos [\lambda (\pi - \gamma)] \sin 2 \lambda \beta^2  \Big] + 3 \cos \left[ \lambda (\pi - \gamma) + 4 \gamma \right] \notag \\ & \times \Big(\sin 4 \beta^2 \thinspace \big[ \mu^2 (\lambda  (\lambda +6)-4 (\lambda -2) \nu^0 -4)- \mu^0  (\lambda  (\lambda +6)- 4 (\lambda -2) \nu^2 - 4) \big] \notag \\ & + 4 \sin [2 (\lambda - 2)\beta^2] \big[ \mu^2 (1 - 2\nu^0) - \mu^0 (1 - 2\nu^2)  \big] \Big) \Bigg) \Bigg]
    \Bigg\}
\end{align}
\end{small}

\subsection{Harmonic potential related domain integrals}

\begin{small}
\begin{align}
    &\int_{\Omega^{2p}} \phi_{,11} \overline{H}_{11}^{2*}\thinspace r^{-\lambda} d\textbf{x}' = |x_1|^{-\lambda}\frac{\pi (\lambda - 1) \csc \lambda \pi]}{4 \mu^0 \mu^2} \Bigg\{ 
    -2 c_1^2 (\lambda -2) \thinspace (\mu^0 -\mu^2) \Big( (\lambda -1) \sin 2 \beta^2 \thinspace \sin \left[ \lambda (\pi - \gamma) + 4\gamma \right] \notag \\ 
    & + \sin \left[ \lambda (\pi -\gamma ) \right] \sin \left[ 2(1 - \lambda) \beta^2 \right]\Big) 
    + 2 c_2^2 (\lambda -2) \thinspace (\mu^0 -\mu^2) \Big( (\lambda -1) \sin 2 \beta^2 \thinspace \cos \left[ \lambda (\pi - \gamma) + 4 \gamma \right] \notag \\ & -\cos \left[ \lambda (\pi - \gamma) \right] \sin [2 (1 - \lambda) \beta^2] \Big)  
    + c_3^2 \Big[ 2 (1 - \lambda) (\mu^0 -\mu^2) \sin [\lambda (\pi - \gamma)] \sin 2 \lambda \beta^2 \thinspace \notag \\ & 
    + \lambda (\lambda -1) \sin 4 \beta^2 \thinspace (\mu^0 -\mu^2) \sin \left[ \lambda (\pi - \gamma) + 4 \gamma \right] - 4 \sin \left[ \lambda (\pi - \gamma) + 2\gamma \right] \Big( (\lambda -1) \sin 2 \beta^2 \thinspace + \sin [2 (1 - \lambda) \beta^2] \Big) \notag \\ & \times (\mu^2 (1 - 2 v) + \mu^0  (2 \nu^2-1)) \Big] 
    + c_4^2 \thinspace \Big[ (\lambda -1) (\mu^0 -\mu^2) \Big( \lambda  \sin 4 \beta^2 \thinspace \cos \left[ \lambda (\pi - \gamma) + 4\gamma \right] + 2 \cos \left[ \lambda (\pi - \gamma) \right] \sin 2 \lambda \beta^2 \Big) \notag \\ &  + 4 \cos \left[ \lambda (\pi - \gamma) + 2 \gamma \right] \big( \sin [2 (1 - \lambda) \beta^2] - (\lambda -1) \sin 2 \beta^2 \big) \big( \mu^2 (1 - 2\nu^0) - \mu^0 (1 - 2\nu^2) \big) \Big]
     \Bigg\} 
\end{align}
\end{small}

\begin{small}
\begin{align}
    &\int_{\Omega^{2p}} \phi_{,11} \overline{H}_{12}^{2*}\thinspace r^{-\lambda} d\textbf{x}' = |x_1|^{-\lambda} \frac{\pi (\lambda - 1) \csc \lambda \pi \thinspace (\mu^0 - \mu^2)}{4 \mu^0 \mu^2} \Bigg\{ 2 c_1^2 (\lambda -2) \Big[ (\lambda -1) \sin 2 \beta^2 \thinspace \cos \left[ \lambda (\pi - \gamma) + 4\gamma \right] \notag \\ & -\cos [\lambda (\pi - \gamma)] \sin [2 (1 - \lambda) \beta^2] \Big]
    + 2 c_2^2 (\lambda -2) \Big[ (\lambda -1) \sin 2 \beta^2 \thinspace \sin \left[ \lambda (\pi - \gamma) + 4 \gamma \right] - \sin [\lambda (\pi - \gamma)] \sin [2 (\lambda - 1) \beta^2] \Big] \notag \\ 
    & - c_3^2 (\lambda -1)  \Big[ \lambda  \sin 4 \beta^2 \thinspace \cos \left[ \lambda (\pi - \gamma) + 4 \gamma \right] + 2 \cos [\lambda (\pi - \gamma)] \sin 2 \lambda \beta^2 \Big] 
    + c_4^2 (\lambda -1) \Big[ \lambda  \sin 4 \beta^2 \sin \left[\lambda (\pi - \gamma) + 4 \gamma \right] \notag \\ & -2 \sin [\lambda (\pi -\gamma)] \sin 2 \lambda \beta^2 \Big]   \Bigg\}
\end{align}
\end{small}

\begin{small}
\begin{align}
    &\int_{\Omega^{2p}} \phi_{,11} \overline{H}_{22}^{2*}\thinspace r^{-\lambda} d\textbf{x}' = |x_1|^{-\lambda} \frac{\pi (\lambda - 1) \csc \lambda \pi \thinspace (\mu^0 - \mu^2)}{4 \mu^0 \mu^2} \Bigg\{ 2 c_1^2 (\lambda -2) \Big[ (\lambda -1) \sin 2 \beta^2 \sin \left[ \lambda (\pi - \gamma) + 4 \gamma \right] \notag \\ & + \sin [\lambda (\pi - \gamma)] \sin [2 (1 - \lambda) \beta^2] \Big] 
    + 2 c_2^2 (\lambda -2) \Big[ \cos [\lambda (\pi - \gamma)] \sin [2 (1-\lambda) \gamma] - (\lambda -1) \sin 2 \beta^2 \thinspace \cos \left[ \lambda (\pi - \gamma) + 4 \gamma\right] \Big] \notag \\ &
    + \frac{c_3^2}{\mu^0 - \mu^2} \Big[ 2 (\lambda -1) (\mu^0 - \mu^2) \sin [\lambda (\pi - \gamma)] \sin 2 \lambda \beta^2 \thinspace - (\lambda -1) \lambda  \sin 4 \beta^2 \thinspace (\mu^0 - \mu^2) \sin \left[ \lambda (\pi - \gamma) + 4 \gamma \right] \notag \\ & + 4 \sin \left[ \lambda (\pi - \gamma) + 2\gamma \right] \big(\sin [2 (\lambda - 1) \beta^2] - (\lambda -1) \sin 2 \beta^2 \big) ( \mu^2 (1 - 2 \nu^0) + \mu^0 (1 - 2\nu^2) ) \Big] \notag \\ &
    + \frac{c_4^2}{\mu^0 - \mu^2} \Big[ 4 \cos \left[ \lambda (\pi - \gamma) + 4\gamma \right] \big(\sin [2 (1- \lambda) \beta^2] - (\lambda -1) \sin 2 \beta^2 \big) \big(\mu^2 (1 - 2\nu^0) - \mu^0 (1 - 2\nu^2) \big) \notag \\ & -  (\lambda -1) (\mu^0 - \mu^2) \Big( \lambda  \sin 4 \beta^2 \thinspace \cos \left[ \lambda (\pi - \gamma) + 4\gamma \right] + 2 \cos [\lambda (\pi - \gamma)] \sin 2 \lambda \beta^2 \thinspace \Big) \Big]
    \Bigg\}
\end{align}
\end{small}

\begin{small}
\begin{align}
    & \int_{\Omega^{2p}} \phi_{,12} \overline{H}_{11}^{2*} r^{-\lambda} \thinspace d\textbf{x}' = |x_1|^{-\lambda} \frac{\pi (\lambda - 1) \csc \lambda \pi}{4 \mu^0 \mu^2} \Bigg\{ 2 c_1^2 (\lambda -2) 
     (\mu^0 -\mu^2) \Big[ (\lambda -1) \sin 2 \beta^2 \thinspace \cos \left[ \lambda(\pi - \gamma) + 4 \gamma \right] \notag \\ & +\cos \left[ \lambda (\pi - \gamma) \right]  \sin [2 (1 -\lambda) \beta^2] \Big] %
     + 2 c_2^2 (\lambda -2) (\mu^0 - \mu^2) \Big[ (\lambda -1) \sin 2\beta^2 \thinspace \sin \left[ \lambda (\pi - \gamma) + 4 \gamma \right] \notag \\ & -\sin [\lambda (\pi - \gamma)] \sin [2 (1 - \lambda) \beta^2] \Big]  
     + c_3^2 \Big[ 4 \cos \left[ \lambda (\pi - \gamma) + 2\gamma \right] \big( (\lambda -1) \sin 2 \beta^2 \thinspace + \sin [2 (1 - \lambda) \beta^2] \big) \notag \\ & \times \big( \mu^2 (1 - 2\nu^0) - \mu^0 (1 - \nu^2) \big) - (\lambda -1) (\mu^0 - \mu^2) \Big( \lambda  \sin 4 \beta^2 \thinspace \cos \left[ \lambda (\pi - \gamma) + 4\gamma \right] - 2 \cos [\lambda (\pi - \gamma)]  \sin 2 \lambda \beta^2  \Big) \Big] \notag \\ & 
     + c_4^2 \Big[ 2 (\lambda -1) (\mu^0 - \mu^2) \sin [\lambda (\pi - \gamma)] \sin 2 \lambda \beta^2 \thinspace + (\lambda -1) \lambda  \sin 4 \beta^2 \thinspace (\mu^0 - \mu^2) \sin \left[ \lambda (\pi - \gamma) + 4 \gamma \right] \notag \\ & - 4 \sin \left[ \lambda (\pi - \gamma) + 2\gamma \right] \big( (\lambda -1) \sin 2 \beta^2 \thinspace + \sin [2 (\lambda - 1) \beta^2] \big) \big( \mu^2 (1 - 2\nu^0) - \mu^0 (1 - 2\nu^2) \big) \Big]   \Bigg\}
\end{align}
\end{small}

\begin{small}
\begin{align}
    & \int_{\Omega^{2p}} \phi_{,12} \overline{H}_{12}^{2*} r^{-\lambda} \thinspace d\textbf{x}' = |x_1|^{-\lambda} \frac{\pi (\lambda - 1) \csc \lambda \pi \thinspace (\mu^0 - \mu^2)}{2 \mu^0 \mu^2} \Bigg\{ 2 c_1^2 (\lambda -2) \Big[ (\lambda -1) \sin 2 \beta^2 \thinspace \sin \left[\lambda (\pi - \gamma) + 4 \gamma \right] \notag \\ & - \sin [\lambda (\pi - \gamma)]  \sin [2 (1 - \lambda) \beta^2]  \Big] 
    + 2 c_2^2 (\lambda -2) \Big[ \cos [\lambda (\pi - \gamma)] \sin [2 (\lambda - 1) \beta^2] \notag \\ & -(\lambda -1) \sin 2 \beta^2 \thinspace \cos \left[ \lambda (\pi - \gamma) + 4\gamma \right] \Big] 
    + c_3^2 (\lambda -1)  \Big[ \lambda  \sin 4
   \beta^2 \thinspace \sin \left[ \lambda(\pi - \gamma) + 4 \gamma \right] - 2 \sin [\lambda (\pi - \gamma)] \sin 2 \lambda \beta^2 \Big] \notag \\ & 
   - c_4^2 (\lambda -1) \Big[ \lambda  \sin 4
   \beta^2 \! \cos \left[ \lambda (\pi - \gamma) + 4 \gamma \right] -2 \cos [\lambda (\pi - \gamma)] \sin 2 \lambda \beta^2 \Big]
    \Bigg\}
\end{align}
\end{small}

\begin{small}
\begin{align}
& \int_{\Omega^{2p}} \phi_{,12} \overline{H}_{22}^{2*} r^{-\lambda} \thinspace d\textbf{x}' = |x_1|^{-\lambda} 
\frac{\pi(\lambda-1)}{4 \,\mu^0 \,\mu^2}\,\csc \lambda \pi \,
\Bigg\{
\;2 c_1^2 (\lambda-2)(\mu^0-\mu^2)\Big[
(\lambda-1)\,\sin\!\big[\lambda(\pi-\gamma)+4\gamma\big]\,\cos 2\beta^2 \notag \\ & +\cos\!\big[2(1-\lambda)\beta^2\big]\,\sin\!\big[\lambda(\pi-\gamma)\big]
\Big]
+ 2 c_2^2 (\lambda-2) (\mu^0-\mu^2) \Big[
-(\lambda-1)\,\sin 2\beta^2 \,\sin\!\big[\lambda(\pi-\gamma)+4\gamma\big]
\notag \\ & +\sin\!\big[ 2(1-\lambda)\beta^2\big]\,\sin\!\big[\lambda(\pi-\gamma)\big]
\Big]
\; + c_3^2 \Big[
-2(\lambda-1)(\mu^0-\mu^2)\,\cos\!\big[\lambda(\pi-\gamma)\big] \, \sin 2 \lambda \beta^2 \notag \\ &
+(\lambda-1)\lambda (\mu^0-\mu^2) \,\cos\!\big[\lambda(\pi-\gamma)+4\gamma\big]\,\sin 4 \beta^2
+4 \big( \mu^2 (1 - 2\nu^0) - \mu^0 (1 - 2\nu^0) \big) \, \notag \\ & \times \cos\!\big[\lambda(\pi-\gamma)+2\gamma\big]\,\sin\!\big[2(1-\lambda)\beta^2\big]
+4(\lambda-1)\big( \mu^2 (1 - 2\nu^0) - \mu^0 (1 - 2\nu^0) \big)\,\cos\!\big[\lambda(\pi-\gamma)+2\gamma\big]\,\sin 2\beta^2
\Big] \notag \\ &
+ c_4^2 \Big[
(\lambda-1)(\mu^0-\mu^2) \Big(
-2\,\sin\!\big[\lambda(\pi-\gamma)\big]\,\sin 2\lambda \beta^2 \thinspace -\lambda\,\sin\!\big[\lambda(\pi-\gamma)+4\gamma\big] \,\sin 4\beta^2 \!
\Big)
\notag \\ &
-4 \big( \mu^2 (1 - 2\nu^0) - \mu^0 (1 - 2\nu^0) \big)\,
\sin\!\big[\lambda(\pi-\gamma)+2\gamma\big]\,
\Big((\lambda-1)\sin 2\beta^2 \! + \sin\!\big[2(1-\lambda)\beta^2\big]\Big)
\Big]\,
\Bigg\}
\end{align}
\end{small}

\begin{small}
\begin{align}
    & \int_{\Omega^{2p}} \phi_{,22} \overline{H}_{11}^{2*} r^{-\lambda} \thinspace d\textbf{x}' = |x_1|^{-\lambda} \frac{\pi (\lambda - 1) \csc \lambda \pi}{4 \mu^0 \mu^2} \Bigg\{ 2 c_1^2
     (\lambda - 2) (\mu^0 - \mu^2) \Big[ (\lambda -1) \sin 2 \beta^2 \thinspace \sin \left[ \lambda (\pi - \gamma) + 4 \gamma \right] \notag \\ & + \sin \left[ \lambda (\pi - \gamma) \right] \sin [2 (1 - \lambda) \beta^2] \Big] 
    + 2 c_2^2 (\lambda -2) (\mu^0 - \mu^2) \Big[ \cos [\lambda (\pi - \gamma)] \sin [2 (1 - \lambda) \beta^2] -(\lambda -1) \sin 2 \beta^2 \notag \\ & \times \cos \left[ \lambda(\pi - \gamma) + 4 \gamma \right] \Big] 
    + c_3^2 \Big[ 2 (\lambda -1) (\mu^0 - \mu^2) \sin [\lambda (\pi - \gamma)] \sin 2 \beta^2  \lambda \thinspace - (\lambda -1) \lambda  
    \sin 4 \beta^2 \thinspace (\mu^0 - \mu^2) \notag \\ & \times  \sin \left[ \lambda (\pi - \gamma) + 4 \gamma\right] + 4 \sin \left[ \lambda (\pi - \gamma) + 2 \gamma\right] \Big((\lambda -1) \sin 2 \beta^2 \thinspace + \sin [2 (1 - \lambda) \beta^2] \Big) \big( \mu^2 (1 - 2\nu^0) - \mu^0 (1 - 2\nu^2) \big) \Big] \notag \\ &
    + c_4^2 \Big[ 4 \cos \left[ \lambda (\pi - \gamma) + 2 \gamma \right]  \big((\lambda -1) \sin 2 \beta^2 \thinspace + \sin \left[ 2 (\lambda - 1) \beta^2 \right] \big) \big( \mu^2 (1 - 2\nu^0) - \mu^0 (1 - 2\nu^2) \big)  \notag \\ & - (\lambda -1) (\mu^0 - \mu^2) \Big(\lambda  \sin 4 \beta^2 \thinspace \cos \left[ \lambda (\pi - \gamma) + 4\gamma \right] +2 \cos [\lambda (\pi - \gamma)] \sin 2 \lambda \beta^2 \Big) \Big]
    \Bigg\}
\end{align}
\end{small}

\begin{small}
\begin{align}
    & \int_{\Omega^{2p}} \phi_{,22} \overline{H}_{12}^{2*} r^{-\lambda} \thinspace d\textbf{x}' = |x_1|^{-\lambda} \frac{\pi (\lambda - 1) \csc \lambda \pi}{4 \mu^0 \mu^2} \Bigg\{ -2 c_1^2 (\lambda - 2) (\mu^0 - \mu^2) \Big[ (\lambda -1) \sin 2 \beta^2 \thinspace \cos \left[ \lambda (\pi - \gamma) + 4 \gamma \right] \notag \\ & -\cos [\lambda (\pi - \gamma)] \sin [2 (1 - \lambda) \beta^2] \Big]
    - 2 c_2^2 (\lambda -2) (\mu^0 - \mu^2) \Big[ (\lambda -1) \sin 2 \beta^2 \thinspace \sin \left[ \lambda (\pi - \gamma) + 4 \gamma \right] \notag \\ & + \sin [\lambda (\pi - \gamma)] \sin [2 (1-\lambda) \beta^2] \Big]
    + c_3^2 (\lambda -1) (\mu^0 - \mu^2) \Big[ \lambda  \sin 4 \beta^2 \thinspace \cos \left[ \lambda (\pi - \gamma) + 4 \gamma \right] +2 \cos [\lambda (\pi - \gamma)] \sin 2 \lambda \beta^2 \Big] \notag \\ &
    - c_4^2 (\lambda -1)^2 (\mu^0 - \mu^2) \Big[ \lambda  \sin 4 \beta^2 \thinspace \sin \left[ \lambda (\pi - \gamma) + 4\gamma \right] - 2 \sin [\lambda (\pi - \gamma)] \sin 2 \lambda \beta^2 \Big]
    \Bigg\}
\end{align}
\end{small}
 
\begin{small}
\begin{align}
    & \int_{\Omega^{2p}} \phi_{,22} \overline{H}_{22}^{2*} r^{-\lambda} \thinspace d\textbf{x}' = |x_1|^{-\lambda} \frac{\pi (\lambda - 1) \csc \lambda \pi}{4 \mu^0 \mu^2} \Bigg\{ - 2 c_1^2 (\lambda -2) (\mu^0 - \mu^2) \Big[ (\lambda -1) \sin 2 \beta^2 \thinspace \sin \left[ \lambda (\pi - \gamma) + 4 \gamma \right] \notag \\ & +\sin [\lambda (\pi - \gamma)] \sin [2 (1 - \lambda) \beta^2] \Big]
    + 2c_2^2 (\lambda -2) (\mu^0 - \mu^2) \Big[(\lambda -1) \sin 2 \beta^2 \thinspace \cos \left[ \lambda (\pi - \gamma) + 4 \gamma \right] \notag \\ & -\cos [\lambda (\pi - \gamma)]  \sin [2 (1 - \lambda) \beta^2] \Big]
    + c_3^2 \Big[ -2 (\lambda -1) (\mu^0 - \mu^2) \sin [\lambda (\pi - \gamma)] \sin 2 \lambda \beta^2 \thinspace + (\lambda -1) \lambda  \sin 4 \beta^2 \thinspace (\mu^0 - \mu^2) \notag \\ & \times \sin \left[ \lambda (\pi - \gamma) + 4\gamma \right] +4 \sin \left[ \lambda (\pi - \gamma) + 2 \gamma \right] \Big( (\lambda -1) \sin 2 \beta^2 \thinspace +\sin [2(1-\lambda) \beta^2] \Big) \big( \mu^2 (1 - 2\nu^0) - \mu^0 (1 - 2\nu^2) \big) \Big] \notag \\ & 
    + c_4^2 \Big[ (\lambda -1) (\mu^0 - \mu^2) \Big( \lambda  \sin 4 \beta^2 \thinspace \cos \left[ \lambda (\pi - \gamma) + 4 \gamma \right] + 2 \cos [\lambda (\pi - \gamma)]  \sin (2 \beta^2  \lambda ) \Big) \notag \\ & + 4 \cos \left[ \lambda (\pi - \gamma) + \gamma \right] \big( (\lambda -1) \sin 2 \beta^2 \thinspace + \sin [2 (\lambda - 1) \beta^2]  \big) \big( \mu^2 (1 - 2\nu^0) - \mu^0 (1 - 2\nu^2) \big) \Big]
    \Bigg\}
\end{align}
\end{small}

\end{document}